\documentclass[12pt]{article}
\pdfoutput=1

\usepackage{cancel,slashed}
\usepackage{bbm}
\usepackage{amssymb}
\usepackage{amsfonts}
\usepackage{amsmath}
\usepackage{graphicx}
\usepackage{cite}

\usepackage{latexsym}
\usepackage{color}
\usepackage{xypic}
\usepackage{cancel,slashed}
\usepackage[hyperfootnotes=false,linktocpage]{hyperref}
\usepackage{color}
\usepackage{transparent}
\usepackage[hang,flushmargin]{footmisc}

\usepackage{blkarray}
\usepackage{multirow}

\newcommand{\bmat}{\left(\begin{array}}
\newcommand{\emat}{\end{array}\right)}

\def\Z{\mathbb{Z}}
\def\R{\mathbb{R}}
\def\C{\mathbb{C}}

\def\P{\mathbb{P}}
\def\CK {{\cal K}}

\def\ov{\overline}

\def\IM{\text{Im}\,}
\def\RE{\text{Re}\,}
\def\ov{\overline}
\def\1{{\bf 1}}
\def\2{{\bf 2}}
\def\3{{\bf 3}}
\def\4{{\bf 4}}
\def\6{{\bf 6}}
\def\OR{\Omega\mathcal{R}}

\def\targ#1#2{\genfrac{[}{]}{0pt}{}{#1}{#2}}
\def\targ2#1#2{\genfrac{}{}{0pt}{}{#1}{#2}}

\definecolor{mygr}{rgb}{0,0.6,0}
\definecolor{mygrey}{rgb}{0,0.1,0.2}
\definecolor{myblue}{rgb}{0,0.5,0.9}
\definecolor{myblue2}{rgb}{0,0.5,0.5}
\definecolor{myblue3}{rgb}{0,0.7,0.9}
\definecolor{myblue4}{rgb}{0,0.6,0.6}
\definecolor{myorange}{rgb}{1,0.5,0}
\definecolor{mypurple}{rgb}{0.6,0,1}
\definecolor{mygolden}{rgb}{1,0.8,0.2}
\definecolor{mycyan}{rgb}{0,1,1}
\definecolor{mymagenta}{rgb}{1,0,1}
\definecolor{mykiwi}{rgb}{0.8,1,0.5}
\definecolor{mybrown}{cmyk}{0.14, 0.42, 0.56, 0.2}
\definecolor{myturq}{cmyk}{0.99, 0, 0.2, 0.4}
\definecolor{myaubergine2}{cmyk}{0.4, 0.5, 0, 0.1}
\definecolor{myaubergine}{cmyk}{0.6,0.85,0,0}
\definecolor{CycleGreen}{cmyk}{0.52,0,1,0}
\definecolor{CycleBrown}{cmyk}{0, 0.4, 0.9, 0.2}

\DeclareFontFamily{U}{rcjhbltx}{}
\DeclareFontShape{U}{rcjhbltx}{m}{n}{<->rcjhbltx}{}
\DeclareSymbolFont{hebrewletters}{U}{rcjhbltx}{m}{n}

\DeclareMathSymbol{\lamed}{\mathord}{hebrewletters}{108}
\DeclareMathSymbol{\mem}{\mathord}{hebrewletters}{109}
\DeclareMathSymbol{\ayin}{\mathord}{hebrewletters}{96}
\DeclareMathSymbol{\tsadi}{\mathord}{hebrewletters}{118}
\DeclareMathSymbol{\qof}{\mathord}{hebrewletters}{113}
\DeclareMathSymbol{\resh}{\mathord}{hebrewletters}{114}
\DeclareMathSymbol{\pe}{\mathord}{hebrewletters}{112}
\DeclareMathSymbol{\pesofit}{\mathord}{hebrewletters}{80}
\DeclareMathSymbol{\samekh}{\mathord}{hebrewletters}{115}
\DeclareMathSymbol{\tav}{\mathord}{hebrewletters}{116}
\DeclareMathSymbol{\vav}{\mathord}{hebrewletters}{119}
\DeclareMathSymbol{\het}{\mathord}{hebrewletters}{120}
\DeclareMathSymbol{\yod}{\mathord}{hebrewletters}{121}
\DeclareMathSymbol{\zayin}{\mathord}{hebrewletters}{122}
\DeclareMathSymbol{\alephdot}{\mathord}{hebrewletters}{128}
\DeclareMathSymbol{\tsadisofit}{\mathord}{hebrewletters}{90}
\DeclareMathSymbol{\shin}{\mathord}{hebrewletters}{152}

\makeatletter
\newsavebox\myboxA
\newsavebox\myboxB
\newlength\mylenA

\newcommand*\xoverline[2][0.75]{%
\sbox{\myboxA}{$\m@th#2$}%
\setbox\myboxB\null
\ht\myboxB=\ht\myboxA%
\dp\myboxB=\dp\myboxA%
\wd\myboxB=#1\wd\myboxA
\sbox\myboxB{$\m@th\overline{\copy\myboxB}$}
\setlength\mylenA{\the\wd\myboxA}
\addtolength\mylenA{-\the\wd\myboxB}%
\ifdim\wd\myboxB<\wd\myboxA%
   \rlap{\hskip 0.5\mylenA\usebox\myboxB}{\usebox\myboxA}%
\else
    \hskip -0.5\mylenA\rlap{\usebox\myboxA}{\hskip 0.5\mylenA\usebox\myboxB}%
\fi}
\makeatother

\begin{document}
\pagestyle{plain}

\makeatletter
\@addtoreset{equation}{section}
\makeatother
\renewcommand{\theequation}{\thesection.\arabic{equation}}

\pagestyle{empty}
\rightline{ IFT-UAM/CSIC-18-115}
\vspace{0.5cm}
\begin{center}
\Huge{{Type IIA Flux Vacua with Mobile D6-branes}
\\[10mm]}
\large{Dagoberto Escobar, Fernando Marchesano, Wieland Staessens \\[10mm]}
\small{
Instituto de F\'{\i}sica Te\'orica UAM-CSIC, Cantoblanco, 28049 Madrid, Spain
\\[8mm]} 
\small{\bf Abstract} \\[5mm]
\end{center}
\begin{center}
\begin{minipage}[h]{15.0cm} 

We analyse type IIA Calabi-Yau orientifolds with background fluxes and D6-branes. The presence of D6-brane deformation moduli redefines the 4d dilaton and complex structure fields and complicates the analysis of such vacua in terms of the effective K\"ahler potential and superpotential. One may however formulate  the F-term scalar potential as a bilinear form on the flux-axion polynomials $\rho_A$ invariant under the discrete shift symmetries of the 4d effective theory. We express the conditions for Minkoswki and AdS flux vacua in terms of such polynomials, which allow to extend the analysis to include vacua with mobile D6-branes. We find a new, more general class of ${\cal N} = 0$ Minkowski vacua, which nevertheless present a fairly simple structure of (contravariant) F-terms. We compute the soft-term spectrum for chiral models of intersecting D6-branes in such vacua, finding a quite universal pattern.

%
%
%
%
%



\end{minipage}
\end{center}
\newpage
\setcounter{page}{1}
\pagestyle{plain}
\renewcommand{\thefootnote}{\arabic{footnote}}
\setcounter{footnote}{0}


\tableofcontents


\section{Introduction}

In the last two decades, models of particle physics and cosmology have thrived in the literature of string compactifications \cite{Ibanez:2012zz,Baumann:2014nda}. Two key ingredients that allowed to build this abundance of phenomenologically interesting models are background fluxes and D-branes \cite{Douglas:2006es,Blumenhagen:2006ci,Marchesano:2007de,Quevedo:2014xia}. On the one hand, fluxes allow to build more general compactifications with fewer and fewer moduli, in which supersymmetry can be spontaneously broken. On the other hand, D-branes allow to construct realistic chiral gauge sectors, and to localise their degrees of freedom in a particular region of the compactification. 

Needless to say, when combining both ingredients in a single compactification one must do it consistently. In  first instance this gives rise to constraints of topological nature, like avoiding Freed-Witten anomalies~\cite{Freed:1999vc,Maldacena:2001xj}. At a finer level of detail, one must ensure to capture the dynamical effects that branes and fluxes exert on each other, as well as on the rest of the compactification. In general, D-branes are known to create potentials for certain closed string moduli, and to contribute to the 4d light degrees of freedom with moduli of their own. Fluxes are known to be sourced by branes, and to create potentials for closed and open string moduli alike. Clearly, in order to properly describe the low energy effective dynamics all of these effects must be taken into account on equal footing.  

The same observations apply when searching for 4d type II orientifold flux vacua. Indeed, in the presence of D-brane moduli these must be considered simultaneously with the closed string moduli when minimising the potential, as opposed to adding them at a later stage of the analysis. This is manifest when using the standard ${\cal N}=1$ recipe for computing the F-term scalar potential in terms of a K\"ahler potential and superpotential. For instance, in the case of Calabi-Yau compactifications the presence of open string moduli redefines the 4d fields that appear in the K\"ahler potential, modifying the K\"ahler metrics non-trivially \cite{Jockers:2004yj,Grimm:2011dx,Kerstan:2011dy,Carta:2016ynn}. In particular, the factorised metric structure between K\"ahler and complex structure moduli, inherited from the unorientifolded ${\cal N}=2$ parent theory, is lost whenever open string moduli are considered~\cite{Carta:2016ynn}. This in turn implies that the no-scale properties of closed-string moduli potentials, a key ingredient to find certain classes of flux vacua \cite{Giddings:2001yu}, may be modified or even lost when open string moduli are present. 

In this work we analyse the properties of flux vacua in the presence of D-brane moduli. In particular we focus our attention on type IIA Calabi-Yau orientifolds with fluxes and D6-branes hosting open string moduli, dubbed mobile D6-branes in the following. For this class of compactifications it has recently been shown that the classical flux potential can be expressed in an alternative form to the standard Cremmer et al. F-term potential \cite{Herraez:2018vae}. In short, one can show that the scalar potential generated by fluxes and D6-branes takes the form $V = Z^{AB}\rho_A\rho_B$, with the index $A$ running over the fluxes of the compactification. Here $\rho_A$ are polynomials of the closed and open string axions of the 4d effective theory, and $Z^{AB}$ is an (inverse) metric that only depends on their saxionic partners. The polynomial coefficients in the different $\rho_A$ are topological quantities of the compactification, like triple intersection numbers or flux quanta, and such that the $\rho_A$ are invariant under the discrete shift symmetries of the 4d effective theory. As we show, one can easily rewrite the conditions for Minkowski and AdS vacua from the closed-string type IIA flux potential in this language, obtaining algebraic equations on the $\rho_A$ that reproduce known results in the literature \cite{DeWolfe:2005uu,Camara:2005dc,Palti:2008mg}. In this context, a particularly interesting set of solutions are the ${\cal N} =0$ Minkowski vacua analysed in \cite{Palti:2008mg}, mirror dual to those in \cite{Giddings:2001yu}.\footnote{In order to correctly establish the duality, one needs to include $\alpha$'-corrections into the analysis of the type IIA K\"ahler potential \cite{Palti:2008mg}. For simplicity, in this paper we ignore such corrections, since they complicate the discussion and do not affect our main results. We relegate the computations that take them into account to \cite{Escobar:2018rna}.} In terms of the bilinear form of the potential, the assumptions taken to construct these vacua imply that $V$ takes a bilinear semi-definite positive form, i.e. a sum of squares. When each of these squares vanishes, one recovers the algebraic conditions of the $\rho_A$ that correspond to such Minkowski vacua. 

One advantage of rewriting the potential as a bilinear is that one can easily incorporate the presence of D6-brane moduli. Indeed, in terms of the expression $V = Z^{AB}\rho_A\rho_B$ this only means that $A$ runs over more fluxes and that $Z$ and the $\rho$'s depend on more fields, but the structure of the potential remains the same. In this way, one may easily add mobile D6-branes to, e.g., the class of flux compactifications analysed in \cite{Palti:2008mg}. Remarkably, for this case we find that the potential can still be written as a sum of squares, which allows us to find new and more general classes of non-supersymmetric Minkowski vacua, now with the open string moduli stabilised at non-trivial vevs. 

While more intricate, flux vacua with mobile D6-branes share a lot of properties similar to their pure-closed-string counterparts. In particular, ${\cal N} =1$ AdS vacua and ${\cal N} =0$ Minkowski vacua have the same value for the 4d gravitino mass as in the absence of mobile D6-branes. In the case of ${\cal N} =0$ Minkowski vacua one can analyse the structure of their F-terms, which can be easily rewritten in terms of the $\rho_A$. Surprisingly, one finds that for these vacua there is only one kind of non-vanishing contravariant F-term, namely those corresponding to the complex structure moduli of the compactification. We therefore dub these ${\cal N} =0$ vacua as complex structure dominated, or CSD vacua for short. This F-term structure simplifies considerably the computation of soft terms developed at gauge sectors of the compactification, like 4d chiral fields localised at D6-brane intersections. In particular we find that to leading order soft terms depend on the gravitino mass and on the complex structure modular weight of the corresponding chiral field, in agreement via mirror symmetry with results in the type IIB literature \cite{Ibanez:2004iv,Camara:2004jj,Font:2004cx,Conlon:2006tj,Conlon:2006wz,Aparicio:2008wh}.

The paper is organised as follows. In section \ref{S:IIAORD6} we review the moduli space and effective theory of type IIA Calabi-Yau orientifolds with mobile D6-branes. In section \ref{S:IIAFluxVacua} we add background fluxes and consider the case without D6-brane moduli. We rewrite the flux potential as a bilinear of axion polynomials and use it to analyse how moduli are stabilised at ${\cal N}=0$ Minkowski and ${\cal N}=1$ AdS vacua. In section \ref{S:FluxVacD6branes} we consider compactifications with both fluxes and mobile D6-branes simultaneously, analyse their potential in bilinear form and use it to find a more general class of ${\cal N}=0$ Minkowski vacua, dubbed CSD vacua. We then turn to analyse the effective gravitino mass and the structure of soft terms on intersecting brane models for such vacua in section \ref{S:SUSYSoftTerms}. We discuss the validity of our approach in section \ref{S:ScalesIIA}, and finally draw our conclusions in section \ref{sec:con}. 

Several technical details have been relegated to the appendices. In appendix \ref{A:MetModSpace} we compute properties and relations for the K\"ahler metrics of type IIA Calabi-Yau compactifications with mobile D6-branes.  Appendix \ref{A:OpenBil} discusses the type IIA superpotential in the presence of mobile D6-branes, and how this allows to deduce the redefinition of complex structure moduli by open string moduli. Appendix \ref{A:TorOrb} describes the K\"ahler metrics for open string fields at D6-branes intersections.


\section{Type IIA Moduli Space with D6-branes}\label{S:IIAORD6}
Upon compactification of type IIA string theory on Calabi-Yau (CY) orientifold backgrounds~${\cal M}_6$, a residual ${\cal N}=1$ supersymmetry survives for the four-dimensional Minkowski spacetime $\R^{1,3}$. The standard IIA orientifold action $\Omega_p (-)^{F_L} {\cal R}$ consists of a worldsheet parity $\Omega_p$, a projection operator $(-)^{F_L}$ counting the number of spacetime fermions in the left-moving sector and an internal anti-holomorphic involution ${\cal R}$ acting on the K\"ahler 2-form $J$ and the CY 3-form $\Omega_3$ as follows:
\begin{equation}\label{Eq:CYGeoOrient}
{\cal R}(J) = - J, \qquad \qquad {\cal R} (\Omega_3) = \ov \Omega_3.
\end{equation}
By evaluating the parity of the RR-forms $C_1$ and $C_3$ and the NS 2-form $B_2$ under the discrete operations $\Omega_p$ and $(-)^{F_L}$, one can straightforwardly derive the parity of the orientifold-invariant states under the involution:\footnote{The RR-forms $C_1$ and $C_3$ are odd under $(-)^{F_L}$, while $C_3$ and $B_2$ are odd under $\Omega_p$.}
\begin{equation}
{\cal R} (C_1) = - C_1, \qquad {\cal R}(C_3) = C_3, \qquad {\cal R} (B_2) = - B_2.
\end{equation} 
The Kaluza-Klein (KK) zero modes of these massless $p$-forms recombine with the metric deformations to form the complex scalar components of ${\cal N}=1$ chiral multiplets. More explicitly, the $B_2$-axions fit together with the K\"ahler deformations into $h^{1,1}_-({\cal M}_6)$ K\"ahler moduli $T^a$ defined through 
\begin{equation}
J_c \equiv B + i\, e^{\frac{\phi}{2}} J = T^a \omega_a, \qquad \quad a \in \{1, \ldots, h^{1,1}_- \}.
\end{equation} 
Here the K\"ahler 2-form is expressed in the Einstein frame, while $\phi$ represents the ten-dimensional dilaton. The basis 2-forms $\ell_s^{-2}\omega_a$ correspond to harmonic representatives of the classes in $H^{2}_-({\cal M}_6, \Z)$ and are dimensionless due to the insertion of the string length $\ell_s = 2 \pi \sqrt{\alpha'}$. These zero modes parameterise the K\"ahler moduli space $\mathfrak{M}_K$ of the Calabi-Yau manifold, which exhibits a K\"ahler structure with K\"ahler potential:
\begin{equation}\label{Eq:KahlerPotKahlerMod}
K_T = - \log \left( \frac{4}{3} \int_{{\cal M}_6} e^{\frac{3\phi}{2}} J \wedge J \wedge J \right) =   - \log \left( \frac{i}{6} {\cal K}_{abc} (T^a - \ov T^a) (T^b - \ov T^b) (T^c - \ov T^c)  \right).
\end{equation} 
The triple intersection numbers ${\cal K}_{abc} = \ell_s^{-6} \int_{{\cal M}_6} \omega_a \wedge \omega_b \wedge \omega_c$ are moduli-independent integers, which allow to express the internal volume ${\cal V} = \frac{1}{6} \ell_s^{-6} \int_{{\cal M}_6}  J \wedge J \wedge J $ as a cubic polynomial in $t^a = \IM(T^a)$. The homogeneity of the function ${\cal G}_T = e^{- K_T}$ with degree three in the geometric  K\"ahler moduli $t^a$ is linked to the no-scale condition for the K\"ahler potential $K_T$:
\begin{equation}
(K_{T})_{a}(K_T)^{a\ov b} (K_{T})_{\ov b} = 3.
\label{noscaleK}
\end{equation}   

The ${\cal N}=1$ supergravity description of type IIA orientifold compactifications with K\"ahler potential \eqref{Eq:KahlerPotKahlerMod} is only reliable for sufficiently large internal volumes. Away from this limit, the K\"ahler potential is modified by the so-called $\alpha'$-corrections. In the regime of moderately large volumes in which the world-sheet instanton corrections can be neglected, the most relevant $\alpha'$-corrections are those that descend from $(\alpha')^3 R^4$ curvature corrections in the ten-dimensional supergravity action. In type IIA compactifications such corrections can be incorporated by means of a pre-potential on the K\"ahler moduli space. This results in a K\"ahler potential of the form
\begin{equation}\label{Eq:alphaCorrectedKaehlerPot}
K_T = - \log \left( \frac{4}{3} {\cal K}_{abc} t^a t^b t^c + 2 K^{(3)} \right) .
\end{equation} 
The presence of $K^{(3)} = - \frac{\zeta(3)}{(2\pi)^3}\,  \chi_{{\cal M}_6}$, with $\chi_{{\cal M}_6}$ the Euler characteristic of the compactification manifold, breaks the no-scale relation \eqref{noscaleK} for generic Calabi-Yau manifolds. As discussed in \cite{Palti:2008mg}, these $\alpha'$-corrections improve the stabilisation of K\"ahler moduli in the presence of background fluxes, allowing to fix them at moderately large values. For the sake of simplicity in this work we will mostly neglect the effect of such $\alpha$'-corrections, leaving their detailed analysis for \cite{Escobar:2018rna}, and only comment on how our results are modified when they are taken into account. 

The $C_3$-axions fit together with the complex structure deformations of the CY metric to form complexified scalars of ${\cal N}=1$ chiral multiplets. The identification of these so-called complex structure moduli is a bit more involved for type IIA orientifolds \cite{Grimm:2004ua}. Typically one first considers the unorientifolded ${\cal N}=2$ parent theory where the holomorphic three-form can be written as $\Omega_3 = {\cal Z}^\kappa \alpha_\kappa - {\cal F}_\kappa \beta^\kappa$, where $({\cal Z}^\kappa, {\cal F}_\lambda)$ are the complex periods with respect to the symplectic basis $(\alpha_\kappa, \beta^\lambda) \in H_3({\cal M}_6, \Z)$. Under the orientifold projection the basis decomposes in a basis of ${\cal R}$-even 3-forms $(\alpha_K, \beta^\Lambda) \in H_+^3({\cal M}_6,\Z)$ and ${\cal R}$-odd 3-forms $(\beta^K, \alpha_\Lambda) \in H_-^3({\cal M}_6,\Z)$, in which the orientifold action~(\ref{Eq:CYGeoOrient}) eliminates half of the degrees of freedom of the original complex periods in $\Omega_3$. To preserve the scale-invariance of the holomorphic three-form $\Omega_3 \rightarrow e^{- {\RE(h)}} \Omega_3$ in the orientifolded theory, a compensator field ${\cal C} \equiv  e^{- \phi} e^{\frac{1}{2}(K_{cs} - K_T)}$ is introduced, where $K_{cs} = - \log \left( i\ell_s^{-6} \int_{{\cal M}_6} \Omega_3 \wedge  \ov \Omega_3 \right)$ transforms as $K_{cs} \rightarrow K_{cs} +  2 \RE( h )$. The geometric components of the complex structure moduli are then encoded in the 3-form $\RE({\cal C} \Omega_3)$, which is turned into the complexified 3-form $\Omega_c$ by adding the RR-form $C_3$:
\begin{equation}
\Omega_c \equiv C_3 + i \, \RE ({\cal C} \Omega_3).
\end{equation}
The ${\cal N}=1$ complex structure moduli can now be properly defined in terms of the ${\cal R}$-odd 3-form basis:
\begin{equation}
N^K_\star = \ell_s^{-3} \int_{{\cal M}_6} \Omega_c \wedge \beta^K, \qquad  U_{\star \, \Lambda} = \ell_s^{-3} \int_{{\cal M}_6} \Omega_c \wedge \alpha_\Lambda.
\end{equation}
The complex structure moduli space $\mathfrak{M}_{cs}$ for an orientifold compactification maintains a K\"ahler structure with K\"ahler potential given by:
\begin{equation}\label{Eq:KaehlerPotCS1}
 K_Q = -2 \log \left( \frac{1}{4} \IM({\cal C} {\cal Z}^\Lambda) \RE({\cal C} {\cal F}_\Lambda) - \frac{1}{4} \RE({\cal C}{\cal Z}^K) \IM({\cal C} {\cal F}_K) \right) = -  \log(e^{-4 D}),
\end{equation}
where $D$ is the four-dimensional dilaton defined through $e^{D} \equiv \frac{e^{\phi}}{\sqrt{ {\cal V}}}$. As is well-known, the periods ${\cal F}_K$ and ${\cal F}_\Lambda$ ought to be considered as homogeneous functions of degree one in the periods ${\cal Z}^K$ and ${\cal Z}^\Lambda$, implying that the function ${\cal G}_Q= e^{-K_Q/2}$ is a homogeneous function of degree two in $n^K_\star = \IM(N^K_\star)$ and $u_{\star \, \Lambda} = \IM(U_{\star\, \Lambda})$. Consequently, the K\"ahler potential~$K_Q$ for the complex structure moduli satisfies a no-scale condition of the form:
\begin{equation}
(K_Q)_{\kappa} (K_Q)^{\kappa \ov \lambda} (K_Q)_{\ov \lambda}= 4,
\end{equation}  
where the indices $\kappa$ and $\lambda$ sum over all complex structure moduli $N^K_\star$ and $U_{\star\, \Lambda}$.

\begin{center}
{\bf Mobile D6-branes}
\end{center}

Calabi-Yau spaces equipped with an anti-holomorphic involution ${\cal R}$ come with a set of special Lagrangian ({\it SLag}) three-cycles $\Pi$ subject to the geometric conditions:
\begin{equation} \label{Eq:SLAG3Cycles}
J\big|_\Pi = 0, \qquad \IM \Omega_3 \big|_{\Pi}  = 0.
\end{equation}
When modding out the anti-holomorphic involution to obtain the Calabi-Yau orientifold, the fixed loci $\Pi_{O6}$ under the involution ${\cal R}$ define the locations of O6-planes wrapping one or more special Lagrangian ({\it SLag}) three-cycles. The O6-plane RR-charges have to be cancelled along the internal directions, which can be achieved by introducing D6-branes wrapping {\it SLag} three-cycles $\Pi_a$ and filling out the four-dimensional spacetime. In the absence of background fluxes, the RR tadpole cancellation conditions can be recast into constraints in homology 
\begin{equation}\label{Eq:RRTadpoleD6}
\sum_\alpha N_\alpha ([\Pi_\alpha] + [{\cal R}\Pi_\alpha]) - 4 [\Pi_{O6}] = 0,
\end{equation}      
where $N_a$ indicates the number of D6-brane in each stack $a$. Since the 3-form $\Omega_3$ is the natural calibration form for the ({\it SLag}) 3-cycles on Calabi-Yau 3-folds, the three-cycle volume for the supersymmetric D6-branes can be expressed as follows \cite{Becker:1995kb} for a chosen point in the Calabi-Yau moduli space: 
\begin{equation}\label{Eq:SLAGVolume}
{\cal C}\Omega\big|_{\Pi_\alpha} = e^{- \frac{\phi}{4}} d{\rm Vol}\big|_{\Pi_\alpha}.
\end{equation}  
Whenever a {\it SLag} three-cycle $\Pi_\alpha$ can be continuously deformed along a normal vector without violating the special Lagrangian condition, a D6-brane wrapped around it can change its embedding or position along its transverse internal directions. As a result it has a non-trivial moduli space, parametrised by one or more open string moduli. More precisely, if we pick a set $\{X_j\}$ of normal vectors  to $\Pi_\alpha$ which preserve the {\it SLag} condition,\footnote{The preservation of the {\it SLag} condition along direction $X_i$ can be expressed through the corresponding Lie-derivative, i.e.~${\cal L}_{X_i} J \big|_{\Pi_\alpha} = 0 =  {\cal L}_{X_i} \Omega_3 \big|_{\Pi_\alpha}$.} McLean's theorem states that the one-forms $\iota_{X_i} J\big|_{\Pi_\alpha}$ are proportional to harmonic one-forms in ${\cal H}^1(\Pi_a,\Z)$. In this sense, a generic, infinitesimal deformation $X = \ell_s X_i \varphi^i$ is expected to yield $b_1(\Pi_\alpha)$ different position moduli $\varphi^i$. In order to properly define the chiral superfields for the open string moduli, we introduce instead the basis of harmonic two-forms $\ell_s^{-2} \rho^i \in {\cal H}^2(\Pi_\alpha, \Z)$, to which we each assign an open string modulus as follows:
\begin{equation}\label{Eq:OpenStringModDef}
\Phi_\alpha^i = {-} \frac{1}{\ell_s^4} \int_{\Pi_\alpha} \left(\frac{\ell_s^2}{\pi} A  - \iota_X J_c\right) \wedge \rho^i =  T^b (\eta_{\alpha\, b})^i{}_{j} \varphi^j - \theta^i_\alpha = \hat  \theta_\alpha^i + i\, \phi^i_\alpha  .
\end{equation}       
In this expression $A$ represents the D6-brane gauge potential, which reduces along the internal directions to Wilson line degrees of freedom $\theta^i_\alpha$. By introducing the constant parameters $(\eta_{\alpha\, b})^i{}_{j}$,
\begin{equation}  
(\eta_{\alpha\, b})^i{}_{j} = \frac{1}{\ell_s^3} \int_{\Pi_\alpha} \iota_{X_j} \omega_b \wedge \rho^i,
\end{equation}
the implicit dependence of the open string moduli on the K\"ahler moduli has been extracted in the right hand side of (\ref{Eq:OpenStringModDef}). When extending the infinitesimal deformation to a finite deformation of the {\it SLag} three-cycle, the functional dependence of the open string moduli on the position moduli $\varphi^i$ will no longer be linear and higher order powers in the position moduli have to be computed through a normal coordinate expansion. Roughly speaking, the term $(\eta_{\alpha\, b})^i{}_{j} \varphi^j$ in (\ref{Eq:OpenStringModDef}) then has to be replaced by a generic function $f_{\alpha\, b}^i (\varphi)$, which can further depend on the closed string geometric moduli $t^a$, $n^K$ and $u_\Lambda$~\cite{Carta:2016ynn}.  The open string modulus then reads
\begin{equation}\label{Eq:OpenStringModDefFin}
\Phi_\alpha^i = T^a f_{\alpha\, a}^i - \theta^i_\alpha = \hat  \theta_\alpha^i + i\, \phi^i_\alpha  .
\end{equation}  

When introducing mobile D6-branes into the type IIA orientifold compactification, the full moduli space of the compactification does generically not correspond to a direct product of the closed string moduli space $\mathfrak{M}_K\times \mathfrak{M}_{cs}$ with the open string moduli space. For small field fluctuations around a chosen point in the moduli space, one can adopt the approach in which the calibration conditions for {\it SLag} three-cycles (\ref{Eq:SLAG3Cycles}) and (\ref{Eq:SLAGVolume}) are evaluated in a particular background with frozen closed string moduli. As such, only those small deformations of the D6-brane that respect the {\it SLag} conditions with respect to this background have to be considered. Even in this approach, the reduction of the ten-dimensional theory induces kinetic mixing between open string and bulk moduli, such that a redefinition of the complex structure moduli is necessary to identify the proper ${\cal N}=1$ chiral superfields. Following the reasoning of appendix \ref{A:RedCSM}, one deduces the following field redefinition for the complex structure moduli:
\begin{equation}\label{Eq:RedefComplexStructure}
N^K = N^K_\star + \frac{1}{2} \sum_\alpha (g_{\alpha i}^K \theta_\alpha^i  - T^a  {\bf H}_{\alpha\, a}^{K}), \qquad  U_\Lambda = U_{\star \,\Lambda} + \frac{1}{2} \sum_\alpha ( g_{\alpha\, \Lambda\, i} \theta^i_\alpha -  T^a  {\bf H}_{\alpha\, \Lambda \, a}),
\end{equation}   
where the real functions ${\bf H}_{\alpha\, a}^{K}$ and ${\bf H}_{\alpha\, \Lambda\, a}^{K} $ are defined through the expressions:
\begin{equation}\label{Eq:HFunction1}
\partial_{\phi_\beta^i} (t^a {\bf H}_{\alpha\, a}^{K} ) =  \delta_{\alpha \beta}\, g_{\alpha i}^K, \qquad  
\partial_{\varphi^j} g_{\alpha i}^K = \ell_s^{-3} \int_{\Pi_\alpha}  \iota_{X_j} \beta^K \wedge \zeta_i,
\end{equation}
and
\begin{equation}\label{Eq:HFunction2}
\partial_{\phi_\beta^i} ( t^a {\bf H}_{\alpha\, \Lambda\, a} ) =  \delta_{\alpha \beta}\, g_{\alpha \, \Lambda\, i}, \qquad \partial_{\varphi^j} g_{\alpha\, \Lambda\, i} = \ell_s^{-3} \int_{\Pi_\alpha}  \iota_{X_j} \alpha_\Lambda \wedge \zeta_i.
\end{equation}
with $\phi_\alpha^i = \IM(\Phi_\alpha^i)$. The functions $g_{\alpha i}^K$ and $g_{\alpha\, \Lambda\, i}$ are chain integrals that allow to write the two-forms $\iota_{X} \beta^K$ and $\iota_{X} \alpha_\Lambda$ on the three-cycle $\Pi_\alpha$ in terms of the more appropriate basis of quantised harmonic two-forms~$\rho^i$, related to the quantified one-forms~$\zeta_i$ as $\ell_s^{-3}\int_{\Pi_\alpha} \zeta_i \wedge \rho^j = \delta_i{}^j$.
As argued in appendix A of~\cite{Carta:2016ynn}, the functions $g_{\alpha i}^K$ and $g_{\alpha\, \Lambda\, i}$ are homogeneous functions of degree zero in the moduli $\{ t^a, n^K, u_\Lambda, \phi_\alpha^i \}$, which implies that also the functions ${\bf H}_{\alpha\, a}^{K} $ and ${\bf H}_{\alpha\, \Lambda\, a}^{K}$ are homogeneous functions of degree zero in the respective moduli. The field redefinition also has repercussions for the K\"ahler potential~(\ref{Eq:KaehlerPotCS1}) depending on the complex structure moduli. More precisely, the function ${\cal G}_Q(n^k,u_\Lambda)$ hidden in the K\"ahler potential~(\ref{Eq:KaehlerPotCS1}), as inherited from the  ${\cal N}=2$ Calabi-Yau compactifications, remains a homogeneous function of degree two in the geometric moduli, but has to be rewritten in terms of the redefined complex structure moduli and the open string moduli:
\begin{equation}  \label{Eq:OpenClosedKaehlerPotential} 
K_Q = - 2 \log \left[ {\cal G}_Q\left( n^K +\frac{1}{2} t^a \sum_\alpha {\bf H}^K_{\alpha\,a},  u_\Lambda +\frac{1}{2} t^a \sum_\alpha {\bf H}_{\alpha\, \Lambda\,\,a} \right)\right].
\end{equation}
An immediate consequence of the moduli redefinition is the explicit dependence of the function ${\cal G}_Q$ on all geometric moduli $\{ t^a,  n^K,  u_\Lambda, \phi_\alpha^i \}$, such that the moduli space obviously no longer factorises for type IIA orientifold compactification with D6-branes. Ignoring $\alpha'$-corrections for $K_T$, the combined K\"ahler potentials $K_T + K_Q = - \log ({\cal G}_T {\cal G}_Q^2) $ still satisfy a no-scale condition:
\begin{equation}
K_A K^{A \ov B} K_{\ov B} = 7,
\end{equation}   
where the indices $A$ and $B$ sum over all closed and open string moduli, in line with the conventions used in appendix~\ref{A:MetModSpace} to express some revelant properties of the full K\"ahler potential.

\section{The Type IIA Flux Landscape}\label{S:IIAFluxVacua}

If type IIA orientifold compactifications ought to provide for vacuum solutions exhibiting the well-known features of our universe, the various open and closed geometric moduli have to be stabilised with sufficiently high masses. Fortunately, the richness of background NS- and RR-fluxes in type IIA offers a controlled, perturbative method to deal with moduli stabilisation for all closed string moduli \cite{Louis:2002ny,Kachru:2004jr,Grimm:2004ua,DeWolfe:2005uu,Camara:2005dc}. This section is devoted to rewriting the known flux stabilisation of closed string moduli in a formalism in which the shift symmetries for the axions are manifest. In this language, also the stabilisation of open string moduli can be dealt with in a much more elegant way, as  we will argue in the next section.
 
\subsection{Fluxes, Freed-Witten anomalies and Axion Polynomials}\label{Ss:FluxFWAPol}
From a ten-dimensional perspective, the democratic formulation of type IIA superstring theory offers the best starting point to capture the physics of string backgrounds with fluxes and D-branes. In this description, all RR gauge potentials $C_{2p-1}$ with $p = 1,2,3,4,5$ are treated on equal footing and are grouped together in a polyform ${\bf C} = C_1 + C_3 + C_5 + C_7 + C_9$. Similarly to the NS 2-form $B_2$ they appear in the bosonic part of the type IIA supergravity action~(\ref{Eq:10dIIA}) through their associated field strengths ${\bf G} = G_0 + G_2 + G_4 + G_6 + G_8 + G_{10}$ and $H_3$. Apart from their equations of motion, these field strengths also have to satisfy the Bianchi identities, which in the absence of D-branes or other external sources read:
\begin{equation}\label{Eq:BianchiIdent}
d (e^{-B_2} \wedge {\bf G} ) = 0, \qquad d H_3 = 0
\end{equation} 
On a compact manifold, the Bianchi identities imply that the polyforms $e^{-B_2} \wedge {\bf G}$ and NS 3-form $H_3$ are closed forms, such that these field strengths can be decomposed in terms of exact  and harmonic forms:\footnote{The chosen form of the Bianchi identities allows to extract the solution for the RR field strengths in terms of the {\bf A}-basis instead of the {\bf C}-basis, which are related to each by a simple $B_2$-transformation, i.e.~${\bf A} = {\bf C}\wedge e^{-B_2}$.} 
\begin{equation}\label{Eq:FieldStrengths}
{\bf G} = e^{B_2}\wedge (d {\bf A} + \ov{\bf G}), \qquad H_3 = dB_2 + \ov H_3. 
\end{equation}
At the same time, the Bianchi identities written in this form allow to argue for the quantisation of the associated Page charge~\cite{Marolf:2000cb},
\begin{equation}
\frac{1}{\ell_s^{2p-1}} \int_{\pi_{2p}} d A_{2p-1} + \ov G_{2p} \in \Z, \qquad  \frac{1}{\ell_s^{2}}\int_{\pi_3} dB_2 + \ov H_3 \in \Z,
\end{equation} 
arising through integration over the non-trivial homological cycles $\pi_{2p}$ with $p=1,2,3$ and~$\pi_3$. The quantisation argument itself relies on the consistency of the field theory on a probe $(2p-2)$-brane wrapping a $(2p-1)$ cycle inside one of the non-trivial homological cycles $\pi_{2p}$ or $\pi_3$. 
In the absence of localised sources such as D-branes, the gauge potentials~${\bf A}$ are well-defined everywhere and  the non-trivial harmonic parts $\ov{G}_{2p}$ with $p=0,1,2,3$ and $\ov H_3$ with legs along the compactification manifold capture the quantised flux. For orientifold compactifications the internal $p$-cycles have to comply with the orientifold projection, such that    
the background flux can be characterised by virtue of flux quanta $(m,m^a,e_a,e_0)$:\footnote{We adhere to the conventions of~\cite{Carta:2016ynn} for the sign of the fluxes.}
\begin{equation} 
\ell_s {\ov G}_0 = m, \qquad \frac{1}{\ell_s} \int_{\tilde \pi^a}{\ov G}_2 = m^a, \qquad \frac{1}{\ell_s^3}\int_{\pi_a} {\ov G}_4 = e_a, \qquad  \frac{1}{\ell_s^5}\int_{{\cal M}_6} {\ov G}_6 = e_0,
\end{equation}   
with $\tilde \pi^a \in H_2^{-}({\cal M}_6, \Z)$ and $\pi_a \in H_4^{+}({\cal M}_6, \Z)$. The internal RR-fluxes ${\ov{\bf G}}$ are known to generate a perturbative superpotential for the K\"ahler moduli \cite{Gukov:1999ya,Taylor:1999ii}:
\begin{equation} \label{Eq:KahlerSuperpotential}
\ell_s W_T = \frac{1}{\ell_s^5} \int_{{\cal M}_6} {\ov{\bf G}} \wedge e^{ J_c} = e_0 + e_a T^a + \frac{1}{2} {\cal K}_{abc} m^a T^b T^c + \frac{m}{6} {\cal K}_{abc} T^a T^b T^c\, .
\end{equation}
The NS 3-form flux ${\ov H}_3$ on the other hand threads the ${\cal R}$-odd three-cycles $(B^K,A_\Lambda) \in H^-_3({\cal M}_6, \Z)$, which are the de Rham duals to the ${\cal R}$-odd three-forms $(\beta^K, \alpha_\Lambda)$ introduced earlier. Similar as for the RR-fluxes, the quantised Page charge for the NS-flux background can be expressed in terms of the integer flux quanta $(h_K, h^\Lambda)$:
\begin{equation} \label{Eq:CpxstSuperpotential}
 \frac{1}{\ell_s^2} \int_{B^K} {\ov H}_3 =  - h_K, \qquad   \frac{1}{\ell_s^2} \int_{A_\Lambda} {\ov H}_3 = h^\Lambda . 
\end{equation}
The NS-fluxes generate in turn a linear superpotential for the complex structure moduli:
\begin{equation} 
\ell_s W_{Q} = \frac{1}{\ell_s^5} \int_{{\cal M}_6} \Omega_c \wedge {\ov H}_3 = h_K N_\star^K  + h^\Lambda U_{\star\, \Lambda}\, . 
\end{equation}
The combination of RR and NS-fluxes suffices to generate a four-dimensional F-term scalar potential for the geometric moduli $(t^a, n_\star^K, u_{\star\, \Lambda})$ and closed string axions $(b^a,\xi_\star^K, \xi_{\star\, \Lambda})$, whose precise shape exhibits a remarkable bilinear form factorising into a geometric moduli part, an axion part and a flux part~\cite{Bielleman:2015ina,Herraez:2018vae}. Namely, we have a structure of the form
\begin{equation}\label{Eq:VFbil0}
V_F  = \frac{1}{8\kappa_4^2}  \rho_A(b, \xi_\star) \, Z^{AB}(t,n_\star,u_\star) \, \rho_B\, (b, \xi_\star),
\end{equation}  
where the $\rho_A$ depend on the flux quanta and the axions, and $Z^{AB}$ only on saxions. In the language of standard ${\cal N}=1$ supergravity this sort of factorisation also exhibits itself in the superpotential, which can be expressed as the product   
\begin{equation}\label{Eq:SuperPotFactForm}
 W_T + W_Q = \vec{\Pi}^t  \cdot \vec{\rho} , \qquad \qquad  \ell_s\vec{\rho}  = (R^{-1})^t \cdot \vec{q},
\end{equation} 
of a saxion vector $\vec{\Pi}^t (t^a, n_\star^K, u_{\star\, \Lambda}) = (1, i t^a, -\frac{1}{2} {\cal K}_{abc} t^b t^c,$ $-\frac{i}{3!} {\cal K}_{abc} t^a t^b t^c, i n^K_\star, i u_{\star \Lambda} )$ and an axion vector $\vec{\rho}$ of components $\rho_A$. The latter is given in terms of an $(2 h_-^{11} + h^{21} + 3)\times (2 h_-^{11} + h^{21} + 3)$ dimensional  axion rotation matrix, 
\begin{equation}
R(b^a,\xi_\star^K, \xi_{\star\, \Lambda}) = \left(\begin{array}{cccccc} 
1 & 0 & 0 & 0 &0 & 0 \\
-b^a & \delta^a{}_b & 0 & 0& 0& 0\\
\frac{1}{2} {\cal K}_{abc}  b^b b^c & - {\cal K}_{abc} b^c & \delta^a{}_{b} & 0 & 0 & 0 \\ 
-\frac{1}{3!} {\cal K}_{abc} b^a b^b b^c & \frac{1}{2} {\cal K}_{abc}  b^b b^c & - b^a & 1 & 0& 0  \\
- \xi^K_\star & 0&  0& 0& \delta^K{}_L& 0 \\
-\xi_{\star \Lambda}& 0 & 0& 0& 0& \delta^{\Sigma}{}_\Lambda
 \end{array}\right),
\end{equation}
and a charge vector $\vec{q}$ consisting of the quantised fluxes, i.e.~$\vec{q} = (e_0, e_a, m^a, m, h_K,  h^\Lambda)^t$. The factorised form of the superpotential enables to expose the multi-branched structure of the vacua for the closed string axions: the periodic shift symmetry of the axions leaves the action, potential and superpotential invariant provided that the flux quanta $\vec{q}$ are shifted simultaneously. Formally, the shift symmetries of the closed string axions are generated by the nilpotent matrices $P_a$, $P_K$ and $P^\Lambda$,  
\begin{equation}\label{Eq:NilGenShiftCS}
\begin{array}{cc}
\multirow{2}{*}{ $P_a = \left( \begin{array}{cccccc} 0 &  -\vec{\delta}_a^t & 0 & 0& 0 &0\\
0& 0 & -{\cal K}_{abc} & 0 & 0 &0 \\
0& 0& 0 & -\vec{\delta}_a & 0& 0\\
0& 0& 0& 0& 0& 0\\
0& 0& 0& 0& 0& 0\\
0&0&0&0&0& 0
  \end{array}\right),$ }
&  P_K = \left( \begin{array}{cccccc}
  0 & 0& 0 & 0& - \vec{\delta}_K^t  &0\\
0& 0 & 0 & 0 & 0 &0 \\
0& 0& 0 &0 & 0& 0\\
0& 0& 0& 0& 0& 0\\
0& 0& 0& 0& 0& 0\\
    \end{array}\right),\\ 
 & P^\Lambda = \left( \begin{array}{cccccc} 
    0 & 0& 0 & 0& 0&  -(\vec{\delta}^L)^t  \\
0& 0 & 0 & 0 & 0 &0 \\
0& 0& 0 &0 & 0& 0\\
0& 0& 0& 0& 0& 0\\
0& 0& 0& 0& 0& 0\\
  \end{array}\right), 
  \end{array} 
\end{equation}
which mutually commute among each other. As such, the axion rotation matrix can be expressed in terms of these matrices through exponentiation:
\begin{equation} \label{Eq:RotMatrixClosedStringAxions}
R^t(b^a  ,\xi^K_\star, \xi_{\star \Lambda}) = e^{b^a P_a + \xi^K_\star P_K + \xi_{\star \Lambda} P^\Lambda}.
\end{equation}
The matrix notation also allows to express elegantly the invariance of the theory under the axionic shift symmetries, which acts on the axion rotation matrix as:
\begin{equation}\label{Eq:AxShiftSymmRotMat}
(R^{-1})^t (b^a +  r^a ,\xi^K_\star + \varpi^K, \xi_{\star \Lambda} + \varpi_\Lambda ) = (R^{-1})^t (b^a, \xi^K_\star,  \xi_{\star \Lambda})\, \cdot e^{ - r^a P_a  - \varpi^K P_K - \varpi_\Lambda P^\Lambda},
\end{equation}
with $r^a, \varpi^K, \varpi_\Lambda \in \Z$. The invariance of the superpotential is manifest provided the charge vector transforms as,
\begin{equation}\label{Eq:ChargeTransShift}
\vec{q}\quad \rightarrow \quad  e^{ r^a P_a  + \varpi^K P_K +  \varpi_\Lambda P^\Lambda} \vec{q} \, .
\end{equation}
The shift symmetry implies the existence of a set of gauge-invariant axion polynomials $\ell_s \vec{\rho} \equiv (R^{-1})^t \cdot \vec{q}$, whose explicit component forms are given by,
\begin{equation}\label{Eq:CSgaugeinvariantrho}
\begin{array}{lcl}
\ell_s \rho_0 &=& e_0 + e_a b^a + \frac{1}{2} {\cal K}_{abc} m^a b^b b^c + \frac{m}{6} {\cal K}_{abc} b^a b^b b^c + h_K \xi_\star ^K  + h^\Lambda \xi_{\star \Lambda} , \\
\ell_s \rho_a &=& e_a + {\cal K}_{abc}  m^b b^c + \frac{m}{2} {\cal K}_{abc} b^b b^c, \\
\ell_s \tilde \rho^a &=& m^a + m b^a , \\
\ell_s \tilde \rho &=& m , \\
\ell_s \hat \rho_K & = & h_K, \\
\ell_s \hat \rho^\Lambda & = &  h^\Lambda. 
\end{array}
\end{equation}
As shown in \cite{Escobar:2018rna}, all of the above statements also hold when taking into account the effect of curvature $\alpha'$-corrections. Indeed, one can still define gauge-invariant axion polynomials that generalise the expressions above.

The invariance under the axion shift symmetries is not coincidental, but relies microscopically on the cancellation of Freed-Witten anomalies for four-dimensional strings in the presence of background fluxes \cite{Herraez:2018vae}. More concretely, each of the axions $(b^a, \xi^K_\star, \xi_{\star\Lambda})$ can be Hodge-dualised in four dimensions to its corresponding two-form coupling to four-dimensional strings. In type IIA backgrounds these axionic strings arise from NS5-branes wrapping the Poincar\'e-dual four-cycles ${\rm PD}(\omega_a)$ ($b$-type axionic strings) and D4-branes wrapping the Poincar\'e-dual three-cycles ${\rm PD}(\alpha_K)$ and ${\rm PD}(\beta_\Lambda)$ respectively ($\xi$-type axionic strings). In the presence of background RR-flux $\ov G_{2p}$ the $b$-type axionic strings develop a Freed-Witten anomaly in case $G_{2p}\big|_{{\rm PD}(\omega_a)}$ is non-trivial in cohomology, which can be mediated by emitting a $D(6-2p)$-brane wrapping the $(4-2p)$-cycle in the Poincar\'e dual class of $G_{2p}\big|_{{\rm PD}(\omega_a)}$. Similarly, the $\xi$-type axionic strings resolve the Freed-Witten anomaly in the presence of $H_3$-flux by emitting $D2$-branes, as summarised in table~\ref{Tab:StringsDomainWalls}. The emitted D-branes form four-dimensional domain walls bounded by axionic strings that separate vacua in the axion moduli space with different RR- and/or NS-fluxes \cite{BerasaluceGonzalez:2012zn}. In this respect the domain walls are unstable under nucleation of holes bounded by axionic strings, which allows the axions to cross the domain wall by virtue of a monodromy generated by the matrices~$P_a$, $P_K$ and $P^\Lambda$. Under the axion monodromies the flux quanta will shift as prescribed in~(\ref{Eq:ChargeTransShift}), such that both effects cancel each other out and all vacua for the axions are equivalent. It is also straightforward to verify that the field strengths in (\ref{Eq:FieldStrengths}) remain invariant under such shift symmetries, which can be seen as a particular subset of gauge transformations.

\begin{table}[h]
\begin{center}
\hspace*{-0.6in}
\begin{tabular}{|c||c||c||c||c|}
\hline
 \multicolumn{2}{|c||}{\bf String} &{\bf Flux} &  \multicolumn{2}{|c|}{\bf Domain Wall}\\
  \hline
  Axion & Brane Set-up & type &  Brane Set-up & Rank\\
  \hline\hline
$B_2 = b^a \omega_a$ & NS5 on $[\pi_a] \in H_4^+({\cal M}_6,\Z)$& $\ov G_0 = m$ & D6 on $[\pi_a]$ & $m$ \\
$B_2 = b^a \omega_a$ & NS5 on $[\pi_a] \in H_4^+({\cal M}_6,\Z)$& $\ov G_2 = m^a \omega_a$ & D4 on $[{\rm PD}({\ov G_2} \wedge\omega_a)]$ & $\int_{\tilde \pi^a}\omega_c = {\cal K}_{abc} m^b  $ \\
$B_2 = b^a \omega_a$ & NS5 on $[\pi_a] \in H_4^+({\cal M}_6,\Z)$& $\ov G_4 = e_a \tilde \omega^a$ & D2 at point in ${\cal M}_6$  & $\int_{\pi_a} \ov{G}_4 = e_a $ \\
\hline 
$C_3 = \xi^K_\star \alpha_K $ & D4 on $[B^K] \in H_3^-({\cal M}_6, \Z)$  & $\ov H_3 = h_K \beta^K $  & D2 at point in ${\cal M}_6$   & $ \int_{B^K} {\ov H}_3 = - h_K$ \\
$C_3 = - \xi_{\star \Lambda} \beta^\Lambda $ & D4 on $[A_\Lambda] \in H_3^-({\cal M}_6, \Z)$ & $\ov H_3 = h^\Lambda \alpha_\Lambda $  & D2 at point in ${\cal M}_6$   &  $\int_{A_\Lambda} {\ov H}_3 =  h^\Lambda$ \\
\hline
\end{tabular}
\caption{Summary of 4d axionic strings with their respective attached domain walls arising from Dp- and NS5-branes wrapping internal cycles on a Calabi-Yau manifold with internal flux. \label{Tab:StringsDomainWalls}}
\end{center}
\end{table}

\subsection{Type IIA Flux Vacua}\label{Ss:IIAFluxVacua}
An important implication of non-trivial background fluxes concerns the stabilisation of closed string moduli at non-vanishing vacuum expectation values. The factorisation of the perturbative superpotential induced by NS- and RR-fluxes, encourages us to understand how moduli stabilisation respects this factorisation and can be formulated in terms of the axion polynomial language. This is precisely the goal of this section, where two well-known examples from the literature, i.e.~non-supersymmetric Minkowski vacua and supersymmetric AdS vacua, are used as toy examples to highlight the general idea.
\begin{center}
{\bf Non-Supersymmetric Minkowski Flux Vacua}
\end{center}
The imaginary self dual (ISD) flux vacua of type IIB can be T-dualised to type IIA flux vacua~\cite{Camara:2005dc,Palti:2008mg} for which all RR-fluxes are switched on and the NS 3-form flux is turned on along only one $\OR$-odd three-cycle. 
Following the symplectic basis choice of~\cite{Palti:2008mg} in which the complex structure moduli $\{N_\star^K\}_{K\neq 0}$ are projected out, we can assume that the four-dimensional dilaton $N^0_\star =S_\star = \xi^0_\star + i\, \IM(S_\star)$ factorises from the other complex structure moduli $U_{\star \Lambda}$ in the K\"ahler potential:
\begin{equation}\label{Eq:KahPotISDFact}
K_Q^{\rm ISD} = -  \log \left[-i (S_\star-\ov S_\star)\right]   -  2 \log \left[ \tilde {\cal G}_Q (u_{\star \Lambda}) \right],
\end{equation}
where $\tilde {\cal G}_Q (u_{\star \Lambda})$ is a homogeneous function of degree $3/2$ with an implicit dependence on the geometric moduli $u_{\star \Lambda}$. More precisely, the functional dependence of~$\tilde {\cal G}_Q$ can be expressed in terms of the rescaled periods $\IM(Z^\Lambda) \equiv 2 \RE({\cal C} Z^0)^{-1/2} \IM({\cal C} {\cal Z}^\Lambda)$ and upon inverting the relation $u_{\star \Lambda} = \partial_{\IM(Z^\Lambda)} \tilde {\cal G}_Q$ the function $ \tilde {\cal G}_Q$ can in principle be written in terms of the geometric moduli $u_{\star \Lambda}$.  
Finally, if we further assume that the only non-vanishing NS-flux is supported along the $\OR$-odd three-form $\beta^0$, we obtain the generic superpotential for ISD fluxes, 
\begin{equation}\label{Eq:IIAISDSuperPot}
\ell_s W_{\rm ISD} = h_0 S_\star + e_0 + e_a T^a + \frac{1}{2} {\cal K}_{abc} m^a T^b T^c + \frac{m}{6} {\cal K}_{abc} T^a T^b T^c,
\end{equation}
which in terms of the axion polynomials reads
\begin{equation}\label{Eq:IIAISDSuperPotaxion}
W_{\rm ISD} = i s_\star \hat{\rho}_0 + \rho_0 + i  t^a \rho_a - \frac{1}{2} {\cal K}_{a} \tilde{\rho}^a  - \frac{i}{6} {\cal K} \tilde{\rho}.
\end{equation}

Given the specific form of the K\"ahler potential \eqref{Eq:KahPotISDFact}, the F-term scalar potential takes the form
\begin{eqnarray}\label{Eq:FTermScalarPotISD}
V_F & = & \frac{e^{K}}{\kappa_4^2} \left[ K^{A \bar B} F_A F_{\bar B}    - 3 \big| W \big|^2   \right] \\ \nonumber
& = & \frac{e^{K}}{\kappa_4^2} \left[ K^{T^a \bar T^b} F_{T^a} F_{\bar T^b}  + K^{S_\star \bar S_\star} F_{S_\star} F_{\bar S_\star} + K^{U_{\star \Lambda} \bar U_{\star \Lambda}} F_{U_{\star \Lambda}} F_{\bar U_{\star \Lambda}}  - 3 \big| W \big|^2   \right] \\ \nonumber
& = & \frac{e^{K}}{\kappa_4^2} \left[ K^{T^a \bar T^b} F_{T^a} F_{\bar T^b}  + K^{S_\star \bar S_\star} F_{S_\star} F_{\bar S_\star}  \right]
\end{eqnarray}  
where in the last line we have used that by assumption $F_{U_{\star \Lambda}} = K_{U_{\star \Lambda}}  W$ and the no-scale relation $K^{U_{\star \Lambda} \bar U_{\star \Lambda}} K_{U_{\star \Lambda}} K_{\bar U_{\star \Lambda}} = 3$ that arises from \eqref{Eq:KahPotISDFact}. Therefore, for these kind of vacua we recover a positive semidefinite flux potential whose absolute minima are reached whenever $F_{S_\star} = F_{T^a} =0$.  In general, the factorisable form~(\ref{Eq:SuperPotFactForm}) of the ISD flux superpotential enables us to simplify the F-terms for the dilaton $S_\star$ and K\"ahler moduli and express them entirely in terms of geometric moduli and the gauge-invariant axion polynomials~(\ref{Eq:CSgaugeinvariantrho}). Focusing first on the F-term for the dilaton we obtain\footnote{To simplify the expressions, we use ${\cal K} = {\cal K}_{abc} t^a t^ b t^c$, ${\cal K}_{a} = {\cal K}_{abc}t^b t^c$, ${\cal K}_{ab} = {\cal K}_{abc} t^c$.} 
\begin{equation}\label{Eq:FtermDilCS}
F_{S_\star}= - i\, \partial_{s_\star} {W}_{\rm ISD} + \frac{i}{2s_\star} {W}_{\rm ISD} = \frac{1}{2 s_\star} \left( i\rho_0 - t^a \rho_a - \frac{i}{2} {\cal K}_a \tilde \rho^a + \frac{1}{6} {\cal K} \tilde \rho + s_\star \hat \rho_0  \right),
\end{equation}
where we have used the holomorphicity of the superpotential, i.e.~$\partial_{\ov S_\star} {W}_{\rm ISD} = 0$, to obtain a first order derivative purely with respect to the four-dimensional dilaton~$s_\star = \IM(S_\star)$. Similar considerations can be made for the F-terms of the K\"ahler moduli, 
\begin{equation}\label{Eq:FtermKaeMod}
\begin{array}{rcl}
F_{T^a} &=& - i\, \partial_{t^a}   {W}_{\rm ISD} + \frac{3i {\cal K}_a}{2{\cal K}}  {W}_{\rm ISD}  \\
  & =&  \rho_a + i {\cal K}_{ab} \tilde \rho^b + \frac{3i {\cal K}_a}{2{\cal K}} \left( \rho_0 + i t^b \rho_b - \frac{1}{2} {\cal K}_b \tilde \rho^b + \frac{i}{6} {\cal K} \tilde \rho  + i s_\star \hat \rho_0 \right).
 \end{array}
\end{equation}
Finally, a more elegant polynomial expression in terms of the geometric moduli and axion polynomials is found in the form of the linear combination $t^a F_{T^a}$, 
\begin{equation}\label{Eq:LinComFTermKaeModCS}
t^a F_{T^a} = \frac{3i}{2} \rho_0 - \frac{1}{2} t^a \rho_a + \frac{i}{4} {\cal K}_a \tilde \rho^a - \frac{3}{2} \left( \frac{1}{6} {\cal K} \tilde \rho + s_\star \hat \rho_0 \right).
\end{equation}
When considering the expressions (\ref{Eq:FtermDilCS}), (\ref{Eq:FtermKaeMod}) and (\ref{Eq:LinComFTermKaeModCS}) as polynomials in $t^a$ simultaneously, the vanishing of the F-terms implies that their coefficients ought to vanish:
\begin{equation}\label{Eq:GenEqISD}
\tilde \rho^a = 0, \quad  \rho_a =0,  \quad  \frac{1}{6} {\cal K} \tilde \rho + s_\star \hat \rho_0 = 0,  \quad  \rho_0 = 0.
\end{equation} 
As we discuss in section \ref{Ss:FluxVacuaD6-branesModStab}, one can easily rederive these conditions from the bilinear form of the potential \eqref{Eq:VFbil0}.
The first set of equations $\tilde \rho^a = 0$ stabilise the K\"ahler axions in terms of the RR flux quanta:
\begin{equation}\label{Eq:KAISD}
b^a = - \frac{m^a}{m},
\end{equation}
while the second set of equations $\rho_a = 0$ represent a set of constraints on the flux quanta:
\begin{equation}\label{Eq:ConstraintFluxesISD}
 2 m e_a - {\cal K}_{abc} m^b m^c  = 0.
\end{equation} 
Upon imposing these set of relations, the third and last equation stabilise the four-dimensional dilaton $\IM(S_\star)$ and its axion $\xi^0_\star$ respectively in terms of flux quanta and the K\"ahler moduli:
\begin{equation}\label{Eq:DilatonISDFlux}
h_0 s_\star = - \frac{m}{6} {\cal K}_{abc} t^a t^b t^c, \qquad h_0 \xi^0_\star = - \frac{1}{m^2}\left( e_0 m^2 - \frac{1}{6} {\cal K}_{abc} m^a m^b m^c  \right).
\end{equation} 

Thus, the analysis of the F-terms for the dilaton and K\"ahler moduli in terms of the axion polynomials allows to easily extract the generic ISD vacua~(\ref{Eq:GenEqISD}), which reproduce the results of section~3.1 in~\cite{Palti:2008mg} represented by the last four relations (\ref{Eq:KAISD})-(\ref{Eq:DilatonISDFlux}). In these vacua, the saxionic parts of the K\"ahler moduli and complex structure moduli remain unstabilised partly due to the no-scale symmetry for the complex structure moduli $U_{\star\Lambda}$.
  This no-scale symmetry combined with the vanishing F-terms for the dilaton and K\"ahler moduli imply a vanishing F-term scalar potential at the ISD vacuum, which corresponds to a non-supersymmetric Minkowski spacetime in four dimensions. 
Supersymmetry is then spontaneously broken by the non-vanishing F-terms of the complex structure moduli $U_{\star \Lambda}$, given that the on-shell superpotential for ISD flux vacua is non-vanishing for arbitrary Romans mass, 
\begin{equation}\label{WISD}
\langle W_{ISD} \rangle = - \frac{i}{3} {\cal K} \tilde \rho. 
\end{equation}  
The structures of the F-terms in the complex structure moduli sector will be further analysed in section~\ref{S:SUSYSoftTerms}, in conjunction with the structures of flux-induced soft terms.

As argued in~\cite{Palti:2008mg}, a more compelling moduli stabilisation scenario is achieved upon inclusion of the $\alpha'$-corrections that deform the K\"ahler potential from \eqref{Eq:KahlerPotKahlerMod} to \eqref{Eq:alphaCorrectedKaehlerPot}. Indeed, in that case one is also able to fix the saxionic component of the K\"ahler moduli. One can see that the presence of such $\alpha'$-corrections is compatible with the simplified form of the scalar potential \eqref{Eq:FTermScalarPotISD}, and that the conditions $F_{S_\star} = F_{T^a} =0$ are equivalent to the following relations among axion polynomials \cite{Escobar:2018rna}
\begin{equation}
\begin{array}{l@{\hspace{0.4in}}l}
\ov \rho_0 = 0, & \frac{1}{6} {\cal K} {\tilde \rho} + s_\star {\hat \rho}_0 = \tilde{\rho} K^{(3)}  \frac{\frac{1}{6} \CK + K^{(3)}}{\frac{4}{3}\CK - K^{(3)}} ,\\
 {\tilde\rho}^a = 0, & \ov \rho_a = \tilde{\rho} K^{(3)}  \frac{\frac{3}{2} \CK_a}{\frac{4}{3}\CK - K^{(3)}}  .
\end{array}
\end{equation}
Here $\ov \rho_0$,  $\ov \rho_a$ are the appropriate redefinition of the axion polynomials $\rho_0$, $\rho_a$ in the presence of $\alpha'$-corrections.\footnote{More precisely, they correspond to substitute $e_0, e_a \rightarrow \ov e_0, \ov e_a$ in such polynomials, where $\ov e_0, \ov e_a \in \mathbb{Z}$ stand for a redefinition of RR flux quanta due to $\alpha'$-corrections of order lower than $K^{(3)}$. See \cite{Escobar:2018rna} for more details.} Since $\ov \rho_a \neq 0$, we do not need to impose the analogue of \eqref{Eq:ConstraintFluxesISD}, and the K\"ahler moduli are stabilised at moderately large, finite values. In particular one finds that the saxions $t^a$ minimise the potential energy at
\begin{equation}
\label{Eq:KahlerModuliAlphaCorr}
 {\cal K}_a   =  \frac{(4 {\cal K} - 3 K^{(3)})}{9m^2 K^{(3)}} \left( 2 m \ov e_a  - {\cal K}_{abc} m^b m^c \right).
\end{equation}
in agreement with the results of section 4.2 in \cite{Palti:2008mg}. 
%

\begin{center}
{\bf Supersymmetric Anti-de Sitter Flux Vacua}
\end{center}
As soon as the no-scale structure for the complex structure moduli $U_{\star \Lambda}$ is broken by the presence of additional NS-fluxes, both the complex structure moduli and K\"ahler moduli can be stabilised to non-trivial values simultaneously. Considering all RR- and NS-fluxes turned on in a type IIA flux compactification, the geometric moduli, K\"ahler axions and one linear combination of complex structure axions can be stabilised supersymmetrically or non-supersymmetrically, yielding a four-dimensional Anti-de Sitter vacuum~\cite{DeWolfe:2005uu,Camara:2005dc}. Once more, the axion polynomials provide a very elegant way to find supersymmetric vacua by analysing the F-terms:
\begin{equation}\label{Eq:SUSYAdSFterms}
\begin{array}{rcl}
F_{N^K_\star}& =&  \hat \rho_K -   i \frac{\IM({\cal C F}_K)}{2{\cal G}_Q} \left( W_T + W_Q\right),  \\
F_{U_{\star \Lambda}}&=&  \hat \rho^\Lambda +   i \frac{ \IM({\cal C Z}^\Lambda)}{2{\cal G}_Q} \left( W_T + W_Q\right),\\
F_{T^a} & =& \rho_a +i {\cal K}_{ab} \tilde \rho^b - \frac{1}{2} {\cal K}_a \tilde\rho  + \frac{3i}{2} \frac{{\cal K}_a}{{\cal K}}\left( W_T + W_Q\right).
\end{array}
\end{equation}
In order to solve for the full set of vanishing F-terms, let us first sum up strategically the complex structure F-terms
\begin{equation}
\sum_{K=0}^{h^{}} n^K_\star F_{N^K_\star} + \sum_{\Lambda=0}^{h^{}} u_{\star \Lambda} F_{U_{\star \Lambda}} = \sum_{K=0}^{h^{}} \hat \rho _K n^K_\star +  \sum_{\Lambda=0}^{h^{}} \hat \rho^\Lambda u_{\star \Lambda}  +2i \left( W_T + W_Q\right) = 0,   
\end{equation}
such that the real part and complex part lead to two separate conditions:
\begin{equation}\label{Eq:SUSYAdS1}
\rho_0 - \frac{1}{2} {\cal K}_{a} \tilde \rho^a = 0, \qquad n^K_\star \hat \rho_K + u_{\star\Lambda} \hat \rho^\Lambda  = \frac{1}{3} {\cal K} \tilde \rho - 2 t^a \rho_a.
\end{equation}
Also the F-terms of the K\"ahler moduli can be summed up as
\begin{equation}
t^a F_{T^a} = \frac{5}{2}  t^a \rho_a - \frac{3}{4} {\cal K} \tilde \rho + \frac{3i}{2} \rho_0 + \frac{i}{4} {\cal K}_a \tilde \rho^a ,
\end{equation}
leading to two more conditions for vanishing F-terms:
\begin{equation}\label{Eq:SUSYAdS2}
 \frac{3}{2} \rho_0 + \frac{1}{4} {\cal K}_a \tilde \rho^a = 0, \qquad \frac{5}{2}  t^a \rho_a - \frac{3}{4} {\cal K} \tilde \rho = 0.
\end{equation}
Combining all four relations allows us to express the stabilisation conditions for the moduli in terms of the axion polynomials:
\begin{equation}\label{Eq:SUSYAdSVacuaRho}
\rho_0 = 0, \quad \tilde \rho^a = 0, \quad \rho_a =\frac{3}{10} {\tilde \rho}\, {\cal K}_{a}.
\end{equation}
The first condition expresses the fact that a linear combination of complex structure axions is stabilised, while the second condition stabilises the K\"ahler axions:
\begin{equation}\label{axionstab}
h_K \xi_\star^K + h^\Lambda \xi_{\star\, \Lambda} = -\frac{e_0 m^2 - m e_a m^a + \frac{1}{3} {\cal K}_{abc} m^a m^b m^c}{m^2}, \qquad b^a = - \frac{m^a}{m}.
\end{equation}
The third condition stabilises the geometric part of the K\"ahler moduli in terms of the fluxes. Inserting the identified solutions back into the F-terms for the complex structure moduli enables to write down the stabilisation conditions for the complex structure moduli in terms of their ``dual" periods and the overall volume ${\cal K}$:
\begin{equation}\label{cpxFterm}
{\cal G}_Q \frac{\hat \rho_K}{\IM({\cal C F}_K)} = - {\cal G}_Q \frac{\hat \rho^\Lambda}{\IM({\cal C Z}^\Lambda)} =    \frac{1}{15}  \tilde \rho\, {\cal K}.
\end{equation}
To arrive at these relations,  we imposed the vacuum expectation value for the superpotential in supersymmetric AdS vacua, which can be obtained by imposing the vacuum constraints on the axion polynomials:
\begin{equation}
\langle W_{\rm AdS} \rangle = - \frac{2i}{15} {\cal K} \tilde \rho.
\end{equation}
One can check that the conditions \eqref{Eq:SUSYAdSVacuaRho} and \eqref{cpxFterm} are equivalent to the vanishing F-term conditions  \eqref{Eq:SUSYAdSFterms}. 
Hence, the vacuum relations found in~\cite{DeWolfe:2005uu} for supersymmetric AdS vacua can be derived very elegantly by virtue of the axion polynomial language.  

Similarly to the ISD flux vacua, the supersymmetric AdS vacua are only realised in the presence of a non-vanishing Romans' mass~$m\neq0$, and are modified when taking into account the effect of $\alpha'$-corrections. This time the modification is less dramatic, because the classical scenario already stabilises all moduli, but their value will be nevertheless shifted from their previous value. In terms of axion polynomials, we have that the vacuum relations \eqref{Eq:SUSYAdSVacuaRho} become
\begin{equation}\label{Eq:SUSYAdSVacuaRhocorr}
\ov\rho_0 = 0, \quad \tilde \rho^a = 0, \quad \ov\rho_a =\frac{3}{10}  {\tilde \rho}\, {\cal K}_{a} \left[  \frac{\CK + 3K^{(3)}}{ \CK + \frac{3}{5} K^{(3)}}\right],
\end{equation}
and \eqref{cpxFterm} turn into
\begin{equation}\label{cpxFtermAdS}
{\cal G}_Q \frac{\hat \rho_K}{\IM({\cal C F}_K)} = - {\cal G}_Q \frac{\hat \rho^\Lambda}{\IM({\cal C Z}^\Lambda)} =    \frac{1}{15}  \tilde \rho\,  \left(\CK + \frac{3}{2} K^{(3)}\right) \left[\frac{\CK-3K^{(3)}}{\CK + \frac{3}{5}K^{(3)}}\right].
\end{equation}
Notice that these deformations shift the value of the saxions but do not affect the stabilisation of the axions, whose vevs still satisfy \eqref{axionstab}.

\begin{center}
{\bf Cosmological Constant in Flux Vacua}
\end{center}
Both classes of vacuum solutions above have been obtained by solving for vanishing F-terms in the four-dimensional ${\cal N}=1$ supergravity description. For non-vanishing F-terms, the vacuum solutions have to be determined by minimising the F-term scalar potential, computed from the closed string K\"ahler potential and superpotential,
\begin{equation}\label{Eq:FTermScalarPot}
V_F = \frac{e^{K}}{\kappa_4^2} \left[ (\partial_A W + K_A W) K^{A \ov B} (\partial_{\ov B} \ov W + K_{\ov B} \ov W  )  - 3 \big| W \big|^2   \right], 
\end{equation}  
where summation over all closed string moduli is assumed. Alternatively, one may consider the bilinear form of the potential 
\begin{equation}\label{Eq:VFbil}
V_F  = \frac{1}{8\kappa_4^2}  \rho_A(b, \xi_\star) \, Z^{AB}(t,n_\star,u_\star) \, \rho_B\, (b, \xi_\star),
\end{equation}  
where the vector of  axion polynomials is given by $\vec{\rho} = \left(\rho_0, \rho_a, \tilde \rho^a, \tilde \rho, \hat \rho_K, \hat \rho^\Lambda \right)$ and the saxion-dependent (inverse) metric $Z^{AB}$ reads
\begin{equation}\label{Eq:ZABMetricFull}
Z^{AB} = 8 e^{K} \left(\begin{array}{cccccc} 4 & \\
& K^{a \ov b} \\
&&\frac{4}{9} {\cal K}^2 K_{a \ov b} \\
&&& \frac{1}{9} {\cal K}^2 & \frac{2}{3} {\cal K} n^I_\star &   \frac{2}{3} {\cal K} u_{\star \Lambda}\\ 
&&&  \frac{2}{3} {\cal K} n^J_\star & K^{IJ} & K^{I \Sigma} \\
&&& \frac{2}{3} {\cal K} u_{\star \Sigma}& K^{\Lambda J} & K^{\Lambda \Sigma}
 \end{array} \right).
\end{equation} 
Instead of solving for vanishing F-terms, vacuum configurations can be determined more generically by requiring that the first order derivatives of the scalar potential with respect to the moduli vanish. Due to the properties of the rotation matrix~(\ref{Eq:RotMatrixClosedStringAxions}) the constraint equations for the axionic directions can be rephrased as orthogonality conditions between the vector $\vec{\rho}$ and its descendants $P_a \vec{\rho}$, $P_K \vec{\rho}$ or $P^\Lambda \vec{\rho}$:   
\begin{equation}
\begin{array}{rcl}
\vec{\rho}^{\, T} Z^{-1} P_a \vec{\rho} &=&4 \rho_0 \rho_a +  K^{c\ov d} {\cal K}_{dab} \rho_c \tilde \rho^b - \frac{4}{9}  K_{\ov b a} {\cal K}^2 \tilde \rho^b \tilde \rho  = 0, \\
\vec{\rho}^{\, T} Z^{-1} P_K \vec{\rho} &=& 4 \rho_0 \hat \rho_K  = 0, \\
\vec{\rho}^{\, T} Z^{-1} P^\Lambda \vec{\rho} &=& 4 \rho_0 \hat \rho^\Lambda = 0.
\end{array}
\end{equation}
These three constraint equations are solved simultaneously for $\rho_0 = 0$ and $\tilde \rho^a = 0$: two constraints on the axion polynomials that are common among the ISD flux vacua and supersymmetric AdS flux vacua, and are responsible for stabilising a linear combination of complex structure axions and all K\"ahler axions in terms of the flux quanta. The three constraint equations have to be supplemented by the vacuum conditions arising along the geometric moduli directions. In the case of ISD flux vacua, the vacuum conditions for the geometric moduli correspond to setting the following equations to zero,
\begin{equation}\label{Eq:VacConSaxSP}
\begin{array}{lcl}
\vec{\rho}^{\, T}  \partial_{t^a} (Z^{-1}) \vec{\rho} & =&  \vec{\rho}^{\, T} (Z^{-1}) \vec{\rho}\; \partial_{t^a} K + 8 e^{K} \left[ \rho_c \partial_{t^a} K^{c \ov d} \rho_d + {\cal K}_a \tilde \rho \left( \frac{2}{3} {\cal K} \tilde \rho + 4 s_\star \hat \rho_0  \right) \right]   , \\
\vec{\rho}^{\, T}  \partial_{s_\star} (Z^{-1}) \vec{\rho} & =&  \vec{\rho}^{\, T} (Z^{-1}) \vec{\rho}\; \partial_{s_\star} K + 8 e^{K} \hat \rho_0 \left[ \frac{4}{3} {\cal K} \tilde \rho + 8s_\star \hat \rho_0 \right]  ,\\
\vec{\rho}^{\, T}  \partial_{u_{\star \Lambda}} (Z^{-1}) \vec{\rho} & =&  \vec{\rho}^{\, T} (Z^{-1}) \vec{\rho}\; \partial_{u_{\star \Lambda}} K  ,
\end{array}
\end{equation}
where the solutions $\rho_0 = 0$ and $\tilde \rho^a = 0$ to the axion constraint equations have already been taken into account on the right-hand side. One can see that the derivative $ \partial_{u_{\star\, \Lambda}} K$ is proportional to the quotient $\IM({\cal C Z}^\Lambda)/{\cal G}_Q$, and therefore a homogeneous function of $u_{\star\, \Lambda}$ of degree $-1$. As a result, the third relation in~\eqref{Eq:VacConSaxSP} vanishes in regions of the moduli space where the supergravity approximation is no longer valid, i.e. vanishing three-cycle volumes ($\IM({\cal C Z}^\Lambda)=0, \forall \Lambda$) or three-cycles with infinite volumes, unless the four-dimensional vacuum energy proportional to $\vec{\rho}^{\, T} (Z^{-1}) \vec{\rho}$ vanishes for the compactification. The vacuum conditions for the K\"ahler moduli sector and 4d dilaton in Minkowski vacua further lead to the constraints $\rho_a = 0$ and $\frac{1}{6} {\cal K} \tilde \rho + s_\star \hat \rho_0 = 0 $, which complete the set of constraint equations~(\ref{Eq:GenEqISD}) for the ISD flux vacua. Clearly, the axion polynomials jargon allows for a more systematic search of perturbative flux vacua, but it also reveals that many such flux vacua are related to each other through the shift symmetries (\ref{Eq:ChargeTransShift}) and should therefore not be counted as independent vacua. 

Identifying the constraints on the axion polynomials for a particular vacuum configuration also allows to determine the perturbative value of the cosmological constant. To extract information about the cosmological constant from the axion polynomials, it is insightful to rewrite the inverse metric $Z^{AB}$ in~\eqref{Eq:ZABMetricFull}  into a block-diagonal form,  
\begin{equation}\label{Eq:DiagInMetZ}
Z^{AB} = 8 e^{K} {\rm diag} \left(4, K^{a\ov b}, \frac{4}{9} {\cal K}^2 K_{a \ov b}, - \frac{{\cal K}^2}{3}, \left(\begin{array}{cc}  K^{IJ} & K^{I\Sigma} \\ K^{\Lambda J} &  K^{\Lambda \Sigma} \end{array}\right)  \right),
\end{equation} 
by rotating the axion polynomials to a new basis of axion polynomials:
 \begin{equation}\label{Eq:AxionPolyBasisDiag}
 \vec{\rho}_{\rm new} = \Big(\rho_0, \rho_a, \tilde \rho^a, \tilde \rho,  \hat \rho_K - \frac{i{\cal K}}{3} K_{N^K_\star}   \tilde \rho, \hat \rho^\Lambda - \frac{i{\cal K}}{3} K_{U_{\star \Lambda}}   \tilde \rho \Big),
\end{equation}
where we have used the homogeneity of the complex structure K\"ahler potential~\eqref{Eq:KaehlerPotCS1}. Taking into account the expression for the F-terms of the complex structure moduli~\eqref{Eq:SUSYAdSFterms}, the vector~\eqref{Eq:AxionPolyBasisDiag} can be reinterpreted in a slightly more suggestive way:
\begin{equation}
 \vec{\rho}_{\rm new} = \left(\rho_0, \rho_a, \tilde \rho^a, \tilde \rho, F_{N^K_\star} - K_{N^K_\star} \left( W_T + W_Q + \frac{i}{3} {\cal K} \tilde \rho \right), F_{U_{\star \Lambda}} - K_{U_{\star \Lambda}} \left( W_T + W_Q + \frac{i}{3} {\cal K} \tilde \rho \right)\right).
\end{equation}
The virtue of this new basis of axion polynomials lies in the possibility to understand each vacuum as a positive, null-like or negative norm with respect to the diagonalised inverse metric. The ISD flux vacua (with vanishing dilaton and K\"ahler moduli F-terms) for instance are characterised by the constraint equations~\eqref{Eq:GenEqISD} on the axion polynomials and are represented by the vector $\vec{\rho}_{\rm new}   = \tilde \rho \left(0,0,0,1,0, - \frac{i{\cal K}}{3} K_{U_{\star \Lambda}}   \tilde \rho \right) = \left( 0,0,0,\tilde \rho, 0, F_{U_{\star \Lambda}} \right)$. This vector corresponds to a null-like vector with respect to the metric $Z^{AB}$, in line with the vanishing vacuum energy for non-supersymmetric Minkowski vacua.\footnote{In fact, as we will see in the next section, the choice of K\"ahler potential \eqref{Eq:KahPotISDFact} together with $\hat\rho_\Lambda =0$ implies a positive semi-definite scalar potential minimised by this $\vec{\rho}_{\rm new}$.}
SUSY AdS vacua, on the other hand, have vanishing F-terms in all sectors. From the relations~\eqref{Eq:SUSYAdSVacuaRho} we obtain the vector $\vec{\rho}_{\rm new} = \left(0, \rho_a, 0, \tilde \rho, - K_{N^K_\star} \left( W_T + W_Q + \frac{i}{3} {\cal K} \tilde \rho \right), -  K_{U_{\star \Lambda}} \left( W_T + W_Q + \frac{i}{3} {\cal K} \tilde \rho \right)  \right)$ $ 
=  \tilde \rho \left( 0, \frac{3}{10} {\cal K}_a, 0, 1,  - \frac{i}{5} {\cal K} K_{N^K_\star}, -\frac{i}{5} {\cal K}  K_{U_{\star \Lambda}}  \right)$, which forms a negative norm vector whose length corresponds to the negative cosmological constant for the AdS minimum:
\begin{equation}
\langle V_{F} \rangle_{\rm AdS} =  - 3 \frac{ e^{K}}{\kappa_4^2} \left( \frac{2}{15} \tilde \rho {\cal K} \right)^2. 
\end{equation}

The same strategy can be applied for $\alpha'$-corrected type IIA flux vacua. There, the analysis is technically more involved, because $\alpha'$-corrections introduce several off-diagonal entries on the block-diagonal matrix \eqref{Eq:ZABMetricFull}, connecting previously independent blocks \cite{Escobar:2018rna}. Nevertheless, by analysing the axion polynomial vectors one obtains a similar picture, with the above quantities modified in terms of $K^{(3)}$. For instance, for SUSY AdS vacua one obtains a negative cosmological constant corresponding to
\begin{equation}
\langle V_{F} \rangle_{\rm AdS} =  - 3 \frac{ e^{K}}{\kappa_4^2} \left( \frac{2}{15} \tilde \rho \right)^2 \left(\CK + \frac{3}{2} K^{(3)}\right)^2 \left[\frac{\CK-3K^{(3)}}{\CK + \frac{3}{5}K^{(3)}}\right]^2 ,
\end{equation}
where $K = K_T + K_Q$ is computed with $K_T$ given by \eqref{Eq:alphaCorrectedKaehlerPot}.

\section{Perturbative Flux Vacua with Mobile D6-branes}\label{S:FluxVacD6branes}

Backgrounds with localised sources such as D6-branes and O6-planes provide a much more intricate picture for type IIA compactifications with fluxes. First, as reviewed in section~\ref{S:IIAORD6}, they introduce a kinetic mixing between open, K\"ahler and complex structure moduli. Second, some open string moduli for mobile D6-branes will contribute to the superpotential through a bilinear coupling with the K\"ahler moduli\footnote{In non-K\"ahler compactifications contributions quadratic on the D6-brane moduli also appear \cite{Marchesano:2006ns}.}
\begin{equation}\label{Eq:OpenClosedSuperPotential}
W= W_T +W_Q + W_{D6}^0 +  \ell_s^{-1}  \sum_\alpha  \Phi_\alpha^i ( n_{{F}\, i}^\alpha  -  n_{a\, i}^\alpha T^a).
\end{equation}
Here $W_T$ is given by \eqref{Eq:KahlerSuperpotential} and $W_Q$ by \eqref{Eq:CpxstSuperpotential} with the replacement $\{N_\star^K, U_{\star\, \Lambda}\} \rightarrow \{N^K, U_{\Lambda}\}$. In addition, $\Phi_\alpha^i$ stands for the $i^{th}$ open string modulus of the D6-brane $\alpha$, defined in terms of a reference three-cycle $\Pi_\alpha^0$. At this reference point in open string field space $\Phi_\alpha^i=0$ and the open string contribution to $W$ is given by $W_{D6}^0$. Also, because there is a non-trivial two-cycle on $\Pi_\alpha$ per each open string modulus we can define two topological invariants. One is $n_{{F}\, i}$,  the corresponding quantum of worldvolume flux and the other is $n_{\, i}^\alpha$, the homological decomposition of this two-cycle in the bulk. The microscopic justification of this superpotential was derived in \cite{Marchesano:2014iea} and is reviewed in Appendix \ref{A:OpenBil}, where we also refer the reader for a detailed definition of all these quantities. 

\subsection{Axion Polynomials with Open String States}\label{Ss:APOSS}

The particular (bi)linear structure of the last term in \eqref{Eq:OpenClosedSuperPotential} allows for the factorisation of the superpotential \eqref{Eq:SuperPotFactForm} into geometric moduli, axions and flux quanta to go through in the presence of open string moduli as well:
\begin{equation}
\ell_s\left( W - W^0_{D6}\right) =  \vec{\Pi}^t  \cdot (R^{-1})^t \cdot \vec{q},
\end{equation}
where the saxion vector $\vec{\Pi}^t (t^a, n^K, u_\Lambda, \phi^i_\alpha) = (1, i t^a, -\frac{1}{2} {\cal K}_{abc} t^b t^c,$ $-\frac{i}{3!} {\cal K}_{abc} t^a t^b t^c, i n^K,  i u_\Lambda,$  $i \phi^i_\alpha, t^a \phi^i_\alpha)$ is now extended with the open string moduli $\phi^i_\alpha$, the charge vector $\vec{q}= (e_0, e_a, m^a,$ $m, h_K, h^\Lambda, n_{Fi}^\alpha, {n_{ai}^\alpha})^t$ is extended with the open string quanta $(n_{Fi}^\alpha, n_{ai}^\alpha)$ and the axion rotation matrix has to be enlarged with open string axions $\hat \theta_\alpha^i$: 
\begin{equation}
R(b^a,\xi^K, \xi_\Lambda, \hat \theta^i_\alpha) = \left(\begin{array}{cccccccc} 
1 & 0 & 0 & 0 &0 & 0 & 0 & 0 \\
-b^a & \delta^a{}_b & 0 & 0& 0& 0 & 0& 0\\
\frac{1}{2} {\cal K}_{abc}  b^b b^c & - {\cal K}_{abc} b^c & \delta^a{}_{b} & 0 & 0 & 0 & 0& 0 \\ 
-\frac{1}{3!} {\cal K}_{abc} b^a b^b b^c & \frac{1}{2} {\cal K}_{abc}  b^b b^c & - b^a & 1 & 0& 0 &0 &0 \\
- \xi^K & 0&  0& 0& \delta^K{}_L& 0 & 0& 0 \\
-\xi_\Lambda& 0 & 0& 0& 0& \delta^{\Sigma}{}_\Lambda &0 &0\\
\hat \theta^i_\alpha & 0 & 0& 0 & 0 & 0& \delta^i{}_j & 0 \\
\hat \theta^i_\alpha b^a  & \hat \theta_\alpha^i \delta^a{}_b & 0& 0 & 0 & 0& b^a \delta^i{}_{j} & \delta^i{}_{j} \delta^a{}_b  \\
 \end{array}\right).
\end{equation}
Also in the presence of open string axions, the rotation matrix can be generated by a set of nilpotent matrices through exponentiation:
\begin{equation}
R^t(b^a, \xi^K, \xi_\Lambda, \hat\theta^i_\alpha) = e^{b^a \P_a + \xi^K \P_K +\xi_\Lambda \P^\Lambda +\hat \theta^i_\alpha \P^\alpha_i}, 
\end{equation}
with the shift-generating matrices $(\P_a,\P_K, \P^\Lambda)$ forming the natural extension of their closed string counterparts (\ref{Eq:NilGenShiftCS}):
\begin{equation}
P_a \rightarrow { \P}_a =  \left(\begin{array}{ccc}  P_a & \vec{0}^t &  \vec{0}^t \\
\vec{0} & 0& \vec{\delta}_j^t\\
\vec{0}  & 0 &0
  \end{array} \right), 
  \, P_K \rightarrow {\P}_K = \left(\begin{array}{ccc} P_K & \vec{0}^t &  \vec{0}^t \\
  \vec{0}& 0 & 0\\
   \vec{0}& 0 & 0
    \end{array}\right), 
   \, P^\Lambda \rightarrow {\P}^\Lambda = \left(\begin{array}{ccc} P^\Lambda & \vec{0}^t &  \vec{0}^t \\
    \vec{0}& 0 & 0\\
   \vec{0}& 0 & 0  \end{array}\right),
\end{equation}
and the only new generator $\P^\alpha_i$ being associated to the shift symmetries of the open string axions:
\begin{equation}
\P^\alpha_i = \left( \begin{array}{cccccccc}
 0& 0& 0& 0& 0 & 0 & \vec{\delta}^t_j & 0 \\
 0 & 0&  0 & 0& 0 & 0& 0 &  \vec{\delta}_a^t \vec{\delta}^t_j\\
 0 & 0&  0 & 0&  0 & 0&  0 & 0\\
  0 & 0&  0 & 0&  0 & 0&  0 & 0\\
   0 & 0&  0 & 0&  0 & 0&  0 & 0\\
    0 & 0&  0 & 0&  0 & 0&  0 & 0\\
     0 & 0&  0 & 0&  0 & 0&  0 & 0\\
      0 & 0&  0 & 0&  0 & 0&  0 & 0\\
 \end{array} \right). 
\end{equation}
Under the shift symmetries of the closed string axions, the rotation matrix keeps its original transformation properties~(\ref{Eq:AxShiftSymmRotMat}), and the addition of open string axions enforces the axion rotation matrix to transform under an additional set of shift symmetries associated to the open string axions, with $\lambda_\alpha^i \in \Z$:
\begin{equation}
(R^{-1})^t (b^a  , \xi^K , \xi_\Lambda , \hat \theta^i_\alpha + \lambda^i_\alpha ) = (R^{-1})^t (b^a , \xi^K, \xi_\Lambda, \hat \theta^i_\alpha ) \, \cdot\,  e^{ -  \lambda^i_\alpha \P^\alpha_i}. 
\end{equation}
Invariance of the superpotential under the combined axion shift symmmetries requires the charge vector to transform as well:
\begin{equation}
\vec{q} \rightarrow e^{ r^a \P_a +  \varpi^K \P_K  \varpi_\Lambda \P^\Lambda +  \lambda^i_\alpha \P^\alpha_i}\,  \cdot\, \vec{q} \, .
\end{equation}
These considerations thus naturally extend the observations reviewed in section~\ref{S:IIAFluxVacua} and allow to identify a set of shift-invariant axion polynomials $ \ell_s\vec{\varrho} \equiv (R^{-1})^t \cdot \vec{q}$ including both closed and open string axions:
\begin{equation}\label{Eq:AxPolCSOS}
\begin{array}{lcl}
\ell_s \varrho_0 &=& e_0 + e_a b^a + \frac{1}{2} {\cal K}_{abc} m^a b^b b^c + \frac{m}{6} {\cal K}_{abc} b^a b^b b^c + h_K \xi^K  + h^\Lambda \xi_{\Lambda} + n_{Fi}^\alpha \hat \theta_\alpha^i  - n_{ai}^\alpha \hat \theta^i_\alpha b^a  , \\
\ell_s \varrho_a &=& e_a + {\cal K}_{abc}  m^b b^c + \frac{m}{2} {\cal K}_{abc} b^b b^c - n^\alpha_{ai} \hat \theta^i_\alpha , \\
\ell_s \tilde \varrho^a &=& m^a + m b^a , \\
\ell_s \tilde \varrho &=& m , \\
\ell_s \hat \varrho_K & = & h_K, \\
\ell_s \hat \varrho^\Lambda & = &  h^\Lambda,\\
\ell_s \varrho^\alpha_i &=& n^\alpha_{Fi} - b^a n^\alpha_{ai},\\
\ell_s \varrho^\alpha_{ai} &=& n^\alpha_{ai}. 
\end{array}
\end{equation}
The microscopic justification for the invariance under the axion shifts now runs~\cite{Herraez:2018vae} through the Hanany-Witten effect, which is in one-to-one correspondence with the Freed-Witten anomaly condition and allows to identify which combinations of flux quanta form invariant directions. Apart from assuring the consistency of four-dimensional axionic strings in flux backgrounds, the Freed-Witten anomaly conditions also serve to verify the microscopic compatibility between background fluxes and the D6-branes wrapping internal {\it SLag} three-cycles. In first instance, the NS-fluxes can induce Freed-Witten anomalies on the D6-brane worldvolume, unless the pullback of the NS 3-form field strength with respect to the wrapped three-cycle is an exact 3-form, see e.g~\cite{Maldacena:2001xj}:
\begin{equation}
\int_{\Pi_\alpha} H_3 = 0.
\end{equation}
On a formal footing, the requirement of vanishing Freed-Witten anomalies in a background $B_2$-field ensures the absence of global worldsheet anomalies in the  fermionic sector of the open superstring attached to the D6-brane~\cite{Freed:1999vc}. At the level of the 4d ${\cal N}=1$ supergravity theory, a vanishing Freed-Witten anomaly implies that only the linear combination $h_K \xi^K + h^\Lambda \xi_\Lambda$ effectively enters in the superpotential, while all orthogonal combinations can be gauged under the open string $U(1)$ symmetries living on D6-branes~\cite{Camara:2005dc,Villadoro:2006ia} without violating gauge invariance.

\subsection{Non-Supersymmetric Flux Vacua with D6-branes}\label{Ss:FluxVacuaD6-branesModStab}

As we have seen, mobile D6-branes modify the 4d effective action both at the level of the K\"ahler and superpotential. One natural question is then which kind of stable type IIA vacua exist in their presence, and in particular if one can construct Minkowski and AdS vacua analogous to the ones considered in section \ref{Ss:IIAFluxVacua}. On the one hand, in the case of ${\cal N} = 1$ AdS vacua the strategy to find such vacua is rather straightforward, as one must look for points in field space where all the F-terms vanish. On the other hand, the search for ${\cal N} = 0$ Minkowski vacua is less obvious. Indeed, just as in \cite{Giddings:2001yu} the pattern of F-terms that corresponds to stable ${\cal N} = 0$ Minkowski vacua relies on having a semi-definite scalar potential. In turn, the latter relies on the absence of certain fluxes in the superpotential 
and in the factorisation of the dilaton, K\"ahler and complex structure moduli in the K\"ahler potential. However, such a factorisation is lost as soon as mobile D6-branes appear in the construction, due to the 4d field redefinition \eqref{Eq:RedefComplexStructure}. Therefore, it is not clear that the no-scale properties of certain type IIA flux vacua can still be maintained in the presence of D6-branes with moduli.\footnote{Notice that the same observation could be made for type IIB compactifications with D3 and D7-branes.} 

In the following we would like to see if one can achieve stable ${\cal N} = 0$ 4d Minkowski vacua in the presence of mobile D6-branes, where the stability is guaranteed by the semi-definiteness of the (classical) scalar potential. Rather than taking the 10d approach of \cite{Lust:2008zd}, we will address this question in terms of the 4d effective theory discussed above. We will first show how to obtain a semi-definite F-term scalar potential by means of its standard 4d supergravity expression and a simple set of assumptions. We will then recover the same result by using the formalism that rewrites the scalar potential as a bilinear of axion polynomials. Finally, in the next section we will analyse the spectrum of soft terms that arises for these kind of vacua. 

\begin{center}
{\bf The standard 4d supergravity perspective}
\end{center}
Let us first consider the standard form of the F-term scalar potential
\begin{equation}\label{Eq:FTermScalarPot2}
V_F = \frac{e^{K}}{\kappa_4^2} \left[ K^{A \bar B} D_A W D_{\bar B} \ov W   - 3 \big| W \big|^2   \right], 
\end{equation}  
with $D_A = \partial_A + K_A$ the usual covariant derivative. As mentioned, the presence of mobile D6-branes creates a non-trivial mixing in the metric between K\"ahler, complex structure and open string moduli. Nevertheless, as pointed out in \cite{Carta:2016ynn} and \cite{Herraez:2018vae}, under certain assumptions the inverse metric $K^{A \bar B}$ displays a simplified structure.\footnote{One can derive eqs.\eqref{invK} by assuming that the zero degree functions {\bf H} in \eqref{Eq:RedefComplexStructure} only depend on the D6-brane position variables $\varphi^i$, as it happens for instance in the case of toroidal orbifolds.} First, even if $\partial_{a}\partial_{\bar{b}} K$ changes in the presence of mobile D6-branes, we have that $K^{\bar b a}$ remains the inverse of the previous K\"ahler moduli metric $\partial_{a}\partial_{\bar{b}} K_K$ (without open string moduli). Second, the rest of the components read:
\begin{subequations}
\label{invK}
\begin{equation}
\label{eq:im01}K^{\bar a i} =  f^i_b K^{b \bar{a}}, 
\end{equation}
\begin{equation}
\label{eq:im02} K^{\bar j i} = G_{\rm D6}^{ij} +  K^{a \bar b}f_a^if_b^j , 
\end{equation}
\begin{equation}
\label{eq:im1}K^{\bar I a} = -  \frac{1}{2} K^{\bar b a}\mathbf{H}^I_b,
\end{equation}
\begin{equation}
\label{eq:im2}K^{\bar I i} = - \frac{1}{2}\left[G_{\rm D6}^{ij} \,g_j^I + K^{\bar b a }f_a^i\,\mathbf{H}_b^I \right],
\end{equation}
\begin{equation}
\label{eq:im3}K^{\bar J I} =\mathbf N^{I J}+ \frac{1}{4} \left[K^{\bar b a}\,\mathbf H_b^J\mathbf H_a^I+ G_{\rm D6}^{ij}\, g_i^I g_j^J\right] ,
\end{equation}
\end{subequations}
where as before the indices $a,b$ label K\"ahler moduli, $I, J$ label dilaton and complex structure moduli and $i,j$ label open string moduli, absorbing the index $\alpha$ for simplicity.  Here the functions ${\bf H}_a^I$, $f_a^i$ and $g_i^I$ are defined as in section \ref{S:IIAORD6}.
Finally, $G_{\rm D6}^{ij}$ is the inverse of the open string metric 
\begin{equation}
G_{ij}^{\rm D6}  =  \frac{3 e^{-\phi/4}}{4 \CK \ell_s^3} \int_{\Pi_\alpha} \zeta_i \wedge *\, \zeta_j ,
\end{equation}
and $\mathbf N^{I J}$ is the inverse of the complex structure metric without mobile D6-branes
\begin{equation}
\mathbf N_{K \Lambda}  = \frac{1}{4} \partial_{n_\star^K} \partial_{u_{\star\, \Lambda}}  K_Q ,
\end{equation}
with $K_Q$ taken as a function of $n_\star^K$, $u_{\star\, \Lambda}$ as in \eqref{Eq:KaehlerPotCS1}. 

The relations \eqref{invK} allow to write the first piece of \eqref{Eq:FTermScalarPot2} as 
\begin{eqnarray}\nonumber
K^{A \bar B} D_ A W D_{\bar B} \overline W & = & K^{a \bar b}  \left[ D_a  + f_a^i D_i -\frac{1}{2} {\bf H}_a^K D_K\right] W \left[ D_{\bar{b}} + f_b^i D_{\bar{i}} -\frac{1}{2} {\bf H}_a^K D_{\bar{K}}\right] \overline W  \\ \nonumber
&+& G_{\rm D6}^{ij} \left[ D_i W - \frac{1}{2}  g_i^K D_K W \right] \left[ D_{\bar \jmath} \overline W - \frac{1}{2} g_j^L D_{\bar L} \overline W \right]\\
& + &  {\bf N}^{I {J}} D_I W D_{\bar J} \overline W 
\label{Nterm}
\end{eqnarray}
which is a sum of positive definite terms. This rewriting is crucial  in order to match the scalar potential derived from dimensional reduction with the one obtained from the standard supergravity formula \cite{Carta:2016ynn,Herraez:2018vae}. If in addition we consider a K\"ahler potential of the form \eqref{Eq:KahPotISDFact}, namely
\begin{equation}\label{hypoKQ}
K_Q = - {\rm log}\, (2s_\star) - K_{\tilde{Q}} (u_{\star\, \Lambda}) ,
\end{equation}
then the entries of $\mathbf N_{K \Lambda}$ mixing the dilaton and the complex structure moduli $u_{\star\, \Lambda}$ will vanish, and the same will hold for its inverse. As a result, the contribution coming from the last line of \eqref{Nterm} will split as
\begin{equation}
 {\bf N}^{I {J}} D_I W D_{\bar J} \overline W  =  {\bf N}^{S {S}} D_S W D_{\bar S} \overline W +  {\bf N}^{\Lambda {\Sigma}} D_\Lambda W D_{\bar \Sigma} \overline W 
\end{equation}
Finally, if we assume that the fields $U_\Lambda$ do not enter into the superpotential and use the corresponding no-scale relation we obtain
\begin{equation}
{\bf N}^{\Lambda {\Sigma}} D_\Lambda W D_{\bar \Sigma} \overline W =  3 |W|^2 ,
\end{equation}
that cancels the second term in \eqref{Eq:FTermScalarPot2}. Therefore, with similar assumptions as for the ISD closed string vacua and the K\"ahler metric relations \eqref{invK}, we obtain a semi-definite positive scalar potential and the corresponding 4d Minkowski vacua. 

The conditions for such vacua amount to imposing the following relations,
\begin{eqnarray}
\label{MinkD6aS} D_S W  & = & 0 , \\ 
\label{MinkD6aO} D_i W  & =   & \frac{1}{2}  g_i^\Lambda D_\Lambda W ,\\  
\label{MinkD6aK} D_a W & = &  \frac{1}{2} \left( {\bf H}_a^\Lambda - f_a^i g_i^\Lambda\right)  D_\Lambda W,
\end{eqnarray}
which is slightly weaker than imposing the cancellation of the F-terms for $S$, $T^a$ and $\Phi^i$. To rewrite these conditions  in a simple form, let us note that by eq.\eqref{Eq:HFunction1} $\partial_{\phi^i} u_{\star\, \Lambda} = \frac{1}{2} g_i^\Lambda$ and that the same assumptions that led to \eqref{invK} imply  $\partial_{t^a} u_{\star\, \Lambda} = \frac{1}{2} ( {\bf H}_a^\Lambda -  f_a^i g_i^\Lambda)$. We then have that they amount to
\begin{eqnarray}
\label{MinkD6bS} D_S W  & = & 0 , \\ 
\label{MinkD6bO} D_i W  & =   &  (\partial_i K_{\tilde{Q}}) W ,\\  
\label{MinkD6bK} D_a W & = &  (\partial_a K_{\tilde{Q}}) W . 
\end{eqnarray}
Alternatively, one may consider the contra-variant expressions of the F-terms
\begin{equation}
F^A \equiv K^{A \ov B} \ov D_{B} \ov W,
\end{equation}
which allow to designate in which moduli sector supersymmetry is broken spontaneously. Indeed, by imposing the vacuum conditions~\eqref{MinkD6aS}-\eqref{MinkD6aK} and using the expressions~\eqref{invK} for the inverse metric on the moduli space, the only non-vanishing on-shell component is the F-term for the complex structure moduli $U_\Lambda$:  
\begin{equation}  
F^\Lambda =N^{\Lambda \ov \Sigma} K_{\ov \Sigma} \ov W_0 = - 2i    u_{\star  \Lambda}  \ov W_0.
\end{equation}
Note that this relation forms the natural extension of the on-shell F-terms in type IIA closed string ISD flux vacua. Also in the presence of open string moduli (associated to mobile D6-branes) supersymmetry is spontaneously broken by the non-vanishing F-terms in the complex structure moduli sector, prompting us to label the class of such non-supersymmetric Minkowski vacua as {\it complex structure dominated} (CSD) vacua. In the next section we will analyse different phenomenological aspects of these ${\cal N}=0$ flux vacua with non-vanishing on-shell F-terms in the complex structure moduli sector, dubbed CSD vacua for short.

To determine the vacuum expectation value of the superpotential~$\ov W_0$, the axion polynomial formalism turns out to be extremely useful once the vacuum conditions~\eqref{MinkD6aS}-\eqref{MinkD6aK} are rewritten in terms of vacuum constraints on the axion polynomials, as we now discuss.

\clearpage

\begin{center}
{\bf The axion polynomial perspective}
\end{center}

While the reasoning used above to obtain ${\cal N} = 0$ Minkowski vacua fits better with the existing literature on string compactifications, there is a more direct approach to analyse the appearance of semi-definite scalar potentials and the corresponding Minkowski vacua. Indeed, instead of describing the scalar potential in terms of a K\"ahler and superpotential one may consider its expression as a bilinear of axion polynomials, as directly obtained from dimensional reduction. As we will see, one can reproduce similar conditions as above for the semi-positive definiteness of $V_F$, except that now no assumption on the K\"ahler metrics must be made.

As a warm up, let first us consider the well-know ISD case without mobile D6-branes, for which the potential can be expressed as in \eqref{Eq:VFbil}. In this language, the assumption \eqref{hypoKQ} translates into the vanishing of the off-diagonal components $K^{I\Lambda}$ in \eqref{Eq:ZABMetricFull}. When switching to the new basis of axion polynomials $ \vec{\rho}_{\rm new}$ in \eqref{Eq:AxionPolyBasisDiag}, this metric becomes
\begin{equation}\label{Eq:DiagInMetZ2}
Z^{AB} = 8 e^{K} {\rm diag} \left(4, K^{a\bar b}, \frac{4}{9} {\cal K}^2 K_{a \bar b}, - \frac{{\cal K}^2}{3}, K^{S\bar{S}},  K^{\Lambda \bar \Sigma} \right),
\end{equation} 
while
 \begin{equation}\label{Eq:AxionPolyBasisDiag2}
 \vec{\rho}_{\rm new} = \Big(\rho_0, \rho_a, \tilde \rho^a, \tilde \rho,  \hat \rho_0 -  \tilde \rho {\cal K}\frac{i}{3}  K_{S}, \hat \rho^\Lambda -  \tilde \rho {\cal K}\frac{i}{3}   K_{\Lambda} \Big).
\end{equation}
Imposing that the complex structure moduli $U_{\star\, \Lambda}$ do not enter the superpotential is equivalent to requiring that $\hat \rho^\Lambda=0$. Then, using the no-scale relation $K^{\Lambda \bar \Sigma} K_{\Lambda} K_{\bar{\Sigma}} = - K^{\Lambda \bar \Sigma} K_{\Lambda} K_{\Sigma} = 3$ one finds an exact cancellation between the contribution of the Romans mass component $\tilde{\rho}$ of \eqref{Eq:AxionPolyBasisDiag2} and the last one. As a result the scalar potential \eqref{Eq:VFbil} reads
\begin{equation}\label{Eq:VFbilISD}
V_F  = \frac{e^K}{\kappa_4^2}  \left(4 \rho_0^2 + K^{a\bar{b}} \rho_a\rho_b + \frac{4}{9} {\cal K}^2 K_{a \bar b}  \tilde \rho^a  \tilde \rho^b + K^{S\bar{S}} \left(\hat \rho_0 -  \tilde \rho {\cal K}\frac{i}{3}  K_{S}\right)^2   \right) ,
\end{equation}  
which is clearly semi-definite positive and vanishes if and only if the conditions \eqref{Eq:GenEqISD} are met. In this way, we directly recover the relations for the axion polynomials obtained in section \ref{Ss:IIAFluxVacua} without having to consider any particular pattern for the F-terms. 

Similarly, we may apply this strategy to the case of CSD vacua (with mobile D6-branes), where now the vector of axion polynomials has the components \eqref{Eq:AxPolCSOS}. From the results of section 3 of~\cite{Herraez:2018vae} adapted to our conventions for quantised fluxes, one obtains that inverse metric takes the diagonalised form
\begin{equation}\label{Eq:DiagInMetZ2b}
Z^{AB} = 8 e^{K} {\rm diag} \left(4, K_K^{a\bar b}, \frac{4}{9} {\cal K}^2 (K_K)_{a \bar b}, - \frac{{\cal K}^2}{3}, {\bf N}^{S\bar{S}},  {\bf N}^{\Lambda \bar \Sigma}, G_{\rm D6}^{ij}, G_{\rm D6}^{ij} \right),
\end{equation} 
in the following basis of axion polynomials 
 \begin{equation}\label{Eq:AxionPolyBasisDiag3}
 \vec{\varrho}_{\rm new} = \Big(\varrho_0, \varrho_a^\prime, \tilde \varrho^{a \prime}, \tilde \varrho,  \hat \varrho_0 -  \tilde \varrho {\cal K}\frac{i}{3}  K_{S}, -  \tilde \varrho {\cal K}\frac{i}{3}   K_{U_\Lambda}, \varrho_i^\prime , t^a \varrho_{ai} \Big).
\end{equation}
Here we have defined
\begin{eqnarray}
\varrho_a^\prime & = & \varrho_a + f^i_a \varrho_i - \frac{1}{2} {\bf H}_a^0 \hat \varrho_0, \\
\tilde \varrho^{a \prime} & = & \tilde \varrho^a - (\CK^{ab}t^cf_c^i+\CK^{ac}t^bf_c^i)\varrho_{bi}, \\
\varrho_i^\prime & = & \varrho_i - \frac{1}{2}g_i^0 \hat \varrho_0
\end{eqnarray}
and we have already imposed that ${\bf N}^{S \Lambda}=0$ and that $\hat{\varrho}^\Lambda = 0$. Again, we find a cancellation between the quadratic terms in the $4^{th}$ and $6^{th}$ entry of \eqref{Eq:AxionPolyBasisDiag3}. This results into a semi-definite positive, bilinear scalar potential of the form
\begin{equation}\label{Eq:VFbilMD6}
V_F  = \frac{e^K}{\kappa_4^2}  \left(4 \varrho_0^2 + K_K^{a\bar{b}} \varrho_a^\prime  \varrho_b^\prime + \frac{4}{9} {\cal K}^2 (K_K)_{a \bar b}  \tilde \varrho^{a \prime}  \tilde \varrho^{b \prime} + {\bf N}^{S\bar{S}} \left(\hat \varrho_0 -  \tilde \varrho {\cal K}\frac{i}{3}  K_{S}\right)^2 + G_{\rm D6}^{ij} \left[\varrho_i^\prime \varrho_j^\prime +  t^at^b \varrho_{ai} \varrho_{bj}\right] \right) ,
\end{equation}  
We then find that the conditions for a Minkowski vacuum are
\begin{subequations}
\label{MinkD6}
\begin{equation}
\varrho_0 = 0, 
\end{equation}
\begin{equation}
\varrho_a =  \frac{1}{2} \left( {\bf H}_a^0 - f_a^i g_i^0\right) \hat \varrho_0 , 
\end{equation}
\begin{equation}
\tilde\varrho^a = \CK^{ab}\phi^i \varrho_{bi}, 
\end{equation}
\begin{equation}
\hat{\varrho}_0 = - \frac{\CK}{6s_\star} \tilde\varrho,
\end{equation}
\begin{equation}
\varrho_i = \frac{1}{2} g^0_i \hat \varrho_0,
\end{equation}
\begin{equation}
\label{MinkD6ai}
t^a \varrho_{ai} = 0 ,
\end{equation}
\end{subequations}
and that whenever they are satisfied the superpotential takes the value
\begin{equation}
W_0 = 2is_\star \hat{\varrho}_0 = -\frac{i\CK}{3} \tilde{\varrho}\, .
\end{equation}
Equivalently, at these vacua we have $ \vec{\varrho}_{\rm new}  =  \Big(0, 0, 0,  \tilde \varrho, 0, F_{U_\Lambda}, 0, 0 \Big)$. 
One can easily check that these relations are equivalent to eqs.\eqref{MinkD6bS}-\eqref{MinkD6bK} if one uses eq.~\eqref{Eq:HFunction1} and assumes that $\partial_{t^a} (t^a{\bf H}_a^0) = {\bf H}_a^0 -  f_a^i g_i^0$. In the next section we will analyse several phenomenological aspects of these CSD vacua.

\section{Fluxed Supersymmetry-Breaking and Soft Terms}\label{S:SUSYSoftTerms}
The ${\cal N} = 0$  Minkowski vacua of the previous section represent examples of string vacua in which supersymmetry is spontaneously broken due to background fluxes. A first manifestation of broken supersymmetry are the non-vanishing F-terms in the complex structure moduli sector, yet the genuinely physical observables resulting from spontaneous supersymmetry-breaking correspond to the gravitino mass and soft terms for the visible sector (chiral matter charged under gauge symmetries). In this section, we compute the gravitino mass and soft terms for the CSD vacua in terms of the axion polynomials of the compactification, in such a way that the vacuum constraints on the axion polynomials suffice to determine whether supersymmetry is broken and how the soft terms relate to the gravitino mass.

\subsection{Fluxed Supersymmetry-Breaking}
The perturbative toolbox in ${\cal N}=1$ supergravity to obtain a supersymmetry-breaking vacuum consists in coupling gravity to chiral multiplets subject to a non-trivial superpotential. The vacuum configuration of the resulting F-term scalar potential then determines the sign and value of the vacuum-energy, indicating whether the vacuum of the four-dimensional theory corresponds to an Anti-de Sitter, Minkowski or de Sitter spacetime. To discriminate supersymmetric from non-supersymmetric vacua it suffices to analyse the F-terms and identify at least one chiral superfield with a non-vanishing F-term in case of non-supersymmetric vacua. In that case, the fermionic partner inside the chiral superfield serves as the massless Goldstino, which is absorbed by the gravitino through the super-Brout-Englert-Higgs mechanism \cite{Cremmer:1982en,Grisaru:1982sr}. The would-be mass of the gravitino in the Lagrangian, also dubbed {\it apparent} gravitino mass in~\cite{Ferrara:2016ntj}, is proportional to the vacuum expectation value of the superpotential
\begin{equation}\label{Eq:AppGravMass}
m_{3/2}^2 = e^{K} |W|^2.
\end{equation}
Note, however, that a non-vanishing apparent gravitino mass does not imply supersymmetry is spontaneously broken, as is the case for the supersymmetric AdS vacua introduced in section~\ref{Ss:IIAFluxVacua}. 
To evaluate whether supersymmetry is spontaneously broken, it is more appropriate to consider an {\it effective} gravitino mass
\begin{equation}\label{Eq:EffGravMass}
\ov m_{3/2}^2 = m_{3/2}^2  + \frac{\kappa_4^2}{3} V_F   =  \frac{1}{3} e^{K}F_{A} K^{A \ov B} F_{\ov B},
\end{equation}
whose scale is set by the (non-vanishing) F-terms of the chiral multiplets. This relation between the effective gravitino mass and the F-terms of the chiral multiplets has been obtained by virtue of the expression for the F-term scalar potential (\ref{Eq:FTermScalarPot}). When evaluating the value of the effective gravitino mass in the vacuum of the theory, its value corresponds to the on-shell apparent gravitino mass corrected by the vacuum energy for curved spacetimes.
The evaluation of these formulae for ISD flux vacua and supersymmetric AdS vacua will follow shortly. For now, we summarise  
the various background vacua that can potentially emerge from an ${\cal N}=1$ supergravity theory coupled to chiral supermultiplets in table~\ref{Tab:N1SUGRAVacua}.
\begin{table}[h]
\begin{center}
\begin{tabular}{c@{\hspace{0.2in}}c@{\hspace{0.4in}}c@{\hspace{0.4in}}c}\hline
background & $m_{3/2}^2$ & $\langle V \rangle$ & $\ov m_{3/2}^2 $  \\
\hline\hline
SUSY Minkowski& 0 & 0 & 0\\
non-SUSY Minkowski & $> 0$ & 0 & $> 0$ \\
SUSY AdS & $ > 0$ & $<0$ & $0$ \\
non-SUSY AdS &$>0$  &  $<0$& $>0$ \\
non-SUSY dS &$> 0$ & $> 0$ & $>0$\\
\hline
\end{tabular}
\caption{Overview of four-dimensional vacuum configurations in ${\cal N}=1$ supergravity coupled to chiral supermultiplets with the corresponding apparent gravitino mass, vacuum energy and effective gravitino mass.\label{Tab:N1SUGRAVacua}}
\end{center}
\end{table}

The 4d low-energy effective field theory for type~IIA orientifold compactifications is (partly) captured by an ${\cal N}=1$ supergravity theory coupled to chiral supermultiplets, with scalar components played by closed and open string moduli. Hence, by studying the vacuum structure of the F-term scalar potential we can both determine the consistency of the compactification as well as the physics of the four-dimensional spacetime. In the previous sections we showed that perturbative flux vacua are easily identified in terms of constraints on the shift-invariant axion polynomials (\ref{Eq:CSgaugeinvariantrho}) or (\ref{Eq:AxPolCSOS}). Our next aim is to forge a connection  between the spontaneous breaking of supersymmetry and these axion polynomials by rewriting the gravitino masses appropriately. Exploiting the factorability of the perturbative flux superpotential the apparent gravitino mass (\ref{Eq:AppGravMass}) can be expressed  in terms of the axion polynomials (\ref{Eq:CSgaugeinvariantrho}) as follows:          
\begin{equation}
m_{3/2}^2 =  e^{K} \rho_A  (\Pi^\dagger \ltimes \Pi )^{AB}  \rho_B,
\end{equation}
where the purely saxion-dependent matrix $\Pi^\dagger \ltimes \Pi$ reads more explicitly
\begin{equation}
\Pi^\dagger \ltimes \Pi  =\left(\begin{array}{cccccc} 1 & 0 & -\frac{1}{2} {\cal K}_a & 0 & 0&0 \\
0& t^a t^b & 0 & - t^a \frac{\CK}{6} & t^a n^K_\star & t^a u_{\star \Lambda}  \\ 
 -\frac{1}{2} {\cal K}_b & 0 & \frac{1}{4} \CK_a \CK_b & 0 & 0 & 0 \\ 
 0 & -  t ^b \frac{\CK}{6} & 0 & \left( \frac{\CK}{6} \right)^2 & - \frac{\CK}{6} n^K_\star& - \frac{\CK}{6} u_{\star\Lambda}\\
 0& t^b n^I_\star & 0 &-  n^I_\star \frac{\CK}{6} & n^I_\star n^K_\star &  n^I_\star  u_{\star\Lambda} \\
  0& t^b u_{\star\Sigma} & 0 &-  u_{\star\Sigma} \frac{\CK}{6} & u_{\star\Sigma} n^K_\star &  u_{\star\Sigma}  u_{\star\Lambda} \\
  \end{array} \right),
\end{equation}
when expressed in the basis of axion polynomials $\vec{\rho} = \left(\rho_0, \rho_a, \tilde \rho^a, \tilde \rho, \hat \rho_K, \hat \rho^\Lambda \right)$. 

Also the effective gravitino mass (\ref{Eq:EffGravMass}) can be expressed in terms of the axion polynomials by working out the F-terms for the K\"ahler and complex structure moduli explicitly. When neglecting open string moduli or considering compactifications without D6-branes, the factorability of the closed string moduli space translates into a factorisation of the F-terms per sector:   
\begin{equation}\label{Eq:EffGravMassb}
\ov m_{3/2}^2  =  \frac{1}{3} e^{K} \vec{\rho}^{\, T} \, \left( \mathbb{F}_T + \mathbb{F}_{UN}  \right) \vec{\rho},
\end{equation}
where the matrix $\mathbb{F}_{UN}$ for the complex structure moduli is given by
\begin{equation}
\mathbb{F}_{UN} = \left( \begin{array}{ccccccc} 
 4 & 0 & -2 {\cal K}_a & 0 & 0 & 0 \\
0& 4 t^a t^b & 0 & -2 t^a \frac{\CK}{3} & 2 t^a n^I_\star & 2 t^a u_{\star\Lambda } \\ 
 -2 {\cal K}_b & 0 & \CK_a \CK_b & 0 & 0 & 0 \\ 
 0 & - 2t ^b \frac{\CK}{3} & 0 &4 \left( \frac{\CK}{6} \right)^2 & - \frac{\CK}{3} n^K_\star &- \frac{\CK}{3} u_{\star\Lambda}  \\
 0& 2 t^b n^I_\star & 0 & - \frac{\CK}{3} n^I_\star &K^{N^I \ov N^K} &  K^{N^I \ov U_\Lambda}  \\ 
  0& 2 t^b u_{\star\Sigma} & 0 & -\frac{\CK}{3} u_{\star\Sigma} &K^{U_\Sigma \ov N^K} & K^{U_\Sigma \ov U_\Lambda}  \\ 
 \end{array}\right),
\end{equation}
and the matrix $\mathbb{F}_T$ for the K\"ahler moduli sector reads
\begin{equation}
\mathbb{F}_T = \left( \begin{array}{ccccccc} 
 3 & 0 & \frac{1}{2} {\cal K}_a & 0 & 0 & 0  \\
0&  t^a t^b -\frac{2}{3} \CK \CK^{ab}  & 0 &  t^a \frac{\CK}{6} &  t^a n^I_\star & t^a u_{\star\Lambda} \\ 
 \frac{1}{2} {\cal K}_b & 0 & \frac{3}{4} \CK_a \CK_b  - \frac{2}{3} \CK \CK_{ab}& 0 & 0 &  0 \\ 
 0 &  t ^b \frac{\CK}{6} & 0 &3 \left( \frac{\CK}{6} \right)^2 &  \frac{1}{2}\CK n^K_\star &   \frac{1}{2}\CK  u_{\star\Lambda} \\
 0&  t^b n^I_\star & 0 & \frac{1}{2}  n^I_\star \CK &3 n^I_\star n^K_\star & 3 n^I_\star u_{\star\Lambda} \\  
  0&  t^b u_{\star\Sigma} & 0 & \frac{1}{2}  u_{\star\Sigma}\CK &3 u_{\star\Sigma} n^K_\star & 3 u_{\star\Sigma} u_{\star\Lambda} \\  
 \end{array}\right),
\end{equation}
both expressed in the basis of axion polynomials $\vec{\rho} = \left(\rho_0, \rho_a, \tilde \rho^a, \tilde \rho, \hat \rho_K, \hat \rho^\Lambda \right)$. 
The expressions for the apparent and effective gravitino mass have only taken into account the chiral multiplets from the closed string sector. As long as the superpotential remains factorisable in the sense of section~\ref{Ss:APOSS} when including open string chiral multiplets, the expressions for the gravitino masses can be straightforwardly generalised, which will be the focus of the last part of this section.

\begin{center}
{\bf Supersymmetric Anti-de Sitter Flux Vacua}
\end{center}
Let us now test these considerations for the supersymmetric AdS vacua from section~\ref{S:IIAFluxVacua}, which are represented by the vector $\vec{\rho}_{\rm AdS} = \tilde \rho \left( 0, \frac{3}{10}{{\cal K}_a}, 0, 1, -\frac{i}{5}\CK  K_{N^I_\star}, -\frac{i}{5}  \CK K_{U_{\star\Lambda}}   \right)$. In this vacuum configuration, the apparent gravitino mass happens to have a non-vanishing value proportional to Romans mass $\tilde \rho$: 
\begin{equation}
m_{3/2}^2 =  e^K \left( \frac{2 {\cal K}}{15} \tilde \rho\right)^2.
\end{equation}
The effective gravitino mass in the supersymmetric AdS vacua vanishes, as can be checked explicitly by evaluating expression~\eqref{Eq:EffGravMassb} for the axion vector $\vec{\rho}_{\rm AdS}$. The vanishing effective gravitino mass should not surprise us at all, as it is fully in line with the vanishing F-terms and the (negative) vacuum energy for the supersymmetric AdS vacua, which equates in absolute value to three times the value of the apparent gravitino mass.

\begin{center}
{\bf Non-supersymmetric Minkowski Flux Vacua (ISD)}
\end{center}
A case study for non-supersymmetric vacua are the backgrounds with ISD fluxes, as discussed in section~\ref{S:IIAFluxVacua}. Considering the factorisation of the dilaton as in~(\ref{Eq:KahPotISDFact}) for the ISD flux set-up, the purely saxion-dependent matrix $\Pi^\dagger \ltimes \Pi$ in the apparent gravitino mass takes the form
\begin{equation}
\Pi^\dagger \ltimes \Pi  = \left(\begin{array}{cccccc} 1 & 0 & -\frac{1}{2} {\cal K}_a & 0 & 0 & 0 \\
0& t^a t^b & 0 & -t^a \frac{\CK}{6} & t^a s_\star & t^a u_{\star \Lambda} \\ 
 -\frac{1}{2} {\cal K}_b & 0 & \frac{1}{4} \CK_a \CK_b & 0 & 0 & 0 \\ 
 0 & - t ^b \frac{\CK}{6} & 0 & \left( \frac{\CK}{6} \right)^2 & -  \frac{\CK}{6} s_\star& -  \frac{\CK}{6} u_{\star \Lambda}\\
 0& t^b s_\star & 0 &-   s_\star  \frac{\CK}{6} & s^2_\star & s_\star u_{\star \Lambda} \\
 0& t^b u_{\star \Sigma} & 0 &-  u_{\star \Sigma} \frac{\CK}{6} & s _\star u_\Sigma & u_{\star \Sigma} u_{\star \Lambda} \\ 
  \end{array} \right).
\end{equation}
The apparent gravitino mass for the ISD flux vacua, represented by the axion vector $\vec{\rho}_{\rm ISD}  = \tilde \rho \left(0,0,0,1, 0, - \frac{i}{3} {\cal K} K_{U_\Lambda}   \right)$ also scales with Romans'  mass $\tilde \rho$:
\begin{equation}
m_{3/2}^2 =   e^{K} \left(\frac{{\cal K}}{3} \tilde \rho \right)^2.
\end{equation}
In these vacua the effective gravitino mass does not vanish, which can be verified explicitly when writing out the F-terms by virtue of the axion polynomials: 
\begin{equation}
\ov m_{3/2}^2  =  \frac{1}{3} e^{K} \vec{\rho}^{\, T} \, \left( \mathbb{F}_T + \mathbb{F}_{S_\star} +  \mathbb{F}_{U_\star}  \right) \vec{\rho} =   \frac{1}{3} e^{K}   \left ( \frac{\cal K}{3} \tilde \rho \right)^2,
\end{equation}
where the matrix $\mathbb{F}_{S_\star}$ for the dilaton sector is given by
\begin{equation}
\mathbb{F}_{S_\star}  = \left( \begin{array}{cccccc} 
 1 & 0 & -\frac{1}{2} {\cal K}_a & 0 & 0 & 0 \\
0& t^a t^b & 0 & -  t^a \frac{\CK}{6} & - t^a s_\star & t^a u_{\star\Lambda} \\ 
 -\frac{1}{2} {\cal K}_b & 0 & \frac{1}{4} \CK_a \CK_b & 0 & 0 & 0 \\ 
 0 & - t ^b \frac{\CK}{6} & 0 &  \left( \frac{\CK}{6} \right)^2 &  \frac{\CK}{6} s_\star& - \frac{\CK}{6} u_{\star\Lambda}\\
 0& -t^b s_\star & 0 &  s_\star \frac{\CK}{6} & s^2_\star &  -s_\star\, u_{\star\Lambda} \\
 0& t^b u_{\star\Sigma} & 0 &-  u_{\star\Sigma} \frac{\CK}{6} & - s_\star\, u_{\star\Sigma} & u_{\star\Sigma} u_{\star\Lambda} \\ 
 \end{array}\right),
\end{equation}
the matrix $\mathbb{F}_{U_\star}$ for the complex structure moduli sector reads
\begin{equation}
\mathbb{F}_{U_\star} =  \left( \begin{array}{cccccc} 
 3 & 0 & -  \frac{3}{2} {\cal K}_a & 0 & 0 & 0 \\
0& 3 t^a t^b & 0 & -  t^a \frac{\CK}{2} & 3  t^a s_\star & t^a u_{\star\Lambda} \\ 
 -\frac{3}{2} {\cal K}_b & 0 & \frac{3}{4} \CK_a \CK_b & 0 & 0 & 0 \\ 
 0 & - t ^b \frac{\CK}{2} & 0 & 3 \left( \frac{\CK}{6} \right)^2 & - \frac{\CK}{2} s_\star& - \frac{\CK}{6} u_{\star\Lambda}\\
 0& 3  t^b s_\star & 0 & -  s_\star \frac{\CK}{2} & 3 s^2_\star &   s_\star\, u_{\star\Lambda} \\
 0& t^b u_{\star\Sigma} & 0 & -  u_{\star\Sigma} \frac{\CK}{6} &  s_\star\, u_{\star\Sigma} & K^{\Lambda \Sigma} -  u_{\star\Sigma} u_{\star\Lambda} \\ 
 \end{array}\right) ,
\end{equation}
and the matrix $ \mathbb{F}_T$ for the K\"ahler moduli takes the form
\begin{equation}
 \mathbb{F}_T = \left( \begin{array}{cccccc} 
3 & 0& \frac{1}{2} \CK_a& 0 & 0 &0 \\     
 0 &  t^a t^b -\frac{2}{3} \CK \CK^{ab} & 0& \frac{1}{6} \CK t^a & t^a s_\star & t^a u_{\star\Lambda} \\   
 \frac{1}{2} \CK_b& 0& \frac{3}{4} \CK_a \CK_b -\frac{2}{3} \CK \CK_{ab}&  0  & 0 & 0  \\  
 0 & \frac{1}{6} \CK t^b& 0& 3 \left(\frac{\CK}{6}\right)^2 & \frac{1}{2} \CK s_\star & \frac{1}{2} \CK u_{\star\Lambda}    \\  
 0 & t^b s_\star & 0& \frac{1}{2} \CK  s_\star & 3 s^2_\star & 3 s_\star u_{\star\Lambda}\\  
  0 & t^b u_{\star\Sigma} & 0& \frac{1}{2} \CK  u_{\star\Sigma} & 3 s_\star u_{\star\Sigma} & 3 u_{\star\Sigma} u_{\star\Lambda}
 \end{array}\right).
\end{equation}
The non-vanishing value for the effective gravitino mass is due to the non-vanishing F-terms for the complex structure moduli in the ISD flux vacua, which can be verified explicitly in the axion polynomial language. The factorability of the moduli sectors allows in this case to clearly extract the $U$-dominated character of the supersymmetry-breaking in type IIA ISD flux vacua.

\begin{center}
{\bf Non-supersymmetric Flux Vacua with D6-branes (CSD Vacua)}
\end{center}
As discussed in section~\ref{Ss:FluxVacuaD6-branesModStab}, mobile D6-branes alter the vacuum structure of the 4d effective theory such that the corresponding non-supersymmetric Minkowski vacua~\eqref{MinkD6aS}-\eqref{MinkD6aK} rely on weaker vacuum constraints than the ISD flux vacua. Subsequently, the pattern of supersymmetry-breaking in the presence of mobile D6-branes needs further exploration to assess how it defers from the pure closed string case. To this end, we first consider the apparent gravitino mass, which can still be factorised in a bilinear form consisting of the purely saxion-dependent matrix $\Pi^\dagger \ltimes \Pi$:
\begin{equation}
\Pi^\dagger \ltimes \Pi  =\left(\begin{array}{cccccccc} 1 & 0 & -\frac{1}{2} {\cal K}_a & 0 & 0&0 & 0 & t^a \phi^i  \\
0& t^a t^b & 0 & - t^a \frac{\CK}{6} & t^a n^K & t^a u_\Lambda &  t^a \phi^i & 0   \\ 
 -\frac{1}{2} {\cal K}_b & 0 & \frac{1}{4} \CK_a \CK_b & 0 & 0 & 0 & 0 &- \frac{1}{2} {\CK}_b  t ^a  \phi^i \\ 
 0 & -  t ^b \frac{\CK}{6} & 0 & \left( \frac{\CK}{6} \right)^2 & - \frac{\CK}{6} n^K& - \frac{\CK}{6} u_\Lambda& -\frac{\CK}{6} \phi^i & 0 \\
 0& t^b n^I & 0 &-  n^I \frac{\CK}{6} & n^I n^K &  n^I  u_\Lambda &  n^I \phi^i & 0  \\
  0& t^b u_\Sigma & 0 &-  u_\Sigma \frac{\CK}{6} & u_\Sigma n^K &  u_\Sigma  u_\Lambda&  u_\Sigma \phi^i  & 0\\
   0& t^b \phi^j & 0&  -\frac{\CK}{6} \phi^j & n^K \phi^j & u_\Lambda \phi^j &  \phi^i \phi^j &0\\
  t^b \phi^j & 0 & - \frac{1}{2} {\CK}_b  t ^a  \phi^j& 0& 0& 0 & 0 &  t^a t^b \phi^i \phi^j  \\
  \end{array} \right),
\end{equation}
expressed in terms of the axion basis $\vec{\varrho}\,{}^T =\left( \varrho_0, \varrho_a, \tilde \varrho^a, \tilde \varrho, \hat \varrho_K, \hat \varrho^\Lambda,\varrho_i, \varrho_{ai}    \right)$. Upon evaluating the apparent gravitino mass for the CSD vacuum conditions in \eqref{MinkD6}, one easily retrieves the same functional dependence as for the ISD flux vacua:
\begin{equation}
m_{3/2}^2 =  e^{K} \left(\frac{{\cal K}}{3} \tilde \varrho \right)^2.
\end{equation}

Nevertheless, the relevant quantity to consider for vacua with (spontaneously) broken supersymmetry is the effective gravitino mass~\eqref{Eq:EffGravMass}, whose explicit bilinear expression in terms of the axion polynomials becomes extremely involved upon inclusion of D6-brane moduli. More precisely, it is the mixing between closed and open string moduli sectors that prevents us from writing down the F-terms as axion polynomial bilinears by virtue of the simple matrices $\mathbb{F}_S$, $\mathbb{F}_U$ and $\mathbb{F}_T$, as in the closed string ISD flux case. Instead we look at the effective gravitino mass as the scalar product between the co-variant and contra-variant F-term vectors, 
\begin{equation}
  \ov m_{3/2}^2 = \frac{1}{3} e^{K} \left( F_a F^a + F_S F^S + F_\Lambda F^\Lambda + F_i F^i \right)
\end{equation}
and express both vectors explicitly in terms of the axion polynomials. The co-variant F-term vectors contain two contributions both linear in the axion polynomials
\begin{subequations}
\begin{equation}
\left( \begin{array}{c} F_a \\
F_S\\
F_\Lambda \\
F_i
 \end{array} \right) = \left(\begin{array}{cccccccc} 0 & \delta_a^b & i {\cal K}_{ab} & -\frac{1}{2}{\cal K}_a & 0 & 0& 0 & -i \phi^j  \\
 0 & 0& 0& 0 &1 &0 & 0& 0\\
  0 & 0& 0& 0 &0 &\delta_{\Lambda}^\Sigma & 0& 0\\
   0 & 0& 0& 0 &0 & 0 & \delta_i^j & -it^a 
   \end{array} \right) \cdot \vec{\varrho} + \left( \begin{array}{c} K_{T^a} \\ K_S \\ K_\Lambda \\ K_{\Phi^i}  \end{array} \right) \vec{\Pi}^t \cdot \vec{\varrho},
\end{equation}
and similarly the contra-variant F-term vector can be written as the sum of two linear terms in the axion polynomials
\begin{equation}\label{Eq:ContraVariantFtermsCO}
\begin{array}{rcl}
\left( \begin{array}{c} F^a \\
F^S\\
F^\Lambda \\
F^i
 \end{array} \right) &=&
 \left( \begin{array}{cccccccc}
0 & K^{a \ov b} & -i  K^{a \ov c} {\cal K}_{cb} &- \frac{1}{2} K^{a \ov c} {\cal K}_c  & K^{a \ov S} & K^{a \ov \Sigma} & K^{a \ov j} & i  K^{a \ov j}  t^b\\
0 & K^{S \ov b} & -i  K^{S \ov c} {\cal K}_{cb} &- \frac{1}{2} K^{S \ov c} {\cal K}_c  & K^{S \ov S} & K^{a \ov \Sigma} & K^{S \ov j} & i  K^{S \ov j}  t^b\\
0 & K^{\Lambda \ov b} & -i  K^{\Lambda \ov c} {\cal K}_{cb} &- \frac{1}{2} K^{\Lambda \ov c} {\cal K}_c  & K^{\Lambda \ov S} & K^{\Lambda \ov \Sigma} & K^{\Lambda \ov j} & i  K^{\Lambda \ov j}  t^b\\
0 & K^{i \ov b} & -i  K^{i \ov c} {\cal K}_{cb} &- \frac{1}{2} K^{i \ov c} {\cal K}_c  & K^{i \ov S} & K^{i \ov \Sigma} & K^{i \ov j} & i  K^{i \ov j}  t^b
\end{array}
 \right) \cdot \vec{\varrho}  \\
 && + \left( \begin{array}{c} -2i t^a \\ -2 is  \\ -2 i u_{ \Lambda}  \\ - 2i \phi^i \end{array} \right)  \vec{ \Pi}^\dagger \cdot \vec{\varrho} \; ,
\end{array}
\end{equation}
\end{subequations} 
where we used the expressions~\eqref{invK} for the inverse metrics on the moduli space and the first order derivatives~\eqref{Eq:1ODerK} of the K\"ahler potential to simplify the second term. An alternative (and more explicit) representation of the contra-variant F-terms can be found in~\cite{Herraez:2018vae}. Upon evaluating the F-term vectors in the CSD vacua~\eqref{MinkD6}
\begin{equation}
(\vec{F_A}){}^t = \left( \frac{1}{2} ({\bf H}_{a \Lambda} - f_a^i g_{i\Lambda})  F_\Lambda, 0, F_\Lambda, \frac{1}{2} g_i^\Lambda  F_\Lambda  \right), \qquad (\vec{F}^A){}^t = \left(0,0,-2 i u_{\star \Lambda} \ov W_0 , 0 \right),
\end{equation}
one can immediately deduce that only the complex structure moduli sector provides a non-vanishing contribution to the effective gravitino mass:
\begin{equation}
  \ov m_{3/2}^2 =  \frac{1}{3} e^{K} F_\Lambda F^\Lambda  =  e^{K} \left(\frac{{\cal K}}{3} \tilde \varrho \right)^2.
\end{equation}
Note that the functional dependence of the effective gravitino mass for these CSD or ${\cal N}=0$ Minkowski vacua is precisely the same as for the pure ISD flux vacua.
\
\subsection{Flux-Induced Soft Terms on D6-branes}

Upon including D6-branes into a type IIA flux vacuum with non-vanishing F-terms in the moduli sectors, the spontaneous supersymmetry-breaking is mediated through gravitational couplings to the D6-brane worldvolumes in the form of soft terms for the open string excitations. To extract the soft terms one usually distinguishes~\cite{Kaplunovsky:1993rd,Brignole:1997dp,Kors:2003wf} between the visible sector composed of the massless open string excitations (with vanishing vacuum expectation values) on the one hand and the hidden sector of closed string moduli on the other hand. Given that the D6-brane displacement moduli provide for more generic vacua in the presence of background fluxes, we choose a more suitable factorisation of the ${\cal N} =1$ chiral multiplets: on the one hand open string excitations transforming in bifundamental representations of the D6-brane gauge theories denoted collectively by~${\cal O}^\alpha$ (and its hermitian conjugate $\ov {\cal O}^{\ov \alpha}$), and on the other hand the ``hidden" sector of closed string moduli and D6-brane displacement moduli denoted by ${\cal H} \in \{ T^a, N^K, U_\Lambda, \Phi^i \}$. Subsequently, the K\"ahler potential and superpotential can then be expanded around the background values of the hidden sector moduli:
\begin{eqnarray}\label{Eq:EffKaehlerPotSuperPot}
K( {\cal H},  \ov{\cal H}, {\cal O}, \ov {\cal O} ) &=& K^0 ({\cal H},  \ov{\cal H}) + K_{\alpha \ov \beta}({\cal H} ,  \ov{\cal H}) {\cal O}^\alpha \ov{\cal O}^{\ov \beta}  + \left[ \frac{1}{2} Z_{\alpha \beta}({\cal H},  \ov{\cal H}) {\cal O}^\alpha {\cal O}^{\beta} + h.c.\right] +\ldots, \notag \\
W({\cal H}, {\cal O}) &=& W_0 ({\cal H}) + \frac{1}{2} \mu_{\alpha \beta}({\cal H})   {\cal O}^\alpha {\cal O}^{\beta} + \frac{1}{6} Y_{\alpha \beta \gamma } ({\cal H}) {\cal O}^\alpha {\cal O}^{\beta} {\cal O}^{\gamma} + \ldots.
\end{eqnarray}
In this expansion, the K\"ahler potential $K^0 = K_T + K_Q$ contains the K\"ahler potentials for the dilaton, K\"ahler moduli, complex structure moduli and open string displacement moduli, while the functions $K_{\alpha \ov \beta}({\cal H}, \ov {\cal H})$ represent the K\"ahler metrics for the open string excitations with vanishing vacuum expectation value (at the level of the supergravity analysis). The superpotential $W_0 ({\cal H})$ encompasses the perturbative RR- and NS-flux superpotential as well as the bilinear superpotential as in (\ref{Eq:OpenClosedSuperPotential}), while the quadratic and Yukawa couplings between the open string modes arise from non-perturbative effects such as worldsheet instantons and potentially D-brane instantons. The soft terms for the open string modes follow by inserting the expansion for the K\"ahler potential and superpotential into the F-term scalar potential (\ref{Eq:FTermScalarPot}), and taking the limit $\kappa_4 \rightarrow \infty$ while keeping the apparent gravitino mass $m_{3/2}$ fixed:
\begin{equation}
V_{\rm soft} = m_{\alpha \ov \beta}^2 {\cal O}^\alpha \ov{\cal O}^{ \ov \beta} + \left[ \frac{1}{6} A_{\alpha \beta \gamma} {\cal O}^\alpha {\cal O}^\beta {\cal O}^\gamma + \frac{1}{2} B_{\alpha \beta} {\cal O}^\alpha {\cal O}^\beta + h.c \right], 
\end{equation} 
where the various soft term parameters depend on the closed string and D6-brane displacement moduli (evaluated at their vacuum expectation value):\footnote{To simplify the formulae for the soft terms, we introduced the notations:
\begin{equation}
\begin{array}{rcl}
D_m Y_{\alpha \beta \gamma} &=&\partial_m Y_{\alpha \beta \gamma} - \left( K^{\delta \ov \rho} \partial_m K_{\ov \rho \alpha} Y_{\delta \beta \gamma}  + (\alpha \leftrightarrow \beta)   + (\alpha \leftrightarrow \gamma)  \right) ,\\
D_n \mu_{\alpha \beta}  &=& \partial_m \mu_{\alpha \beta} -  \left(K^{\delta \ov \rho} \partial_m K_{\ov \rho \alpha} \mu_{\delta \beta} + (\alpha \leftrightarrow \beta)   \right),\\
D_n Z_{\alpha \beta}  &=& \partial_m Z_{\alpha \beta} -  \left(K^{\delta \ov \rho} \partial_m K_{\ov \rho \alpha} Z_{\delta \beta} + (\alpha \leftrightarrow \beta)   \right).
\end{array}
\end{equation}
}
\begin{eqnarray}
 m_{\alpha \ov \beta}^2 & = & (m_{3/2}^2 + \frac{V_0}{M_{Pl}^2})  K_{\alpha \ov \beta}  - e^{K^0/M_{Pl}^2} \ov F^{\ov m} \left(  \partial_{\ov m } \partial_n K_{\alpha 
 \ov \beta} - \partial_{\ov m}  K_{\alpha \ov \gamma } K^{\ov \gamma \delta } \partial_n  K_{\delta \ov \beta}   \right) F^n, \notag\\
 A_{\alpha \beta \gamma}  & = & \frac{\ov{\cal W}_0}{|{\cal W}_0|} e^{K^0/ M_{Pl}^2} F^m \left[ \partial_m  K^0\,  Y_{\alpha \beta \gamma} + D_m Y_{\alpha \beta \gamma}  \right], \label{Eq:SoftTermsComplicated} \\
 B_{\alpha \beta} & = & \frac{\ov{\cal W}_0}{|{\cal W}_0|} e^{K^0/ 2M_{Pl}^2} \left\{e^{K^0/ 2M_{Pl}^2}  F^m \left[ \partial_m  K^0\,\mu_{\alpha \beta} + D_m\mu_{\alpha \beta}  \right] - m_{3/2}  \mu_{\alpha \beta} + (2m^2_{3/2} + \frac{V_0}{M_{Pl}^2})    Z_{\alpha \beta}  \right. \notag\\
 && \qquad \qquad \left. - m_{3/2} e^{K^0/ 2M_{Pl}^2} \ov F^{\ov m} \partial_{\ov m}  Z_{\alpha \beta} + m_{3/2} e^{K^0/ 2M_{Pl}^2} F^m D_m   Z_{\alpha \beta} 
 - e^{K^0/ M_{Pl}^2} \ov F^{\ov m} F^n D_n \partial_{\ov m}  Z_{\alpha \beta}  
  \right\}. \notag
\end{eqnarray}
The soft terms depend both on universal data, such as the F-terms\footnote{Note that the expression for the F-terms in this paper differs by a factor $e^{-K^0/2M_{Pl}^2}$ from the expressions usually encountered in the literature. This deliberate choice allows to extract an overall exponential factor $e^{K^0/M_{Pl}^2}$ from the non-universal contribution to the soft terms, in line with the factorisation of the scalar potential~\eqref{Eq:FTermScalarPot2} and the gravitino mass~\eqref{Eq:AppGravMass}.} and the K\"ahler-potential $K^0$, and on model-dependent input data captured through the moduli-dependent K\"ahler metrics $K_{\alpha \ov\beta}$ and coupling parameters $Z_{\alpha \beta}$, $\mu_{\alpha \beta}$, and $Y_{\alpha \beta \gamma}$. 

In the previous section it was shown that the factorability of the closed string and D6-brane displacement moduli in terms of shift-invariant axion polynomials and geometric moduli can be extended to the expressions for the gravitino masses, which serve as order parameters for flux-induced supersymmetry-breaking. Given the structure of the soft terms it is very tempting to expose their factorable character by rewriting them in terms of the shift-invariant axion polynomials and geometric moduli as well. To this end, we consider the orientifold projection suited for the ISD flux vacua with closed string moduli $(T^a, S, U_\Lambda)$ and turn to their respective (contra-variant) F-terms depending linearly on the axion polynomials as denoted in~\eqref{Eq:ContraVariantFtermsCO}.
At this point it suffices to insert the F-term expressions back into the soft terms~\eqref{Eq:SoftTermsComplicated} in order to relate the soft terms to the axion polynomials. Nonetheless, these soft terms do not correspond to the physical parameters as long as the kinetic terms for the open string states are not written in their canonical form. To eliminate the closed string moduli dependence from the open string kinetic terms, an appropriate field redefinition of the open string excitations is required. In case the kinetic terms are all diagonal, i.e.~$K_{\alpha \ov \beta} = K_\alpha \delta_{\alpha \beta}$, such a field redefinition is rather straighforward:
\begin{equation}
{\cal O}^{\alpha} \rightarrow \hat {\cal O}^{\alpha}  =  K_\alpha^{1/2} {\cal O}^{\alpha}.  
\end{equation}
By virtue of this field redefinition, the physical soft terms for the physical open string excitations $\hat {\cal O}^{\alpha}$ reduce to a much simpler form:
\begin{eqnarray} \label{soft-masses}
m_{\alpha}^{2}&=&(m_{3/2}^{2}+V_0)- e^{K^0}F^{\ov m}F^{n}\partial_{\ov m}\partial_n \, \log{K}_\alpha, \notag \\
\hat{ A}_{\alpha\beta\gamma}&=&\hat{Y}_{\alpha\beta\gamma}  F^{m}\left( \partial_m{K}^0+\partial_m \log\,Y_{\alpha\beta\gamma}-\partial_m \log ({K}_{\alpha}\, {K}_{\beta}\, {K}_{\gamma} )
\right), \label{Eq:A-terms}\\
\hat{B}_{\alpha\beta}&=&\hat{\mu}_{\alpha\beta} \left[ e^{K^0/ 2} F^m\left( \partial_m{K}^0+\partial_m \log\,\mu_{\alpha\beta}-\partial_m\,\log ( {K}_\alpha{K}_\beta)\right)-m_{3/2}\right], \notag\\
M_i&=&\frac{1}{2}({\rm Im}\,f{}^{-1}) e^{K^0/ 2} F^{m}\partial_m\,f_{}, \notag
\end{eqnarray}
where we now also included the soft gaugino masses and introduced the physical Yukawa couplings and $\mu$-terms: 
\begin{eqnarray}\label{couplinngs1} 
\hat{Y}_{\alpha\beta\gamma}=\frac{\hat{W}^{*}}{|\hat{W}|}\,e^{{K}^0}\,\left({K}_\alpha {K}_\beta {K}_\gamma\right)^{-1/2}Y_{\alpha\beta\gamma},\qquad\hat{\mu}_{\alpha\beta}=\frac{\hat{W}^{*}}{|\hat{W}|}\,e^{{K}^0}\,\left( {K}_\alpha {K}_\beta\right)^{-1/2} \mu_{\alpha\beta},
\end{eqnarray}
apart from setting $Z_{\alpha \beta} = 0$. In this setting the soft terms can be written quite elegantly by using the factorisation in terms of geometric moduli and axion polynomials. 

\begin{center}
\subsubsection*{Soft Masses}
\end{center}
Focusing first on the soft masses $m_\alpha^2$, we employ the results of the previous section to rewrite them in a matrix notation: 
\begin{equation}\label{Eq:SoftMassesDKM}
m_{\alpha}^{2}
= e^{K^0} \varrho_A\left(\left(\Pi^\dagger \ltimes \Pi \right)^{AB}+\frac{1}{8}Z^{AB}-\left(\mathbb{M}^\dagger\,\qof\,{\mathbb{M}}\right)^{AB}\right)\varrho_B
\end{equation}
where the K\"ahler metric matrix $\qof$
\begin{equation}
\qof=\left(\begin{array}{cccc} 
\partial_{\ov{T}^a}\partial_{T^b} \log {K}_\alpha& \partial_{\ov{T}^a}\partial_S \log {K}_\alpha&\partial_{\ov{T}^a}\partial_{U_\Sigma} \log {K}_\alpha & \partial_{\ov{T}^a}\partial_{\Phi^j_\alpha} \log {K}_\alpha\\
\partial_{\ov{S}}\partial_{T^b} \log {K}_\alpha&\partial_{\ov{S}}\partial_S \log {K}_\alpha&\partial_{\ov{S}}\partial_{U_{\Sigma}} \log {K}_\alpha & \partial_{\ov{S}}\partial_{\Phi^j_\alpha} \log {K}_\alpha \\
\partial_{\ov{U}_\Lambda}\partial_{T^b} \log {K}_\alpha&\partial_{\ov{U}_\Lambda}\partial_S \log {K}_\alpha&\partial_{\ov{U}_\Lambda}\partial_{U_\Sigma} \log {K}_\alpha&\partial_{\ov{U}_\Lambda}\partial_{\Phi^j_\alpha} \log {K}_\alpha\\
\partial_{\ov{\Phi}^i_\alpha}\partial_{T^b} \log {K}_\alpha&\partial_{\ov{\Phi}^i_\alpha}\partial_{S} \log {K}_\alpha&\partial_{\ov{\Phi}^i_\alpha}\partial_{U_\Sigma} \log {K}_\alpha&\partial_{\ov{\Phi}^i_\alpha}\partial_{\Phi^j_\alpha} \log {K}_\alpha\\
 \end{array}\right),
\end{equation}
is introduced to capture the model-dependent\footnote{The epithet ``model-dependent" refers to the freedom of choice regarding the D6-brane configuration once a Calabi-Yau orientifold background is chosen.} contributions to the soft masses and the matrix $\mathbb{M}$ collects all saxion-dependent terms appearing in the contra-variant F-term vector~\eqref{Eq:ContraVariantFtermsCO}. For generic Calabi-Yau manifolds the explicit expressions for the K\"ahler metrics is beyond the scope of present-day computational technology, such that the model-dependent contributions seem to remain unknown at first sight. Nevertheless, closer inspection of the F-term expressions and the K\"ahler metric matrix $\qof$ suggest that it is sufficient to know the scaling behaviour of the K\"ahler metrics $K_\alpha$ to fully determine the model-dependent part of the soft masses. Let us clarify this bold statement by evaluating the soft masses in the CSD vacua represented by the constraints~\eqref{MinkD6}. In these CSD vacua, supersymmetry is broken by the F-terms of the complex structure moduli, i.e.~$ (\vec{F}^A)^t = \left( 0 , 0, F^{U_\Lambda},0\right)$, such that the model-dependent part of the soft terms reduces to:
\begin{equation}\label{Eq:SoftMassISD}
\vec{\varrho}{}^t \cdot \mathbb{M}^T\;\qof\;\ov{\mathbb{M}} \cdot \vec{\varrho} = e^{K_0} |W_0|^2  u_{\star \Lambda} u_{\star \Sigma} \partial_{u_{\star \Lambda}} \partial_{u_{\star \Sigma}} \log K_\alpha  .
\end{equation} 
Under the assumption that the K\"ahler metrics on generic Calabi-Yau manifolds can be locally approximated by their counterparts on toroidal orbifolds discussed in appendix~\ref{A:TorOrb}, we consider the K\"ahler metrics $K_\alpha$ to be homogeneous functions of degree $n_\alpha$ in the complex structure moduli $u_{\star \Lambda}$. Hence, it follows straightforwardly that $u_{\star \Lambda} u_{\star \Sigma} \partial_{u_{\star \Lambda}} \partial_{u_{\star \Sigma}} \log K_\alpha = - n_\alpha$, which leads to a simple expression for the soft masses~\eqref{Eq:SoftMassesDKM} in terms of the gravitino mass:
\begin{equation}
m_\alpha^2 = m_{3/2}^2 ( 1 + n_\alpha).
\end{equation}
 To find the scaling dimension (or modular weight) $n_\alpha$ for an open string state ${\cal O}^\alpha$ we further exploit the knowledge of K\"ahler metrics for intersecting D6-branes on toroidal orbifold compactifications. Similarly to the toroidal orbifold set-up, we distinguish two different sectors based on the origin of the charged open string state:

\begin{itemize}      

\item[(i)] Vector-like/Non-chiral matter:\\
Whenever two supersymmetric D6-branes intersect on a continuous subspace along the internal Calabi-Yau orientifold, their intersection number follows by computing the Euler characteristic of the intersection space.\footnote{When calculating the intersection number for two overlapping surfaces, one of the surfaces has to be deformed along normal directions~\cite{Brunner:1999jq,Blumenhagen:2002wn}. Due to the special Lagrangian property of the cycles considered in this paper, the normal deformations can be mapped to vector fields in the tangent bundle of the intersection space by McLean's theorem. The intersection number is then computed as the number of simple zeros for non-vanishing sections of the tangent bundle, which is equal to the Euler characteristic of the intersection space by the Poincar\'e-Hopf index theorem.} Thus, in case of a codimension 5 intersection with topology $S^1\simeq \R\P^1$, their intersection number is zero due to the vanishing Euler characteristic. Yet the intersection of two such D6-branes can provide for vector-like pairs of ${\cal N}=1$ chiral multiplets. To our knowledge a systematic study of vector-like matter at intersecting D6-branes has not yet been undertaken for generic Calabi-manifolds and the K\"ahler metrics for such states are therefore unknown. Though, we expect that the K\"ahler metrics for vector-like matter can be modelled locally around the intersection locus by homogeneous functions of the closed string moduli and that they exhibit the same scaling behaviour as their counter-parts on toroidal orbifolds. Under this assumption, we can exploit the structure of the K\"ahler metric~\eqref{Eq:KaehlerBifundN2Munich} for vector-like matter on toroidal orbifolds and distinguish between two cases:
the K\"ahler metrics are homogeneous functions of degree $-1$ in the complex structure moduli (thus with modular weight $n_\alpha = -1$), in which case the vector-like matter states do not acquire soft masses. The other option occurs for K\"ahler metrics that are homogeneous of degree $-1/2$ in the complex structure moduli and $-1/2$ in the dilaton (with modular weight $n_\alpha  = - \frac{1}{2}$), for which the vector-like matter does acquire a soft mass $m_\alpha^2 = \frac{m_{3/2}^2}{2} $. 

\item[(ii)] Chiral Matter:\\
Two supersymmetric D6-branes can intersect in a single point of the internal space, in which case a chiral ${\cal N}=1$ supermultiplet in the bifundamental representation is supported at the codimension 6 intersection. Also for these chiral matter states the K\"ahler metrics on generic Calabi-Yau manifolds are unknown, but a modellisation in terms of homogeneous functions depending on the closed string moduli is undoubtedly possible around the intersection locus. As such, we expect the K\"ahler metrics for chiral matter states to exhibit to same scaling behaviour as their counterparts~\eqref{Eq:KaehlerBifundN1Munich} computed for toroidal orbifolds with modular weight $n_\alpha = - \frac{3}{4}$. This implies that the chiral matter states always acquire soft masses in CSD flux vacua of the order $m_\alpha^2 = \frac{m_{3/2}^2}{4}$.     
\end{itemize}

\begin{table}[h]
\begin{center}
\begin{tabular}{c@{\hspace{0.4in}}lc}
\hline
\multicolumn{3}{c}{Soft Terms in Type IIA non-SUSY Minkowski vacua with D6-branes}\\
\hline
\hline
Soft masses & $m_\alpha^2 = m_{3/2}^2 ( 1 + n_\alpha)$ \\
A-terms &$\hat A_{\alpha\beta\gamma} =  \hat{Y}_{\alpha\beta\gamma} m _{3/2} \left(3  + n_\alpha + n_\beta + n_\gamma \right)$ \\
B-terms & $  \hat B_{\alpha\beta} =\hat{\mu}_{\alpha\beta} m _{3/2} \left(2  + n_\alpha + n_\beta \right)$ \\
Gaugino masses& $M_i = m_{3/2}$\\
\hline
\end{tabular}
\caption{Summary of the soft terms in CSD vacua represented by the constraints~\eqref{MinkD6}. A coefficient $n_{\alpha}$ represents the modular weight (degree of the complex structure moduli in the K\"ahler metrics) for the open string excitation ${\cal O}^\alpha$.  \label{Tab:SoftTermsnonSUSYMink}}
\end{center}
\end{table}

\begin{center}
\subsubsection*{A-terms, B-terms and Gaugino Masses}
\end{center}
In type IIA compactifications, Yukawa or cubic interactions involving chiral matter states arise from worldsheet instantons {$\alpha'$-corrections}, which correspond to two-dimensional surfaces with boundaries along the intersecting three-cycles \cite{Aldazabal:2000cn,Cremades:2003qj}. The holomorphic character of the two-dimensional surfaces, with the topology of a disc, ensures that the cubic couplings contribute to the superpotential. The amplitude $Y_{\alpha \beta \gamma}$ of the three-point coupling in \eqref{Eq:EffKaehlerPotSuperPot} is an exponential function depending on the surface area, which can be expressed in terms of K\"ahler moduli. The amplitude $Y_{\alpha \beta \gamma}$ can also include holomorphic couplings to the open string moduli encoding the D6-brane position and Wilson line, such that ${\cal H} \in \{ T^a, \Phi_\alpha^i \}$ for cubic interactions. The fact that {the complex structure moduli do not enter in the holomorphic piece of the Yukawa interactions} has immediate consequences for the flux-induced A-terms in~\eqref{Eq:A-terms}, which can be similarly written in matrix notation by virtue of the matrix $\mathbb{M}$:  
\begin{equation}\label{A-termsrho-basis}
\hat A_{\alpha\beta\gamma} = -i \hat{Y}_{\alpha\beta\gamma}\;  \left(  \partial_{\vec{{\cal H}}}K^{0\; t}  +  \vec{\tsadisofit}{}^t \right)  \cdot {\mathbb{M}} \cdot  \vec{\rho},
\end{equation}
allowing to expose the dependence on the axion polynomials. In this expression we distinguish between a model-independent contribution presented by the vector $ \partial_{\vec{{\cal H}}}K^{0\; t} \equiv \left(\partial_{T^a} K^0,  \partial_S K^0, \partial_{U_\Lambda} K^0, \partial_{\Phi^i_\alpha} K^0   \right)$ and a model-dependent contribution in terms of the vector $\vec{\tsadisofit}$: 
\begin{equation}
\vec{\tsadisofit} = \left( \begin{array}{c} \partial_{T^a} \log\,Y_{\alpha\beta\gamma}-\partial_{T^a} \log ({K}_{\alpha}\, {K}_{\beta}\, {K}_{\gamma})\\
\partial_S \log\,Y_{\alpha\beta\gamma}-\partial_S \log ({K}_{\alpha}\, {K}_{\beta}\, {K}_{\gamma})\\
\partial_{U_\Lambda} \log\,Y_{\alpha\beta\gamma}-\partial_{U_\Lambda} \log ({K}_{\alpha}\, {K}_{\beta}\, {K}_{\gamma})\\
\partial_{\Phi^i_\alpha} \log\,Y_{\alpha\beta\gamma}-\partial_{\Phi^i_\alpha} \log ({K}_{\alpha}\, {K}_{\beta}\, {K}_{\gamma})\\
\end{array}
 \right). 
\end{equation}
The structure of the vector $\vec{\tsadisofit}$ implies that it is sufficient to know the functional dependence of the Yukawa-coupling $Y_{\alpha \beta \gamma}$ on the hidden  sector moduli ${\cal H}$ and the modular weights of the K\"ahler metrics to determine the model-dependent contribution to the A-terms. Once again such a strong statement can be best clarified with the CSD vacua~\eqref{MinkD6} as an example. In these ${\cal N}=0$ vacua with F-term vector $ (\vec{F}^A)^t = \left( 0 , 0, F^{U_\Lambda},0\right)$, there are only contributions from the complex structure moduli sector to the A-terms:
\begin{equation}\label{Eq:YukWorldInst}
\begin{array}{rcl}
\hat A_{\alpha\beta\gamma}&=&\hat{Y}_{\alpha\beta\gamma} \left(  \frac{\partial_{u_{\star \Lambda}}\tilde{\cal G}_{Q}}{\tilde{\cal G}_Q} - \frac{1}{2} \partial_{u_{\star \Lambda}} \log\,Y_{\alpha\beta\gamma} + \frac{1}{2}\partial_{u_{\star \Lambda}} \log ({K}_{\alpha}\, {K}_{\beta}\, {K}_{\gamma})    \right) e^{K_0/2} \frac{2}{3} {\cal K} \tilde \varrho u_{\star \Lambda} \\
&=&  \hat{Y}_{\alpha\beta\gamma} m _{3/2} \left(3  + n_\alpha + n_\beta + n_\gamma \right).
\end{array} 
\end{equation}
To arrive at the last step, we used  that $\tilde{\cal G}_Q$ is a homogeneous function of degree 3/2 in the complex structure moduli, that the K\"ahler metrics $K_\alpha$ are also homogeneous functions of degree $n_\alpha$ in the complex structure moduli, and that holomorphic Yukawa couplings generated by worldsheet instantons do not depend on the complex structure moduli.

In a similar fashion quadratic couplings in the superpotential~\eqref{Eq:EffKaehlerPotSuperPot} might result from worldsheet instantons \cite{Marchesano:2007de}, and these will again be independent from the complex structure moduli. In non-supersymmetric vacua the quadratic couplings give rise to physical B-terms, which can be decomposed in model-independent and model-dependent pieces:
\begin{equation}\label{B-termsrho-basis}
\hat B_{\alpha\beta}=\hat{\mu}_{\alpha\beta} \left[ -i \left( \partial_{\vec{{\cal H}}}K^{0\; t} + \vec{\beth}{}^t \right) \cdot {\mathbb{M}} \cdot \vec{\rho}   - m_{3/2}   \right],
\end{equation}
where the only model-dependent contribution is encoded in the vector $ \vec{\beth}$:
\begin{equation}
\vec{\beth} = \left( \begin{array}{c}
\partial_{T^a} \log\,\mu_{\alpha\beta}-\partial_{T^a}\,\log ( {K}_\alpha{K}_\beta) \\
\partial_S \log\,\mu_{\alpha\beta}-\partial_S\,\log ( {K}_\alpha{K}_\beta)  \\
\partial_{U_\Lambda} \log\,\mu_{\alpha\beta}-\partial_{U_\Lambda}\,\log ( {K}_\alpha{K}_\beta) \\
\partial_{\Phi^i_\alpha} \log\,\mu_{\alpha\beta}-\partial_{\Phi^i_\alpha}\,\log ( {K}_\alpha{K}_\beta)  \\
\end{array}
\right).
\end{equation}
Also in this case, the knowledge about the modular weights of the K\"ahler metrics and the functional dependence of the coupling $\mu_{\alpha \beta}$ on the closed string moduli, i.e.~$\log \mu_{\alpha \beta}$ is a homogeneous function of degree 0, are sufficient to determine the physical B-terms. Using the CSD vacua~\eqref{MinkD6} as an explicit example, we obtain the following expressions:
\begin{equation}\label{Eq:QuadWorldInst}
\begin{array}{rcl}
\hat B_{\alpha\beta}&=&\hat{\mu}_{\alpha\beta} \left(  \frac{\partial_{u_{\star \Lambda}}\tilde{\cal G}_{Q}}{\tilde{\cal G}_Q} - \frac{1}{2} \partial_{u_{\star \Lambda}} \log\,\mu_{\alpha\beta} + \frac{1}{2}\partial_{u_{\star \Lambda}} \log ({K}_{\alpha}\, {K}_{\beta})    \right) e^{K_0/2} \frac{2}{3} {\cal K} \tilde \varrho \, u_{\star \Lambda} - \hat{\mu}_{\alpha\beta}  m_{3/2}  \\
&=&  \hat{\mu}_{\alpha\beta} m _{3/2} \left(2  + n_\alpha + n_\beta \right).
\end{array} 
\end{equation}

In order for worldsheet instantons to contribute to the superpotential, the associated quadratic and cubic couplings of open string states in the superpotential~\eqref{Eq:EffKaehlerPotSuperPot} have to form singlets under the full gauge group supported by  the D6-branes. In case this field theory selection rule is violated for massive Abelian gauge groups by a coupling, it will not result from a worldsheet instanton, but there exist a completely different set of non-perturbative corrections that can generate such couplings, namely D-brane instantons \cite{Blumenhagen:2006xt,Ibanez:2006da,Blumenhagen:2007zk,Blumenhagen:2009qh}. These Euclidean D2-branes wrap completely along internal special Lagrangian three-cycles and are non-perturbative in the string coupling. Furthermore, the amplitude of a D-brane instanton correction depends holomorphically on complex structure moduli. In that case, the functional dependence of the D-brane instanton will provide for an additional model-dependent contribution to the A-terms and B-terms.\footnote{In principle both quadratic and cubic couplings in the superpotential can arise from D-brane instantons and subsequently give rise to B-terms and A-terms that differ from~\eqref{Eq:QuadWorldInst} and~\eqref{Eq:YukWorldInst}  respectively. More precisely, due to the exponential structure of such instanton amplitudes one can immediately deduce that $\log \mu_{\alpha \beta}$ and $\log Y_{\alpha \beta \gamma}$ are homogeneous functions of degree 1 in the complex structure moduli (when poly-instanton corrections are neglected), such that the respective B-terms and A-terms take the form:
\begin{equation}
\begin{array}{rcl}
\hat B_{\alpha\beta} &=&\hat{\mu}_{\alpha\beta} m _{3/2} \left(2  + n_\alpha + n_\beta - \log \mu_{\alpha \beta} \right),\\
\hat A_{\alpha\beta\gamma} &=& \hat{Y}_{\alpha\beta\gamma} m _{3/2} \left(3  + n_\alpha + n_\beta + n_\gamma  - \log Y_{\alpha \beta\gamma} \right),
\end{array}
\end{equation}  
and acquire a moduli-dependent contribution.
}   

Last but not least, also gaugino masses are expected to arise from spontaneous supersym-metry-breaking in the moduli sector with non-vanishing F-terms. In order to compute these gaugino mass, the functional dependence of the holomorphic gauge kinetic function is indispensable. The gauge kinetic functions $f_\alpha$ for gauge theories on D6-branes follow directly from the dimensional reduction of the D-brane Chern-Simons and Dirac-Born-Infeld action~\cite{Kerstan:2011dy,Grimm:2011dx}. For a D6-brane wrapping a three-cycle $\Pi_\alpha$, the (tree-level) gauge kinetic function $f_{\alpha}$ is a linear, holomorphic function of the dilaton and/or the complex structure moduli:\footnote{The tree-level expression for the gauge coupling follows directly from the dimensional reduction of the DBI-action. However, such a KK reduction does not offer a fully holomorphic expression for the gauge kinetic function in the presence of open string D-brane moduli. Only one-loop corrections to the gauge kinetic functions~\cite{Berg:2004ek} allow for a proper holomorphic gauge kinetic function, depending on the redefined complex structure moduli. Such a computation goes beyond the scope of this paper.}
\begin{equation}\label{Eq:HoloGaugeKin}
f_\alpha =  c_{\alpha} S_\star + \sum_{\Lambda} d^\Lambda_\alpha U_{\star \Lambda},
\end{equation}    
where the integers $c_{\alpha}$ and $d^\Lambda_\alpha$ encode information about the three-cycle geometry. To arrive at the gaugino masses, we first rewrite their expression in matrix form by virtue of the F-term factorisation~\eqref{Eq:ContraVariantFtermsCO}:
\begin{equation}
M_\alpha = \frac{1}{2} e^{K^0/ 2} \IM(f_\alpha{}^{-1})  (\partial_{\vec{{\cal H}}} f_{\alpha})^t \cdot {\mathbb{M}} \cdot \vec{\varrho},
\end{equation}
where we introduced the vector $(\partial_{\vec{{\cal H}}} f_{\alpha}{})^t = \left(\partial_{T^a} f_{\alpha}, \partial_S f_{\alpha} , \partial_{U_\Lambda} f_{\alpha }, \partial_{\Phi^i_\alpha} f_{\alpha} \right)$ as a shorthand notation. The linear dependence on the complex structure moduli in~\eqref{Eq:HoloGaugeKin} is sufficient knowledge to determine the gaugino masses in a supersymmetry-breaking vacua. Evaluating the gaugino masses for D6-branes with $c_{\alpha} = 0$ in the CSD vacua~\eqref{MinkD6}, for instance, leads to the familiar expression:
\begin{equation}
M_\alpha =  \frac{1}{2\, \IM(f_\alpha)}  \sum_{\Lambda }d^\Lambda_\alpha u_{\star \Lambda} \frac{2}{3}  {\cal K} \tilde \varrho e^{K^0/2}   = m_{3/2},
\end{equation}
that equates the gaugino mass and the gravitino mass.

A summary of the soft terms in CSD vacua is offered by table~\ref{Tab:SoftTermsnonSUSYMink}. Our results generalise previous results in the literature, in the sense that they also apply to vacua with open string moduli. Indeed, typical soft-term scenarios in type IIB ISD flux vacua correspond to spontaneously broken supersymmetry with non-vanishing F-terms in the K\"ahler moduli sector~\cite{Ibanez:2004iv,Camara:2004jj,Font:2004cx}, which corresponds via mirror symmetry to non-vanishing F-terms in the complex structure moduli sector for Type IIA ISD flux vacua. We find that CSD vacua have the same structure of contravariant F-terms as ISD flux vacua. Therefore, upon assuming that the chiral fields K\"ahler metrics are homogeneous polynomials, we obtain a similar soft term structure. Modelling the K\"ahler metrics for the chiral open string states as homogeneous polynomials in the geometric moduli is mostly inspired by the known results for toroidal models as summarised in appendix~\ref{A:TorOrb}, yet it has been adopted as a standard practice in the literature~\cite{Conlon:2006tj,Conlon:2006wz,Aparicio:2008wh} to parameterise the K\"ahler metrics for generic Calabi-Yau manifolds. Here, we fully exploit the scaling behaviour of the K\"ahler metrics to simplify the model-dependent contributions to the soft terms as much as possible.

\section{Validity of the Type IIA Flux Landscape}\label{S:ScalesIIA}
The previous sections have been devoted to deriving the vacuum structure, spontaneous supersymmetry-breaking and soft terms for perturbative flux vacua in terms of the shift-invariant axion polynomials. A hidden premise behind this approach is the consideration that the low-energy effective description for flux compactifications (with D6-branes) is captured by a four-dimensional ${\cal N}=1$ supergravity theory. To asses the validity of the premise and guarantee the overall consistency of a flux compactification (with D6-branes), one has to determine the geometric scales at which distinct particle states acquire their mass and argue for an adequate separation of scales. 

The first geometric scale to determine in terms of the compactification data is the string mass scale, which follows upon comparison between the Einstein-Hilbert action and the four-dimensional effective field theory arising from the dimensional reduction of the ten-dimensional type IIA supergravity action. More precisely, we start from the kinetic terms for the massless bosonic type IIA string states in the string frame: 
\begin{equation}\label{Eq:10dIIA}
{\cal S} = -\frac{1}{2 \kappa_{10}^2} \int e^{-2 \phi} \left[ \mathfrak{R}\star_{10} \1 - 4 d\phi \wedge \star_{10} d\phi + \frac{1}{2} H_3 \wedge \star H_3\right] - \frac{1}{8\kappa_{10}^2} \int \sum_{p=0}^5 G_{2p} \wedge \star_{10} G_{2p},
 \end{equation} 
where $\mathfrak{R}$ corresponds to the ten-dimensional Ricci scalar, $H_3$ to the NS 3-form field strength and $G_{2p}$ to the RR-form field strengths as introduced in section~\ref{S:IIAFluxVacua}. The conversion to the Einstein frame requires a rescaling of the ten-dimensional metric, i.e.~$G^{(10)} \rightarrow G^{(10)}_E= e^{(\phi - \phi_0)/2} G^{(10)}$, while an overall rescaling of the four-dimensional metric in the form $g^{(4)}_E \rightarrow \frac{{\cal V}^0_E}{{\cal V}_E} g^{(4)}_E $ sneaks into the six-dimensional volume-dependence of the string mass scale:
\begin{equation}
M_{\rm string}^2 = \frac{g_s^2}{4\pi} \frac{M_{Pl}^2}{{\cal V}^0_E} .
\end{equation}
In this expression the string coupling constant $g_s = e^{\phi_0}$ is related to the {\it vev} of the ten-dimensional dilaton and ${\cal V}^0_E$ corresponds to the (dimensionless) volume of the Calabi-Yau orientifold evaluated at the vacuum for the geometric moduli in the Einstein frame. 

In the presence of background fluxes along the internal dimension a perturbative potential~\eqref{Eq:VFbil} for the geometric moduli and axions arises upon the dimensional reduction of the ten-dimensional supergravity action~\eqref{Eq:10dIIA} to four dimensions. This scalar potential matches precisely the F-term scalar potential from the ${\cal N}=1$ supergravity analysis with the K\"ahler potentials given by~\eqref{Eq:KahlerPotKahlerMod} and~\eqref{Eq:KaehlerPotCS1} and the superpotential by \eqref{Eq:SuperPotFactForm} for the pure closed string sector. As we reviewed in previous sections, the inclusion of (mobile) D6-branes into the compactification can be easily mediated through a redefinition of the complex structure moduli whose K\"ahler potential is subsequently given by~\eqref{Eq:OpenClosedKaehlerPotential}, while the superpotential is extended by the bilinear term~\eqref{Eq:OpenClosedSuperPotential}. This supergravity analysis is valid for small string coupling and large internal volume, for which the string mass scale obviously lies below the Planck mass scale. As a second criterion for the validity of the supergravity analysis one has to ensure that the tower of massive Kaluza-Klein states decouples from the massless KK-modes, such that the effective field theory below the KK-scale consists purely of the (massless) ${\cal N}=1$ chiral multiplets containing the K\"ahler moduli, complex structure moduli and open string moduli (as well as other massless open string excitations). Strictly speaking, it is unknown how to determine the KK mass scale for compactifications on generic Calabi-Yau manifolds, yet an adequate approximation follows~\cite{Conlon:2005ki} from toroidal compactifications with characteristic radius size $R = R_s \ell_s$. If we use the dimensionless radius $R_s$ as a proxy for the internal volume ${\cal V}_s^0$, i.e.~${\cal V}_s^0 = (2\pi R_s)^6$ expressed in the string frame, we find a Kaluza-Klein mass scale of the order
\begin{equation}
M_{KK} \sim 2\pi \frac{M_{\rm string}}{({\cal V}_s^0)^{1/6}} \sim   \frac{g_s \sqrt{\pi} M_{Pl}}{({\cal V}_E^0)^{2/3}}.
\end{equation}  
Thus, the ${\cal N}=1$ supergravity analysis represents the effective field theory description of four-dimensional type IIA compactifications for energy scales below the KK-mass scale, and other mass generating effects should yield masses below this scale. For instance, the moduli masses induced by perturbative NS-fluxes take the following form,  
\begin{equation}
M_{\rm mod} \sim \frac{N_{\rm flux}}{4\pi} \frac{M_{\rm string}}{\sqrt{{\cal V}_E^0}} \sim \frac{N_{\rm flux}}{4\pi} \frac{g_s M_{Pl}}{{\cal V}_E^0},
\end{equation}
and lie below the KK-scale for large internal volumes ${\cal V}_s^0 >1$. This scaling of the moduli masses in perturbative type IIA flux vacua can be obtained following the same reasoning as in~\cite{Conlon:2005ki}: the rescaling of the ten-dimensional metric considered above allows to express all relevant quantities, such as the K\"ahler potential and superpotential, in the Einstein frame, after which the scaling with the internal volume can be deduced for the physical moduli masses in the vacuum configuration.

For closed string ISD flux vacua and the CSD vacua in~\eqref{MinkD6}, supersymmetry is spontaneously broken in the complex structure moduli sector and a non-vanishing gravitino mass is induced:
\begin{equation}
m_{3/2} = \ov m_{3/2} \sim g_s |{\cal W}_0| \frac{M_{\rm string}}{{\cal V}_E^0} \sim g_s^2 |{\cal W}_0| \frac{M_{Pl}}{({\cal V}_E^0)^{3/2}},
\end{equation}
where ${\cal W}_0 = \ell_s W_0$ is dimensionless. This gravitino mass clearly lies below the KK mass scale for large internal volumes. Moreover, as we have shown in the previous section and summarised in table~\ref{Tab:SoftTermsnonSUSYMink}, all soft terms in such vacua are proportional to the gravitino mass, such that also the soft masses for the chiral open string excitations lie below the KK-scale. Hence, ${\cal N}=0$ Minkowski vacua with (partly) stabilised moduli through perturbative background fluxes easily satisfy the na\"ive mass hierarchy that is required to justify a Wilsonian effective field theory approach.  Furthermore, in the supergravity limit one can also argue from the ten-dimensional equations of motion that the ten-dimensional dilaton is bounded from above, such that the perturbative type IIA flux vacua with non-vanishing Romans mass are inherently weakly coupled in the string coupling~\cite{Aharony:2010af}.

A more profound worry about the validity of type IIA flux vacua with Romans mass $m\neq 0$ concerns~\cite{McOrist:2012yc} their proper existence as solutions of ten-dimensional supergravity. In first instance, it is not a priori clear whether a Calabi-Yau manifold can be considered a proper compactification background in the presence of internal fluxes. In the case of type IIA ISD flux vacua this worry seems unfounded, as we expect the fluxes to be diluted at large volume such that warping or other back-reaction effects on the Calabi-Yau metric can be neglected to first order, similarly to the mirror dual ISD flux vacua in type~IIB. The supersymmetric AdS vacua on the other hand require a more careful treatment to ensure that they are genuine ${\cal N}=1$ supersymmetric backgrounds with an $SU(3)$ structure. To solve the ten-dimensional equations of motion for Minkowski or Anti-de Sitter compactifications it suffices~\cite{Lust:2004ig} to solve for the supersymmetry variations of the dilatini and gravitini, alongside the Bianchi identities for the RR- and NS-field strengths. By virtue of the pure spinor formulation of generalized complex geometry one can then show that supersymmetric AdS vacua solve the supersymmetry variations with a constant dilaton and form a special subclass of the L\"ust-Tsimpis AdS vacua~\cite{Lust:2004ig,Koerber:2010bx}.

Secondly, to obtain a full-fledged 10d supergravity solution also the Bianchi identities have to be satisfied in the presence of sources. In the case of the RR two-form flux~$G_2$ solving the Bianchi identity might be more involved due to the presence of sources: the NS-three-form acts as a magnetic source for $G_2$ in the presence of a non-vanishing Romans mass. Apart from background fluxes the Bianchi identity for $G_2$ one also has to take into account the RR-charges of the D6-branes and O6-planes, as presented in equation~\eqref{Eq:BianchiIdentGeneral}. As the smooth $H$-flux distribution cannot be cancelled against the localized charges of the O6-planes, it is impossible to solve this Bianchi identity for a two-form flux~$G_2$ consisting only of a harmonic and exact component.\footnote{In the literature smeared O6-planes were proposed~\cite{Acharya:2006ne} as a solution to solve the Bianchi identities for the RR two-form flux. However, it is not a priori clear~\cite{Blaback:2010sj} that smearing O-planes offers consistent approximate solutions to the string theory equations with localised O-planes. Fortunately, solutions with localised O6-planes do exist in massive type IIA supergravity~\cite{Saracco:2012wc}, such that the search for consistent, global type IIA vacua with fluxes, O-planes and D-branes is a well-defined scientific problem.} Adding D6-branes to the mix can help to alleviate the RR tadpoles along the internal directions, but do not help to mediate the non-closedness of the $G_2$-flux. In order to see how the addition of mobile D6-branes alters the type IIA compactifications with ISD fluxes, we included them in section~\ref{S:FluxVacD6branes} and observed that they give rise to  ${\cal N}=0$ CSD vacua, with physically observable features such as a gravitino mass and soft masses. The similarities between the pure ISD flux vacua and the CSD vacua invite to add mobile D6-branes to the known supersymmetric AdS vacua  and search for full-fledged 10d supergravity solutions on Calabi-Yau orientifold or more generic $SU(3)\times SU(3)$ structure backgrounds, such that the supersymmetry variations for the dilatini and gravitini still vanish in the modified vacuum structure with D6-branes.

\section{Conclusions}\label{sec:con}
This paper offers a novel perspective on perturbative type IIA flux vacua with (partly) stabilised moduli and their physical properties at the level of four-dimensional ${\cal N}=1$ supergravity. These four-dimensional vacua correspond to local minima of the four-dimensional scalar potential arising from the dimensional reduction of the tree-level ten-dimensional IIA supergravity action on Calabi-Yau orientifolds with background RR-fluxes, NS-fluxes and D6-branes. Earlier studies of this scalar potential revealed its very simple structure consisting of a symmetric matrix depending solely on the geometric moduli and acting as a metric on the space of axion polynomials. These axion polynomials capture the axionic partners together with the flux quanta into shift-invariant combinations whose precise shapes are intimately connected to Freed-Witten anomaly cancelation. This bilinear structure of the scalar potential in terms of the axion polynomials even persists in the presence of D6-branes accompanied with displacements moduli, referred to as mobile D6-branes in this paper, albeit with the proper addition of open string moduli and axions. Similarly, the shape of the open string axion polynomials can be related to the Hanany-Witten effect.    

At large volume the four-dimensional scalar potential can equally be obtained from the F-term scalar potential of an ${\cal N}=1$ supergravity coupled to chiral multiplets consisting of K\"ahler moduli, complex structure moduli and open string moduli. The background fluxes yield a perturbative superpotential for the closed and open string moduli, such that its form can be expressed as a linear function of the axion polynomials with saxion-dependent  coefficients. It is precisely the complete factorisation of the superpotential in terms of geometric moduli and axion polynomials that lies at the heart of our search for vacuum configurations of the four-dimensional ${\cal N}=1$ supergravity. By solving the F-terms in terms of the axion polynomials we are able to recover the ${\cal N}=1$ supersymmetric AdS vacua and the ${\cal N}=0$ Minkowski vacua with ISD fluxes for purely closed string compactifications.

In the presence of mobile D6-branes, the search for (local) minima of the scalar potential appears at first sight to be much more energy-consuming, as the mixing between closed and open string moduli provides for an extra level of complexity. However, the language of axion polynomials allows to treat these cases in the same way as the pure closed string vacua. More precisely, when generalising the ISD flux set-up by adding D6-branes one can still take advantage of the no-scale symmetry in the complex structure moduli sector to rewrite the scalar potential as a positive semidefinite function, under mild assumptions about the functional dependence of the K\"ahler potential on closed and open string moduli. This positive semidefinite scalar potential has a local ${\cal N}=0$ Minkowski minimum, in which the F-terms for the dilaton, K\"ahler moduli and open string moduli satisfy relations that are weaker than the ISD case. Yet, to expose which sectors break supersymmetry spontaneously, it suffices to look at the contra-variant F-terms in the complex structure moduli sector, which are the only non-vanishing ones for these vacuum configurations and thereby earned them the name complex structure dominated (CSD) vacua. Alternatively, these CSD vacua can also be derived by exploiting the bilinear structure of the open-closed string scalar potential, in which case the vacuum conditions are formulated in terms of the axion polynomials. Once again, the elegant language of the axion polynomials allows to expose the equivalence between the F-term conditions and the axion polynomial vacuum conditions.         

Determining the on-shell F-terms is a necessary step to understand whether a four-dimensional vacuum preserves supersymmetry or not. To assess physically whether supersymmetry is spontaneously broken in the vacuum, it suffices to evaluate the (effective) gravitino mass on-shell. A simple method to do precisely that takes advantage of the off-shell expression for the gravitino mass, which exhibits a bilinear form in the axion polynomials, similarly to the scalar potential. This factorisation in terms of geometric moduli and axion polynomials can also be extended to the soft terms for massless open string excitations located at the intersections of two distinguishable D6-branes. These soft-terms, resulting from the background fluxes through gravity mediation, also take on a (bi)linear expression in terms of the axion polynomials. Hence, this implies that gravitino masses and soft terms are universal for flux vacua that are related through each other by the axion shift symmetries, which is displayed explicitly in terms of the axion polynomials. Here, we have extended the analysis for the soft terms to the CSD vacua, yet their on-shell values exhibit similar scalings with the gravitino mass as the well-studied ISD flux vacua. This similarity suggests a universal pattern for the soft terms in vacua with complex structure dominated supersymmetry breaking.   

A proper look at the ISD flux vacua and the CSD vacua shows that only part of the moduli is stabilised. The no-scale property in the complex structure moduli sector implies that they remain flat directions in this type of vacua. Hence, additional stabilising effects have to be introduced in the compactification to obtain a stable vacuum configuration. In first instance, one may take into account the $\alpha'$ corrections in the K\"ahler moduli sector, which allow to look for vacua in the moduli space regions where the internal volume is only moderately large. Subsequently, one could also take into account various non-perturbative contributions to the superpotential (and K\"ahler potential), such as worldsheet instantons and D-brane instantons, which would however manifestly break the bilinear description in terms of the axion polynomials. It would be illuminating to develop a formalism that combines the perturbative and non-perturbative contributions to the superpotential and allows for elegant methods to determine the vacua of the compactification, in a similar fashion as we explained here for the axion polynomial language.     

It would also be interesting to extend our results to include more general classes of type IIA flux vacua. On the one hand one could consider flux compactifications on non-Calabi-Yau geometries \cite{Behrndt:2004km,Behrndt:2004mj,Villadoro:2005cu,House:2005yc,Camara:2005dc,Grana:2006kf,Aldazabal:2007sn,Tomasiello:2007eq,Koerber:2008rx,Lust:2008zd,Andriot:2016ufg,Blaback:2018hdo}. On the other hand one may consider compactification with more general open string sectors, like models containing coisotropic D8-branes \cite{Font:2006na,Koerber:2009he,Sevrin:2008tp,Sevrin:2009na}. In particular, it would be interesting to see if one can generalise the CSD vacua of section \ref{Ss:FluxVacuaD6-branesModStab} to any of these cases, and then compute the corresponding spectrum of soft terms. Since we have addressed the latter from a 4d effective theory approach, it would be important to develop a microscopic picture of the generation of such soft terms, equivalent to the microscopic computations made in the context of type IIB/F-theory flux backgrounds \cite{Camara:2003ku,Grana:2003ek,Camara:2004jj,Marchesano:2004yn,Lust:2005bd,Jockers:2005zy,Gomis:2005wc,Burgess:2006mn,Lust:2008zd,Camara:2013fta,Camara:2014tba}. One may then compare such soft terms with the results of table \ref{Tab:SoftTermsnonSUSYMink}, and use this to either confirm or correct our Ansatz for the K\"ahler metrics of the chiral open string modes. It would also be interesting to explore the implications of our findings for the phenomenological applications of type IIA flux vacua like, e.g., revisit the cosmological scenarios in \cite{Escobar:2015fda,Escobar:2015ckf}. In any event, we expect that our results help to achieve a wider understanding of type IIA compactifications with fluxes and D-branes and, eventually, a better overview of the landscape of flux vacua.

\bigskip

\bigskip

\centerline{\bf \large Acknowledgments}

\bigskip
We would like to thank Ralph Blumenhagen, Michael Haack, Luis Ib\'a\~nez, Eran Palti, Rafaelle Savelli and Kepa Sousa for useful discussions.  This work is supported by the Spanish Research Agency (Agencia Estatal de Investigaci\'on) through the grant IFT Centro de Excelencia Severo Ochoa SEV-2016-0597, by the grant FPA2015-65480-P from MINECO/FEDER EU, by the grant  IJCI-2015-24908 from MINECO, and by the ERC Advanced Grant SPLE under contract ERC-2012-ADG-20120216-320421. D.E. is supported through the FPI grant SVP-2014-068283.


\appendix


\section{K\"ahler Potentials in Type IIA CY orientifolds} \label{A:MetModSpace}

\subsection{K\"ahler Potentials and Moduli Space Metrics}
Type IIA compactifications on Calabi-Yau orientifolds naturally come with moduli spaces parameterised by K\"ahler moduli and complex structure moduli. The moduli spaces inherit a K\"ahler geometry from the ${\cal N}=2$ compactifications on the Calabi-Yau manifolds before orientifolding, with the K\"ahler metric given by the second order derivative of the K\"ahler potential:
\begin{equation}
K = K_T + K_Q = - \log ({\cal G}_T {\cal G}_Q^2).
\end{equation}
The product ${\cal G} = {\cal G}_T {\cal G}_Q^2$ is a homogeneous function of degree seven in the geometric moduli $\psi^A \in \{ t^a, n^K, u_\Lambda \}$ of the closed string sector:
\begin{equation}\label{Eq:HomogenPrePot}
\psi^A \partial_A {\cal G} = \left(t^a \partial_{t^a} + n^K \partial_{n^K} + u_\Lambda \partial_{u_\Lambda} \right)  {\cal G} = 7 {\cal G},
\end{equation}
indicating that the moduli form homogeneous coordinates on the moduli space subject to the scaling transformations,
\begin{equation}
t^a \rightarrow \lambda\, t^a, \qquad n^K \rightarrow \tilde \lambda\, n^K, \quad u_\Lambda \rightarrow \tilde \lambda u_\Lambda.
\end{equation} 
In the absence of D6-branes the moduli space corresponds to the direct product of the K\"ahler and complex structure moduli space, which allows for an independent scaling transformation on both sectors with $\lambda \neq \tilde \lambda \in \C$. In the presence of D6-branes wrapping {\it SLag} three-cycles $\Pi_\alpha$ with $b^1(\Pi_\alpha) \neq 0$, a redefinition of the complex structure moduli induces a mixing between all closed and open string moduli, as discussed in section~\ref{S:IIAORD6}, such that the scaling symmetries acting on the K\"ahler and complex structure moduli are identified $\lambda = \tilde \lambda$. Nonetheless, ${\cal G}$ is still a homogeneous function of degree seven in terms of the geometric moduli $\psi^A \in \{ t^a, n^K, u_\Lambda, \phi^i_\alpha \}$. From these homogeneous functions the K\"ahler metric can be determined straightforwardly,
\begin{eqnarray}
K_{A} &=& - \frac{1}{2i} \frac{\partial_A {\cal G}}{{\cal G}}, \label{Eq:HomogenPrePotRel1}\\
K_{A \ov B}&=& - \frac{1}{4} \left( \frac{\partial_A \partial_B {\cal G}   }{{\cal G}}  - \frac{\partial_A {\cal G} \partial_B {\cal G}}{{\cal G}^2}\right). 
\end{eqnarray}
The homogeneous property of the function $\cal G$~(\ref{Eq:HomogenPrePot}) implies some additional relations, such as  
\begin{equation}\label{Eq:NoscalePsi}
K^{A \ov B} K_{\ov B} = -2 i \psi^A,
\end{equation}
and the no-scale relation,
\begin{equation}
K^{A \ov B} K_{A} K_{\ov B} = 7,
\end{equation}
and also allows to extract a simple relation for the inverse metric,
\begin{equation}
K^{A \ov B} = \frac{2}{3} \psi^A \psi^B - 4 {\cal G} {\cal G}^{A B},
\end{equation}
with ${\cal G}^{AB}$ the inverse of $\partial_A \partial_B{\cal G}$.


\subsection{K\"ahler metrics with mobile D6-branes}

Let us now specify these relations in the presence of $n$~D6-branes wrapping {\it SLag} three-cycles $\Pi_{\alpha \in \{1, \ldots, n \}}$ and the symplectic basis choice with $\{N^K\}_{K\neq 0} = 0$, as considered in sections~\ref{S:IIAFluxVacua} and~\ref{S:FluxVacD6branes}, such that the K\"ahler potential for the type IIA orientifold compactification reads:
\begin{eqnarray}
K_T &=& - \log \left( \frac{4}{3} {\cal K}_{abc} t^a t^b t^c  \right), \\
K_Q &=&- \log \left[  s + \frac{1}{2} t^a  {\bf H}^0_{\alpha\, a}    \right]
- 2 \log \left[\tilde {\cal G}_Q\left(  u_\Lambda + \frac{1}{2} t^a {\bf H}_{\alpha\, \Lambda\,\,a} \right) \right].
\end{eqnarray}
To obtain analytic relations for the metric, we will further assume that the functions ${\bf H}_{\alpha\, a}^{K}$ and ${\bf H}_{\alpha\, \Lambda\, a}^{K}$ depend only on the geometric moduli $\{t^a, \phi^i_b\}$. Such a functional dependence is characteristic for toroidal backgrounds, but is also expected to be a good approximation in the large volume and large complex structure regions of the moduli space for more generic Calabi-Yau manifolds. Under this assumption the first order derivatives of the K\"ahler potential are given by
\begin{equation}\label{Eq:1ODerK}
\begin{array}{ll}
K_{S} = \frac{i}{2 s + t^a {\bf H}_{\alpha \, a}^0  }, \qquad \qquad K_{ U_\Lambda} = i \frac{1}{\tilde {\cal G}_Q} \partial_{u_\Lambda} \tilde {\cal G}_Q, \\
 K_{T^a} = \frac{3i {\cal K}_{abc} t^b t^c}{2 {\cal K}} + \frac{i}{4 s + 2 t^b  {\bf H}_{\alpha \, b}^0  } \partial_{t^a} (t^c  {\bf H}_{\alpha \, c}^0 ) + \frac{i}{2 \tilde{\cal G}_Q}  \partial_{ u_\Lambda} (\tilde{\cal G}_Q )  \partial_{t^a} ( t^c {\bf H}_{\alpha\, \Lambda\,\,c} ), \\
K_{\Phi^i_\alpha} = \frac{i}{4  s + 2  t^b {\bf H}_{\alpha \, b}^0 } \partial_{\phi^i_\alpha}( t^a {\bf H}^0_{\alpha \, a})  + \frac{i}{ 2\tilde {\cal G}_Q} \partial_{ u_\Lambda} (\tilde{\cal G}_Q)  \partial_{\phi^i_\alpha}( t^a {\bf H}_{\alpha \, \Lambda \, a}).  
\end{array}
\end{equation}
Upon introducing the row vectors
\begin{eqnarray}
{\bf H}_T^0 &=& \left( \frac{1}{2} \partial_{t^a} (t^c  {\bf H}_{\alpha \, c}^0 ) \right), \quad {\bf H}_{\Lambda \,T} = \left( \frac{1}{2}  \partial_{t^a} ( t^c {\bf H}_{\alpha\, \Lambda\,\,c} ) \right), \\
 {\bf H}_\Phi^0 &=& \left( \frac{1}{2} \partial_{\phi^i_\alpha} (t^c  {\bf H}_{\alpha \, c}^0 ) \right), \quad  {\bf H}_{\Lambda\, \Phi} = \left( \frac{1}{2}  \partial_{\phi^i_\alpha} ( t^c {\bf H}_{\alpha\, \Lambda\,\,c} ) \right),
\end{eqnarray}
and the matrices
\begin{eqnarray}
K_{\hat S\ov {\hat S}} & = & \frac{1}{\left( 2 \hat s + t^a {\bf H}_{\alpha \, a}^0  \right)^2},\\
K_{\hat U_\Lambda \ov{\hat U}_M} & =& \frac{1}{2} \left( \frac{ \partial_{\hat u_\Lambda} \tilde {\cal G}_Q \partial_{\hat u_M} \tilde {\cal G}_Q }{\tilde{\cal G}^2_Q}  - \frac{\partial_{\hat u_\Lambda} \partial_{\hat u_M} \tilde{\cal G}_Q}{\tilde{\cal G}_Q}\right),\\
\Xi_{T^a \ov T^b} &=& -\frac{3}{2} \left( \frac{{\cal K}_{ab}}{{\cal K}} - \frac{3}{2} \frac{{\cal K}_a {\cal K}_b}{{\cal K}^2} \right) +   \frac{i}{4} K_{\hat S}\, \partial_{t^a} \partial_{t^b} (t^c  {\bf H}_{\alpha \, c}^0 ) \notag\\
&& \qquad  + \frac{i}{4} K_{\hat U_\Lambda}\,  \partial_{t^a}  \partial_{t^b} ( t^c {\bf H}_{\alpha\, \Lambda\,\,c} ), \\
\Xi_{T^a \ov \Phi^j_\beta} & =&   \frac{i}{4} K_{\hat S}\,  \partial_{t^a} \partial_{\phi^j_\beta} (t^c  {\bf H}_{\alpha \, c}^0 )  +\frac{i}{4} K_{\hat U_\Lambda}\, \partial_{t^a}  \partial_{\phi^j_\beta} ( t^c {\bf H}_{\alpha\, \Lambda\,\,c} ), \\
\Xi_{\Phi^j_\alpha \ov \Phi^j_\beta} & =&   \frac{i}{4} K_{\hat S}\,  \partial_{\phi^i_\alpha} \partial_{\phi^j_\beta} (t^c  {\bf H}_{\alpha \, c}^0 )    + \frac{i}{4} K_{\hat U_\Lambda}\, \partial_{\phi^i_\alpha}  \partial_{\phi^j_\beta} ( t^c {\bf H}_{\alpha\, \Lambda\,\,c} ), \label{Eq:KahlerMetricOSMpart1}
\end{eqnarray}
the K\"ahler metric ${\bf K}_{AB}$ on the full moduli space can be written in an elegant way: 
\begin{equation}
{\bf K}_{AB} = \left( \begin{array}{cccc}  
\1 & 0 & 0 & 0 \\
0  & \1 & 0 & 0 \\
({\bf H}_T^0)^t  & ({\bf H}_{\Lambda \,T} )^t & \1 & 0 \\
 ({\bf H}_\Phi^0)^t  & ({\bf H}_{\Lambda \,\Phi} )^t & 0& \1
\end{array}\right) 
 \left( \begin{array}{cccc} 
 K_{\hat S \ov{\hat S}} & 0 & 0  & 0 \\
 0 & K_{\hat U_\Lambda \ov{\hat U}_M}& 0 & 0 \\
 0& 0&  \Xi_{T^a \ov T^b}  & \Xi_{T^a \ov \Phi^j_\beta} \\
 0& 0& \Xi_{\Phi^i_\alpha \ov T^b } & \Xi_{\Phi^j_\alpha \ov \Phi^j_\beta} 
\end{array}\right) 
\left( \begin{array}{cccc}  
\1 & 0 & {\bf H}_T^0  &  {\bf H}_\Phi^0     \\
0  & \1 & {\bf H}_{\Lambda \,T} &  {\bf H}_{\Lambda \,\Phi} \\
 0 &0 & \1 & 0 \\
 0 & 0& & \1
\end{array}\right).
\end{equation}
From this expression we can straightforwardly determine the inverse K\"ahler metric ${\bf K}^{AB}$:
\begin{equation}
{\bf K}^{AB} =
\left( \begin{array}{cccc}  
\1 & 0 &  -{\bf H}_T^0  &  -{\bf H}_\Phi^0     \\
0  & \1 & -{\bf H}_{\Lambda \,T} & - {\bf H}_{\Lambda \,\Phi} \\
 0 &0 & \1 & 0 \\
 0 & 0& 0 & \1
\end{array}\right)
 \left( \begin{array}{cccc} 
 K_{\hat S \ov{\hat S}}^{-1} & 0 & 0  & 0 \\
 0 & K_{\hat U_\Lambda \ov{\hat U}_M}^{-1}& 0 & 0 \\
 0& 0& \multicolumn{2}{c}{\multirow{2}{*}{\Large ${\ \Xi}^{-1}$}}\\
 0& 0& 
\end{array}\right)
\left( \begin{array}{cccc}  
\1 & 0 & 0 & 0 \\
0  & \1 & 0 & 0 \\
-({\bf H}_T^0)^t  &- ({\bf H}_{\Lambda \,T} )^t & \1 & 0 \\
- ({\bf H}_\Phi^0)^t  & - ({\bf H}_{\Lambda \,\Phi} )^t & 0& \1
\end{array}\right)  
,
\end{equation}
where $\Xi^{-1}$ denotes the inverse of the matrix with entries $\Xi_{T^a \ov T^b}$, $\Xi_{T^a \ov \Phi^j_\beta}$, $ \Xi_{\Phi^i_\alpha \ov T^b }$ and $\Xi_{\Phi^i_\alpha \ov \Phi^j_\beta}$.

\section{Superpotentials with mobile D6-branes}\label{A:OpenBil}

When considering orientifold compactifications with D6-branes and their orientifold images, one has to be aware that their RR-charges act as magnetic sources for the field strength $G_2$, such that the Bianchi identities (\ref{Eq:BianchiIdent}) have to be modified accordingly:     
\begin{equation}\label{Eq:BianchiIdentGeneral}
\ell_s^2 \left( d G_2  -  m H_3 \right) =  -\sum_\alpha N_\alpha \Big[ \delta^3 (\Pi_\alpha^0) + \delta^3 ({\cal R}\Pi_\alpha^0) \Big] + 4 \delta^3 (\Pi_{O6}),
\end{equation}
where the right-hand side considers the bump-like delta-function currents sourced by the D6-branes wrapping reference three-cycles $\Pi_\alpha^0$  their respective orientifold images ${\cal R}\Pi_{\alpha}^0$, and the O6-planes. The field strength $G_2$ is globally well-defined upon imposing the modified RR tadpole cancellation conditions in the presence of  NS 3-form flux and Romans mass $m$:
\begin{equation}\label{Eq:RRTadpoleD6Flux}
\sum_\alpha N_\alpha ([\Pi_\alpha^0] + [{\cal R}\Pi_\alpha^0]) - 4 [\Pi_{O6}]  - m [\Pi_{H_3}]= 0,
\end{equation}
where $[\Pi_{H_3}]$ corresponds to the Poincar\'e-dual three-cycle of the NS-flux $H_3$. Note that in the absence of $H_3$-flux, the RR tadpole condition implies the existence of a four-chain ${\cal C}_4^0$ connecting the D6-branes and their orientifold images to the O6-planes, i.e.~$\partial {\cal C}_4^0  = \sum_\alpha N_\alpha \left( \Pi_\alpha^0 + {\cal R}\Pi_\alpha^0  \right) -  4 \Pi_{O6}$. 

The Lagrangian condition~(\ref{Eq:SLAG3Cycles}) also has to be modified in the presence of worldvolume fluxes including the $U(1)$ field strength $F=dA$:
\begin{equation} 
J_c \big|_{\Pi_\alpha} - \frac{\ell_s^2}{2\pi} F = 0.
\end{equation}
In regions of the closed string moduli space where this condition is violated, a non-vanishing contribution to the superpotential arises that is capable of breaking the ${\cal N}=1$ supersymmetry in four dimensions,
\begin{equation}\label{Eq:DeltaWGen}
\Delta W = \frac{1}{\ell_s^5}\int_{{\cal C}_4^\alpha}   \left( J_c  - \frac{\ell_s^2}{2\pi} \tilde F_\alpha\right) \wedge  \left( J_c  - \frac{\ell_s^2}{2\pi} \tilde F_\alpha\right),
\end{equation}
where the four-chain ${\cal C}_4^\alpha$ connects a three-cycle $\Pi_\alpha$ that is a homotopic deformation of the reference three-cycle $\Pi_\alpha^0$, in line with the philosophy of section~\ref{S:IIAORD6}. The field strength $\tilde F_\alpha$ is the extension of the D6-brane worldvolume field strength to the four-chain. Microscopically, there exist two separate effects that yield a non-vanishing superpotential $\Delta W$ as a function of the open string moduli associated to the three-cycle deformations. The first effects comes from turning on a worldvolume flux:
\begin{equation}
\frac{\ell_s^2}{2\pi} F_\alpha 
= \frac{\ell_s^2}{2\pi} d A_\alpha + n_{{F}i}^\alpha \, \rho^i, \qquad n_{{F}i}^\alpha\in \Z,
\end{equation} 
such that the evaluation of (\ref{Eq:DeltaWGen}) leads to a superpotential containing a linear term in the open string moduli:
\begin{equation}\label{Eq:SUSYBreakD6Flux}
\ell_s \Delta W^{(1)}  = n_{{F} i}^\alpha \Phi^i_\alpha.
\end{equation}
A second contribution is due to the backreaction on the closed string fluxes following the homotopic deformation of a {\it SLag} three-cycle $\Pi_\alpha^0 \rightarrow \Pi_\alpha$. More precisely, after the deformation the backreacted RR-fluxes ${\bf G} = {\bf G}^0 + q_\alpha \Delta_\alpha {\bf G}$  can be decomposed into a component ${\bf G}^0$ that satisfies the Bianchi identities in the reference configuration (with vanishing worldvolume flux) 
\begin{equation} \label{Eq:BianchiIdentitiesFull} 
\ell_s^2 d_H {\bf G}^0 =  - \left( \sum_\alpha N_\alpha \left( \delta^3 (\Pi_\alpha^0) + \delta^3 ({\cal R}\Pi_\alpha^0) \right) - 4 \delta^3 (\Pi_{O6}) \right) \wedge e^B,
\end{equation}
and a component $\Delta_\alpha {\bf G}$ capturing the change in fluxes under the deformation:
\begin{equation}  
\ell_s^2 d_H \Delta_\alpha {\bf G}^0 =     N_\alpha  \delta^3 (\Pi_\alpha^0) \wedge e^B - N_\alpha  \delta^3 ( \Pi_\alpha ) \wedge e^{B- \frac{\ell_s^2}{2\pi} F}.
\end{equation}
\subsection{Open-Closed Superpotentials}\label{A:OCSuper}
In the absence of $H_3$-flux, both of these equations can be solved~\cite{Herraez:2018vae} in terms of bump delta-functions associated with the appropriate four-chains:
\begin{equation}\label{Eq:SolRRformsH0}
{\bf G}^0 = -  \frac{1}{\ell_s} \delta^2({\cal C}_4^0) \wedge e^B , \qquad  \Delta_\alpha {\bf G} = - \frac{1}{\ell_s} \delta^2({\cal C}_4^\alpha) \wedge e^{B- \frac{\ell_s^2}{2\pi} F}.
\end{equation} 
The four-chain ${\cal C}_4^0$ has been introduced above for the reference configuration, while the second four-chain ${\cal C}_4^\alpha$ connects the deformed three-cycle and reference three-cycle such that the delta-function satisfies $\ell_s d \, \delta^2 ({\cal C}_4^\alpha) = N_\alpha \delta^3 (\Pi_\alpha) - N_\alpha \delta^3 (\Pi_\alpha^0)$. In the reference configuration the polyforms $e^{-B} \wedge {\bf G}_0$ still allow to define quantised Page charges, but the harmonic pieces of ${\bf G}_0$ are tied to their co-exact components resulting from the presence of localised sources. Similarly, the back-reacted polyforms $e^{-B}\wedge {\bf G}$ ought to allow for the definition of conserved Page charges upon deformation, which implies that the harmonic parts of $\Delta_\alpha {\bf G}$ are completely determined by their co-exact piece. The presence of a harmonic component for $\Delta_\alpha G_2$ can give rise to a superpotential contribution $\Delta W$ involving open string moduli. To see how this precisely happens, we follow the same logic as in~\cite{Marchesano:2014bia,Marchesano:2014iea} and consider the integral of $\Delta_\alpha G_2$ wedged with the closed four-form $J\wedge \omega_2$:            
\begin{equation}
\int_{{\cal M}_6}  \Delta_\alpha G_2 \wedge J \wedge \omega_2 = \int_{{\cal C}_4^\alpha} J \wedge \omega_2,
\end{equation}
which is non-vanishing for a harmonic two-form $\omega_2$. For an infinitesimal deformation~$X$ of the {\it SLag} three-cycle as in section~\ref{S:IIAORD6}, the chain integral reduces to an integral over the three-cycle,
\begin{equation}
\int_{{\cal C}_4} J \wedge \omega_2 = \int_{\Pi_\alpha} \iota_X J \wedge \omega_2,
\end{equation}
which implies the existence of a non-trivial two-cycle in $H_2(\Pi_\alpha, \Z)$, Poincar\'e dual to the one-form~$ \iota_X J$, for non-vanishing values. By using the more appropriate basis of one-forms~$\zeta^i$ from section~\ref{S:IIAORD6}, the condition can be written out more explicitly through the D6-brane displacement parameters $n_{ai}^\alpha$, 
\begin{equation}
n_{ai}^\alpha = \frac{1}{\ell_s^3} \int_{\Pi_\alpha} \omega_a  \wedge \zeta_i \quad \in \Z.
\end{equation}
If at least one of the parameters $n_{ai}^\alpha \neq 0$, the evaluation of (\ref{Eq:DeltaWGen}) gives rise to a superpotential consisting of a bilinear term mixing open string moduli and K\"ahler moduli:
 \begin{equation}  \label{Eq:D6BilSuper} 
\ell_s \Delta W^{(2)}  = - n_{ai}^\alpha \Phi^i_\alpha T^a. 
\end{equation}
Consequently, the most generic four-dimensional effective superpotential for type IIA flux compactifications with (non-rigid) D6-branes includes an additional supersymmetry-breaking term mixing open string moduli and K\"ahler moduli as in equation~(\ref{Eq:OpenClosedSuperPotential}). 
In this expression, ${W}_{D6}^0$ denotes the constant contribution to the D6-brane superpotential evaluated for the reference three-cycles~$\Pi_\alpha^0$:
\begin{equation}
W^0_{D6} = \frac{1}{2\ell_s^5}  \int_{{\cal C}_4^0}   \left( J_c  - \frac{\ell_s^2}{2\pi} \tilde F_\alpha\right) \wedge  \left( J_c  - \frac{\ell_s^2}{2\pi} \tilde F_\alpha\right),
\end{equation}
in the absence of $H$-flux. 



\subsection{Superpotentials and Redefined Complex Structure Moduli}\label{A:RedCSM}
For flux compactifications with non-vanishing $H_3$-flux, the Bianchi identities (\ref{Eq:BianchiIdentitiesFull}) and RR tadpole conditions (\ref{Eq:RRTadpoleD6Flux}) no longer imply the existence of a four-chain ${\cal C}_4^0$ connecting the full set of D6-branes and O6-planes for the reference configuration. Instead the solutions (\ref{Eq:SolRRformsH0}) of the Bianchi identities have to be adjusted appropriately, as derived for the first time in Appendix B.1 of~\cite{Herraez:2018vae}. Here, we review and extend the reasoning that led to eq.(B.11) there, which allowed to deduce the expression for the redefined complex structure moduli $N^K$ in term of the open string moduli. More precisely, we extend this result in the sense that we consider both kinds of complex structure moduli $(N^K, U_\Lambda)$ considered in the type IIA orientifold literature. 

Following \cite{Herraez:2018vae} we first consider the type IIA flux superpotential
%
\begin{equation}
- iW\, =\,\frac{1}{\ell_s^{6}} \int_{{\cal M}_6}  e^{-\phi} \RE\,  \Omega_3 \wedge H - i {\bf G} \wedge e^{iJ} 
\label{supoflux}
\end{equation}
which is manifestly gauge invariant and globally well-defined. Then one can split the RR flux background {\bf G} into two pieces 
\begin{equation}
{\bf G} \, =\, {\bf G}^0 +  \sum_\alpha \Delta_\alpha {\bf G}
\label{splitG}
\end{equation}
with ${\bf G}^0$ satisfying the Bianchi identities and quantisations conditions for the reference configuration, and $\Delta_\alpha {\bf G}$ representing the change in {\bf G} as we replace the D6-brane at $\Pi_\alpha^0$ with the one at $\Pi_\alpha$. We find that
\begin{equation}
{\bf G}^0 = - j_0- H \wedge C_3 + e^{B} \wedge \bar{\bf G} + \dots
\end{equation}
and
\begin{equation}
\Delta_\alpha {\bf G} \, \simeq\,  \frac{1}{\ell_s^2} \delta (\Pi_\alpha) \wedge \left( \sigma A - \frac{1}{2} \sigma^2 A \wedge F\right)  \wedge e^{B}  - \frac{1}{\ell_s}  \delta ({{\cal C}_4^\alpha}) \wedge \left( e^{B} - \varpi_4\right) 
\label{shiftG2}
\end{equation}
where ${\cal C}_4^\alpha$ is a four-chain such that $\partial {\cal C}_4^\alpha = \Pi_\alpha - \Pi_\alpha^0$, and $\varpi_4$ is the co-exact form such that $d\varpi_4 = H \wedge B$. Replacing this into \eqref{supoflux} one obtains
\begin{equation}
W\, =\,  \frac{1}{\ell_s^6} \int_{{\cal M}_6}  \Omega_c \wedge H + \bar{\bf G} \wedge e^{J_c} + \frac{2}{\ell_s^4} \int_{\Pi_\alpha} \sigma A \wedge \left( J_c - \sigma F\right) - \frac{1}{\ell_s^5} \int_{{\cal C}_4^\alpha} J_c^2 - \varpi_4  + W_0\, .
\label{Wmodu}
\end{equation}
From this last expression one can extract the closed and open-string moduli dependence of the superpotential. We are mainly interested in the terms proportional to the H-flux quanta, which are defined by
\begin{equation}
H \, =\, h_K \beta^K + h^\Lambda \alpha_\Lambda \, .
\label{Hfluxap}
\end{equation}
Then we have that the first piece of \eqref{Wmodu} contributes as
\begin{equation}
\frac{1}{\ell_s^6} \int_{{\cal M}_6}  \Omega_c \wedge H \, =\, h_K N^K_\star + h^\Lambda U_{\star \,\Lambda}\, .
\end{equation}
To evaluate the remaining dependence on the H-flux quanta we split the B-field on the four-chain ${{\cal C}_4^\alpha}$ as
\begin{equation}
B|_{{\cal C}_4^\alpha}\, =\, b^a \omega_a + \tilde{B}
\label{splitB}
\end{equation}
with $\tilde{B}$ the co-exact piece of the B-field satisfying $d\tilde{B} = H|_{{\cal C}_4^\alpha}$. Given this split one can see that $\varpi_4|_{{\cal C}_4^\alpha} = \frac{1}{2} \tilde{B} \wedge \tilde{B}|_{{\cal C}_4^\alpha}$. We then find that the third and fourth terms in \eqref{Wmodu} contain the terms
\begin{equation}\nonumber
 -\frac{1}{\ell_s^4}  \int_{{\cal C}_4^\alpha} J_c \wedge  \tilde{B}   + \frac{2}{\ell_s^4} \int_{\Pi_\alpha} \sigma A \wedge \tilde{B}  \, =\, - \frac{1}{2} \aleph_{a\, \alpha} T^a + \frac{1}{2} \left( h_K g^K_{i\alpha\, i} + h^\Lambda g_{\alpha\, \Lambda\, i}\right)  \theta^i_\alpha 
\end{equation}
where
\begin{equation}
g_{\alpha\, i}^{K} \, =\, \frac{2}{\ell_s^{4}} \int_{{\cal C}_4^\alpha}  \beta^K \wedge \tilde{\zeta}_i \qquad \text{and} \qquad g_{\alpha\, \Lambda\, i} \, =\, \frac{2}{\ell_s^{4}} \int_{{\cal C}_4^\alpha}  \alpha_\Lambda \wedge \tilde{\zeta}_i .
\label{defgs}
\end{equation}
with $\tilde{\zeta}_i$ the extension of the one-form $\zeta_i$ of $\Pi_\alpha$ to ${{\cal C}_4^\alpha}$, and
\begin{align}
\aleph_{a \, \alpha} = \frac{2}{\ell_s^4} \int_{{\cal C}_4^\alpha} \tilde B \wedge \omega_a \, . 
\end{align}
Finally, generalising the computation below eq.(A.31) of \cite{Herraez:2018vae} to a background flux of the form \eqref{Hfluxap} one easily deduces that
\begin{equation}
\aleph_{a\, \alpha} \, =\, \frac{1}{2} \left( h_K {\bf H}_{\alpha\, a}^{K} + h^\Lambda {\bf H}_{\alpha\, \Lambda \, a}\right)
\label{realepH}
\end{equation}
with the definitions of $ {\bf H}_{\alpha\, a}^{K}$ and ${\bf H}_{\alpha\, \Lambda \, a}$ given in the main text.  

Therefore, putting all these results together one finds that the superpotential depends on the H-flux quanta as
\begin{eqnarray}
W & = & h^K \left[ N^K_\star + \frac{1}{2} \sum_\alpha (g_{\alpha i}^K \theta_\alpha^i  - T^a  {\bf H}_{\alpha\, a}^{K}) \right]\\ \nonumber
  & + & h_\Lambda \left[U_{\star \,\Lambda} + \frac{1}{2} \sum_\alpha ( g_{\alpha\, \Lambda\, i} \theta^i_\alpha -  T^a  {\bf H}_{\alpha\, \Lambda \, a})\right] + \dots 
\end{eqnarray}
obtaining the following redefinition for the complex structure moduli of the compactification
\begin{equation}\label{Eqap:RedefComplexStructure}
N^K = N^K_\star + \frac{1}{2} \sum_\alpha (g_{\alpha i}^K \theta_\alpha^i  - T^a  {\bf H}_{\alpha\, a}^{K}), \qquad  U_\Lambda = U_{\star \,\Lambda} + \frac{1}{2} \sum_\alpha ( g_{\alpha\, \Lambda\, i} \theta^i_\alpha -  T^a  {\bf H}_{\alpha\, \Lambda \, a})\, .
\end{equation}

\section{Toroidal Orbifolds and K\"ahler Metrics}\label{A:TorOrb}
A typical set of backgrounds suited to test the ideas developed in this paper consist of the orientifold version of $T^2 \times K^3$ (considered at an orbifold point in moduli space) and toroidal orientifolds (or their $\Z_2 \times \Z_2$ orbifolded version) with a factorisable ambient six-torus $T^6$. Each of the three two-tori $T^2_{(i)}$ is parameterised by periodic coordinates $(x^i, y^i) \sim (x^i + 1, y^i +1)$ and characterised by a modular parameter $\tau_i$. The ambient space can be equipped with a set of basis three-forms which splits up into a symplectic basis of $\OR$-even $(\alpha_0, \beta^j) \in H^3_+(T^6/\OR, \Z)$ and $\OR$-odd $(\beta^0, \alpha_i) \in H^3_-(T^6/\OR,\Z)$ three-forms:
\begin{equation}
\begin{array}{l@{\hspace{0.4in}}l}
\alpha_0 = dx^1 \wedge dx^2 \wedge dx^3, & \beta^0 = - dy^1 \wedge dy^2 \wedge dy^3,\\
\beta^1 = dx^1 \wedge dy^2 \wedge dy^3, & \alpha_1 = dy^1 \wedge dx^2\wedge dx^3,\\  
\beta^2 = dy^1 \wedge dx^2 \wedge dy^3, & \alpha_2 = dx^1 \wedge dy^2\wedge dx^3,\\
\beta^3 = dy^1 \wedge dy^2 \wedge dx^3, & \alpha_3 = dx^1 \wedge dx^2\wedge dy^3,\\    
\end{array}
\end{equation}
under the orientifold projection ${\cal R}:(x^i, y^i) \rightarrow (x^i,-y^i)$. In this basis the holomorphic Calabi-Yau three-form $\Omega_3$ reads
\begin{equation}
\begin{array}{rcl}
\Omega_3 &=& (dx^1 + i \tau_1 dy^1) \wedge (dx^2 + i \tau_2 dy^2) \wedge (dx^3 + i \tau_3 dy^3) \\
&=& \alpha_0 - \tau_2 \tau_3 \beta^1 - \tau_1 \tau_3 \beta^2 - \tau_1 \tau_2 \beta^3 + i \tau_1 \tau_2 \tau_3 \beta^0 + i \tau_1 \alpha_1 + i \tau_2 \alpha_2  + i \tau_3 \alpha_3,
\end{array}
\end{equation} 
yielding the ${\cal N}=2$ K\"ahler potential $K_{cs} = - \log \left( i\int \Omega_3 \wedge \ov \Omega_3 \right) = - \log(8 \tau_1 \tau_2 \tau_3)$ in terms of the modular parameters. The basis of $\OR$-odd two-forms $\omega_a \in H^{1,1}_-(T^6/\OR, \Z)$ and their Poincar\'e dual $\OR$-even four-forms $\tilde \omega^a$ are given by
\begin{equation}
\omega_a  = \delta_{ai} dx^i \wedge dy^i, \qquad  \omega_a \wedge \omega_b = {\cal K}_{abc}  \tilde \omega^c ,
\end{equation}
with ${\cal K}_{abc} = {\cal K}_{123} =1$ (and permutations thereof) the only non-vanishing triple intersection numbers. Each volume of the three two-tori is measured by the geometric part of the corresponding K\"ahler moduli and the overall volume of the internal space is the product of the two-tori volumes, i.e.~${\cal V} =  t_1 t_2 t_3$. The geometric part of the complex structure moduli are given by the periods of ${\cal C} \Omega_3$:
\begin{equation}
S_\star = \int \Omega_c \wedge \beta_0 = \xi^0 + i \frac{e^{-D}}{\sqrt{8 \tau_1 \tau_2 \tau_3}} , \qquad U_{\star i} = \int \Omega_c \wedge \alpha_i = \xi^1 + i \frac{e^{-D}}{\sqrt{8 \tau_1 \tau_2 \tau_3}} \tau_j \tau_k.
\end{equation}
with the compensator field ${\cal C} = \frac{e^{-D}}{\sqrt{8 \tau_1 \tau_2 \tau_3}}$ following from the definition in the main text.
For the factorable toroidal orientifolds, the K\"ahler potential on the K\"ahler moduli space and the complex structure moduli space are given respectively by the well-known expressions:
\begin{equation}
K_T =  - \sum_{a=1}^3 \log\left[-i (T^a-\ov T^a) \right] , \qquad K_Q =  - \log\left[ -i(S_\star-\ov S_\star)\right]- \sum_{i=1}^3 \log \left[ -i (U_{\star i} -\ov U_{\star i}) \right].
\end{equation}

With each $\OR$-even basis three-form $(\alpha_0, \beta^j)$ in $H^3_+(T^6/\OR, \Z)$, we can introduce its de Rahm dual $\OR$-even three-cycle $(\rho_0, \rho_i)$:
\begin{equation}
\begin{array}{lc@{\hspace{0.4in}}lc}
\OR-\text{even three-cycle}& \text{P.D.}& \OR-\text{odd three-cycle}& \text{P.D.}\\
\hline
\rho_0 = \pi_1 \otimes \pi_3 \otimes \pi_5 & \beta^0 & \sigma_0 = \pi_2 \otimes \pi_4 \otimes \pi_6, &\alpha_0\\
\rho_1 = \pi_1 \otimes \pi_4 \otimes \pi_6 & - \alpha_1 & \sigma_1 = \pi_2 \otimes \pi_3 \otimes \pi_5& \beta^1\\
\rho_2 = \pi_2 \otimes \pi_3 \otimes \pi_6& - \alpha_2 & \sigma_2 = \pi_1 \otimes \pi_4 \otimes \pi_5 & \beta^2\\
\rho_3 = \pi_2 \otimes \pi_4 \otimes \pi_5& - \alpha_3& \sigma_3 = \pi_1 \otimes \pi_3 \otimes \pi_6& \beta^3
\end{array}
\end{equation}
and repeat the exercise for their $\OR$-odd counterparts, which provide four $\OR$-odd three-cycles $(\sigma_0, \sigma_i)$. The choice of the symplectic basis of three-cycles from above also determines the Poincar\'e dual (P.D.) three-forms for each of the three-cycles. A generic, factorisable three-cycle with topology $S^1 \times S^1 \times S^1$ on $T^6$ can now be decomposed in terms of this three-cycle basis:
\begin{equation}\label{Eq:3CycleFact}
\begin{array}{rcl}
\Pi^{\rm fact}_\alpha &=& (n_\alpha^1 \pi_1 + m_\alpha^1 \pi_2) \otimes (n_\alpha^2 \pi_3 + m_\alpha^2 \pi_4) \otimes (n_\alpha^3 \pi_5 + m_\alpha^3 \pi_6) \\
&=& n_\alpha^1 n_\alpha^2 n_\alpha^3 \, \rho_0  + n_\alpha^1 m_\alpha^2 m_\alpha^3 \,\rho_1  + m_\alpha^1 n_\alpha^2 m_\alpha^3 \,\rho_2  + m_\alpha^1 m_\alpha^2 n_\alpha^3\, \rho_3  \\
&& \quad  + m_\alpha^1 m_\alpha^2 m_\alpha^3 \, \sigma_0  + m_\alpha^1 n_\alpha^2 n_\alpha^3 \, \sigma_1  + n_\alpha^1 m_\alpha^2 n_\alpha^3 \, \sigma_2  + n_\alpha^1 n_\alpha^2 m_\alpha^3 \, \sigma_3,
\end{array}
\end{equation}
 by virtue of the torus wrapping numbers $(n^i_\alpha, m^i_\alpha)_{\alpha=1,2,3}$, which encode the one-cycle geometry on the two-torus $T_{(i)}^2$. As reviewed in section~\ref{S:IIAORD6}, four-dimensional type IIA orientifold compactifications have to be equipped with spacetime filling D6-branes wrapping such three-cycles fulfilling the special Lagrangian conditions~\eqref{Eq:SLAG3Cycles}, such that their combined RR charges cancel the RR charges of the O6-planes. Massless open string excitations arise at the intersection points of two distinct D6-branes wrapping supersymmetric three-cycles and fill out supermultiplets of the supersymmetry algebra generated by the mutually unbroken supercharges. Furthermore, on toroidal orbifold backgrounds the K\"ahler metrics for these open string states can be computed as a function of the closed string moduli~\cite{Lust:2004cx,Bertolini:2005qh,Blumenhagen:2007ip,Honecker:2011sm}. The type of matter (and subsequently the functional dependence of the K\"ahler metrics) depends on the codimension of the intersection $\Pi_\alpha \cap \Pi_\beta \neq 0 $ in the ambient space $T^6$: 
\begin{itemize}
\item[(i)] Codimension 3 intersection:\\
D6-branes wrapping the three-cycles that coincide along each one-cycle on $T_{(i)}^2$ give rise to one ${\cal N}=1$ chiral supermultiplet $\Phi^i$ per two-torus. The complex scalar within such a multiplet consists of the three-cycle deformation modulus complexified by the Wilson line along the $S^1$ cycle on $T_{(i)}^2$, as described in equation~\eqref{Eq:OpenStringModDef}. The three chiral ${\cal N}=1$ supermultiplets transform in the adjoint representation of the gauge group and combine with the ${\cal N}=1$ vector multiplet into an ${\cal N}=4$ vector multiplet, compatible with the maximal number of supercharges preserved by this D6-brane configuration. Two examples of such highly (super)symmetric configurations are depicted in figure~\ref{Fig:Codim3Intersection}. 
\begin{figure}[h]
\begin{center}
\vspace{0.4in}
\begin{tabular}{c@{\hspace{0.4in}}c@{\hspace{0.4in}}c}
\includegraphics[scale=0.6]{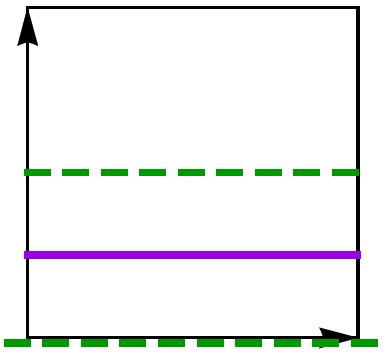} \begin{picture}(0,0) \put(-45,76){$T_{(1)}^2$} \put(0,0){$\pi_1$} \put(-85,55){$\pi_2$} \end{picture} & \includegraphics[scale=0.6]{3Cycle1}  \begin{picture}(0,0) \put(-45,76){$T_{(2)}^2$} \put(0,0){$\pi_3$} \put(-85,55){$\pi_4$} \end{picture}  & \includegraphics[scale=0.6]{3Cycle1}   \begin{picture}(0,0) \put(-45,76){$T_{(3)}^2$} \put(0,0){$\pi_5$} \put(-85,55){$\pi_6$} \end{picture} \\
\vspace{0.2in}& & \\
\includegraphics[scale=0.6]{3Cycle1}  \begin{picture}(0,0) \put(-45,76){$T_{(1)}^2$} \put(0,0){$\pi_1$} \put(-85,55){$\pi_2$} \end{picture}  & \includegraphics[scale=0.6]{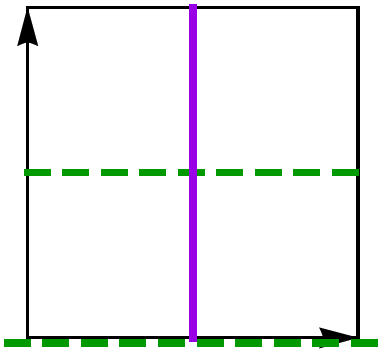}  \begin{picture}(0,0) \put(-45,76){$T_{(2)}^2$} \put(0,0){$\pi_3$} \put(-85,55){$\pi_4$} \end{picture}  & \includegraphics[scale=0.6]{3Cycle2}   \begin{picture}(0,0) \put(-45,76){$T_{(3)}^2$} \put(0,0){$\pi_5$} \put(-85,55){$\pi_6$} \end{picture} 
\end{tabular}
\caption{D6-brane configurations with codimension 3 intersection preserve a local ${\cal N}=4$ supersymmetry: an example of three-cycles with torus wrapping numbers $(1,0;1,0;1,0)$ (above) and an three-cycle example with torus wrapping $(1,0;0,1;0-1)$ (below). The O6-planes are represented by the dashed, green lines.\label{Fig:Codim3Intersection}}
\end{center}
\end{figure}

\noindent The K\"ahler metric for an open string modulus $\Phi^i$ along two-torus $T_{(i)}^2$ can be written (at leading order) as a rational function of the closed string moduli: 
\begin{equation}\label{Eq:KaehlerAdjMunich}
K_{\Phi^i \ov \Phi^i} = - \frac{\delta_{a i} \delta_{\Lambda i}}{(T^a -\ov{T}{}^a) (U_{\star \Lambda}- \ov U_{\star \Lambda})} \left|\frac{(n^j + i\, \tau_j m^j) (n^k + i\, \tau_k m^k)}{n^i + i\, \tau_i m^i} \right|,
\end{equation}
where the last term captures the model-dependent contribution determined by the three-cycle position. In this respect, the model-dependent part of the K\"ahler metric will be constrained by the special Lagrangian conditions~\eqref{Eq:SLAG3Cycles} imposed on the wrapped three-cycle. More precisely, for the two examples in figure~\ref{Fig:Codim3Intersection}, the K\"ahler metrics for the two distinguishable D6-brane configurations take the form: 
\begin{equation}
\begin{array}{llll}
\text{ex.~1:} & K_{\Phi^i \ov \Phi^i} =  \frac{-1}{(T^i -\ov T{}^i) (U_{\star i} -\ov U_{\star i})}, & \text{ex.~2:} & K_{\Phi^i \ov \Phi^i} = \left\{ \begin{array}{l}  \frac{-1}{(T^1 -\ov T{}^1) (S_\star -\ov S_\star)} \quad (i=1),\\
 \frac{-1}{(T^2 -\ov T{}^2) (U_{\star 3} -\ov U_{\star 3})} \quad (i=2), \\
  \frac{-1}{(T^3 -\ov T{}^3) (U_{\star 2} -\ov U_{\star 2})}  \quad (i=3). 
\end{array}\right.
\end{array}
\end{equation}
The main conclusion that one can draw from these examples is that the K\"ahler metric for a deformation modulus $\Phi^i$ is a homogeneous function of degree $-1$ in the K\"ahler moduli and of degree $-1$ in the complex structure moduli (including the dilaton). This statement is true in general for the K\"ahler metric~\eqref{Eq:KaehlerAdjMunich}, since the model-dependent part is independent of the K\"ahler moduli and a homogeneous function of degree zero in the complex structure moduli (upon inclusion of the dilaton).\footnote{The same scaling properties can be found in the K\"ahler metrics for the deformation moduli of non-factorisable three-cycles. }   
\item[(ii)] Codimension 5 intersection:\\
D6-brane stacks wrapping two distinct three-cycles $\Pi_\alpha$ and $\Pi_\beta$ that coincide on a one-cycle $S^1$ along one of the three two-tori and intersect at a point along the remaining four-torus,  give rise to a non-chiral pair of ${\cal N}=1$ chiral supermultiplets. The chiral multiplets transform in bifundamental representation and are each others conjugate, such that they combine into a ${\cal N}=2$ hypermultiplet. This feature is a remnant of the local ${\cal N}=2$ supersymmetry preserved by the D6-brane configuration, for which an explicit example is presented in figure~\ref{Fig:Codim5Intersection}. 
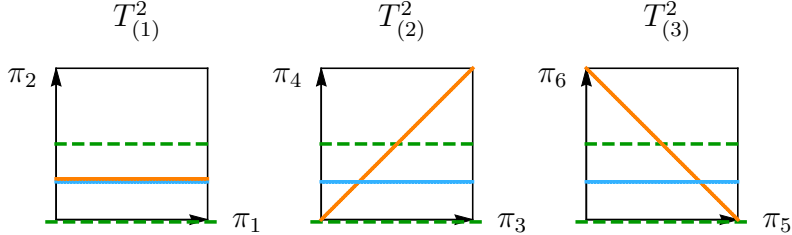
\begin{figure}[h]
\begin{center}
\vspace{0.4in}
\begin{tabular}{c@{\hspace{0.4in}}c@{\hspace{0.4in}}c}
\includegraphics[scale=0.6]{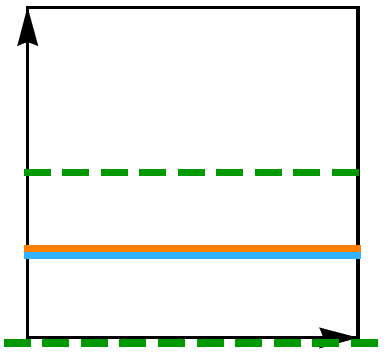}  \begin{picture}(0,0) \put(-45,76){$T_{(1)}^2$} \put(0,0){$\pi_1$} \put(-85,55){$\pi_2$} \end{picture}  & \includegraphics[scale=0.6]{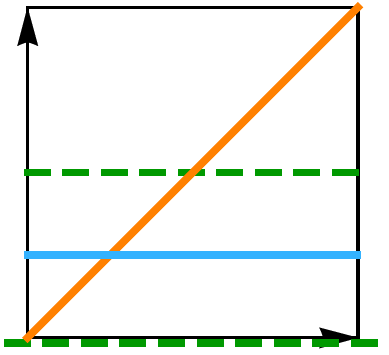}  \begin{picture}(0,0) \put(-45,76){$T_{(2)}^2$}\put(0,0){$\pi_3$} \put(-85,55){$\pi_4$} \end{picture}  & \includegraphics[scale=0.6]{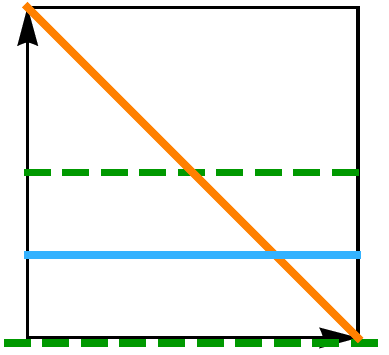}  \begin{picture}(0,0) \put(-45,76){$T_{(3)}^2$}\put(0,0){$\pi_5$} \put(-85,55){$\pi_6$} \end{picture} 
\end{tabular}
\caption{D6-brane configurations with codimension 5 intersection preserve a local ${\cal N}=2$ supersymmetry. The O6-planes are represented by the dashed, green lines.\label{Fig:Codim5Intersection}}
\end{center}
\end{figure}

\noindent The K\"ahler metric for such an ${\cal N}=2$ hypermultiplet is given (at leading order) by a (non-rational) function of the geometric part of the closed string moduli:
\begin{equation}\label{Eq:KaehlerBifundN2Munich}
K_{\alpha \ov\beta} = \frac{|n^i + i\, \tau_i m^i|}{\sqrt{(U_{\star \Lambda} - \ov U_{\star \Lambda})(U_{\star \Sigma} - \ov U_{\star \Sigma})(T^j - \ov T{}^j)(T^k - \ov T{}^k)}},
\end{equation}
where $(n^i, m^i)$ denote the wrapping numbers along the two-torus $T_{(i)}^2$ where the two three-cycles coincide on an $S^1$.
The K\"ahler metric allows for a factorisation in terms of the complex structure moduli and the K\"ahler moduli, such that it is a homogeneous function of degree $-1$ in the complex structure moduli (upon inclusion of the dilaton) and a homogeneous function of degree $-1$ in the K\"ahler moduli. This case also applies to the K\"ahler metrics for chiral matter in the symmetric or antisymmetric representation located at the intersection of a D6-brane with its orientifold image, whenever the three-cycle is parallel (or orthogonal) to the O6-plane along one single two-torus.    
\item[(iii)] Codimension 6 intersection:\\
D6-brane stacks wrapping two distinct three-cycles $\Pi_\alpha$ and $\Pi_\beta$ that intersect point-wise in the ambient space provide for a chiral ${\cal N}=1$ supermultiplet  at each independent intersection point of the six-dimensional compactification space. A simple example of a D6-brane configuration for which the intersection set has codimension 6 is presented in figure~\ref{Fig:Codim6Intersection}. The chiral multiplet transforms in the bifundamental representation and its K\"ahler metric takes the following form:\footnote{In the literature on K\"ahler metrics one might also stumble on expressions in which the exponents of the K\"ahler moduli obtain an additional contribution from the angles $\vartheta^i$ between the two three-cycles. The (potentially) modified exponents are related to a (potential) four-dimensional field redefinition of the K\"ahler moduli to arrive at their proper supergravity equivalents. Given the confusion within the literature itself about these $\vartheta^i$-dependent corrections and the fact they do not alter the overall scaling properties of the K\"ahler metrics, we have decided not to take them into account explicitly.} 
\begin{equation}\label{Eq:KaehlerBifundN1Munich}
K_{\alpha \ov\beta} = \frac{1}{\sqrt[4]{(S_\star  - \ov S_\star) (U_{\star 1} - \ov U_{\star 1}) (U_{\star 2} - \ov U_{\star 2}) (U_{\star 3} - \ov U_{\star 3}) } } \prod_i \frac{C_{\alpha \beta}^{(i)}}{(T^i - \ov T{}^i)^{\frac{1}{2} }}
\end{equation}
with the model-dependent coefficients $C_{\alpha \beta}^{(i)}$ per two-torus defined as, 
\begin{equation}
C_{\alpha \beta}^{(i)} = \left(\frac{\Gamma(|\vartheta^i|)}{\Gamma(1-|\vartheta^i|)} \right)^{\lambda_i}.
\end{equation}
The parameter $\vartheta^i$, chosen in the range $0<|\vartheta^i|<1$, measures the angle between the two intersecting one-cycles on two-torus $T_{(i)}^2$ (in units of $\pi$), while the constant $\lambda_i = \pm 1$ takes into account the sign of $\vartheta^i$. 
\begin{figure}[h]
\begin{center}
\vspace{0.4in}
\begin{tabular}{c@{\hspace{0.4in}}c@{\hspace{0.4in}}c}
\includegraphics[scale=0.6]{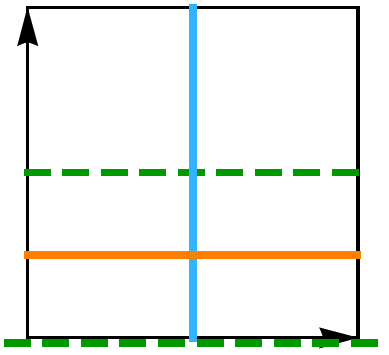}  \begin{picture}(0,0) \put(-45,76){$T_{(1)}^2$}\put(0,0){$\pi_1$} \put(-85,55){$\pi_2$} \end{picture}  & \includegraphics[scale=0.6]{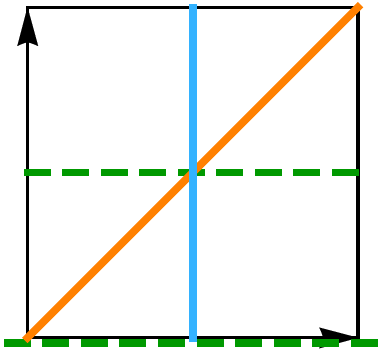}  \begin{picture}(0,0) \put(-45,76){$T_{(2)}^2$}\put(0,0){$\pi_3$} \put(-85,55){$\pi_4$} \end{picture}  & \includegraphics[scale=0.6]{3Cycle4}  \begin{picture}(0,0) \put(-45,76){$T_{(3)}^2$} \put(0,0){$\pi_5$} \put(-85,55){$\pi_6$}\end{picture}  
\end{tabular}
\caption{D6-brane configurations with codimension 6 intersection preserve a local ${\cal N}=1$ supersymmetry. The O6-planes are represented by the dashed, green lines.\label{Fig:Codim6Intersection}}
\end{center}
\end{figure}

\noindent In this case, the K\"ahler metric factorises into a homogeneous function of degree $-1$ in the complex structure moduli (upon inclusion of the dilaton) and a homogeneous function of degree $-\frac{3}{2}$ in the K\"ahler moduli. The model-dependent coefficients $C_{\alpha \beta}^{(i)}$ are homogeneous functions of degree 0 in the complex structure moduli and the K\"ahler moduli. When a three-cycle intersects with its orientifold image at three non-trivial angles, the corresponding K\"ahler metrics for the chiral matter states in the symmetric or antisymmetric representation take the same form as~\eqref{Eq:KaehlerBifundN1Munich}.
\end{itemize}

\bibliographystyle{JHEP2015}
\bibliography{refs_BilSUSYDFSZ}

\end{document}